\documentclass[11pt,a4paper]{article}
\usepackage[margin=24mm]{geometry}
\usepackage[T1]{fontenc}
\usepackage{lmodern,amsmath,amssymb,booktabs,longtable,array,graphicx,microtype}
\usepackage[unicode,hidelinks]{hyperref}
\usepackage[section]{placeins}
\title{Serial Rug-Pull-Like Exits on Pump.fun:\\Team-Level Evidence and Its Sensitivity to Team Construction}
\author{Fail Gafarov\\ Institute of Computational Mathematics and Information Technologies \\
Kazan Federal University \\
Kazan, Russia \\
fail.gafarov@kpfu.ru }
\date{October 2026}
\begin{document}
\maketitle
\begin{abstract}
We study repeated, team-associated selling around sharp collapses of Pump.fun tokens in several deliberately selected data windows, separating the measurement from claims of fraud, common ownership or realized profit. Because token prices derived from trade amounts are not fully reliable, we compare a price-based screen with a screen that uses only the bonding-curve reserve and SOL amounts; most price-based episodes are also found by the reserve screen. Episodes concentrate in groups well beyond a within-hour label-shuffle benchmark, and most of them are collapses sold by a single creator wallet. The recorded own-token cash flow of the identified groups is positive in most windows and negative in a few, but the sign of the negative results is fragile: it changes with the screen, with the definition of a group, with the length of the window and with trading costs of about two per cent per side. A daily SOL price covariate shows no association with the incidence of episodes. An agent-based simulation of a team exit on a bonding curve, extended with a latent attention to each token and calibrated to marginal moments of the data, shows that the team's result depends mainly on the attention that tokens attract, on fees and on the team's own trading, that continual trading by the team is costly and that reacting to the inflow of other traders adds little. The model reproduces the sign of group-level results but not the early exits that are common in the data. Missing token transfers, verified migration labels and human-labelled validation prevent identifying operator profit or fraud prevalence.
\end{abstract}

\section{Introduction}
Rapid token issuance makes the sequence of launches associated with a set of wallets a natural object of study. A single token can attract purchases, experience a large price increase and collapse shortly after early holders sell. The same creator or a funding-linked set of creators may repeat parts of this pattern on later tokens. Examining the sequence can reveal persistence that is invisible in a token-level chart. It can also expose losses on launches that would disappear from a narrative restricted to dramatic exits.

This paper asks three questions. First, what observable evidence supports a description of repeated early-holder exits around rapid market collapses? Second, what economic quantities can actually be recovered from the trading and SOL-transfer records? Third, how do the prevalence, concentration and cash-flow characteristics of such patterns differ across independently observed periods?

We use the term \emph{rug-pull-like exit} for a precisely defined observational pattern. The term does not establish deception, a legal classification, withdrawal of liquidity-provider capital, common wallet ownership or the absence of a legitimate protocol transition. Pump-and-dump selling, a change of trading venue and a liquidity withdrawal are different mechanisms. Their price charts can resemble one another when the available records contain only trades and a field described as bonding-curve reserve. Accordingly, a candidate episode in this paper is evidence of the measured pattern, not a confirmed instance of each of those mechanisms.

The empirical unit is a group \emph{within one data window}. We neither match identities between periods nor carry group history across windows. Every window has its own launch universe, funding graph, group identifiers, episode labels and financial aggregates. Cross-period comparison begins only after those separate analyses are complete. In particular, there is no pooled investment path or pooled operator-profit estimate.

The study builds on descriptive single-token measures such as transaction counts, turnover, price changes, reserve trajectories and wallet sales minus purchases. We add explicit grouping, price validation, censoring, finite-liquidity accounting and retrospective detector rules, and an agent-based simulation of a team launching a token on a bonding curve, which is the base model of the computational experiments in Section~\ref{sec:sim}.

The central empirical finding has two parts. First, selected exits and a group's wider recorded activity differ: positive net selling is part of the episode definition, positive cash flow on \emph{every other launch of that group} is not, and the latter is negative in aggregate in three periods of the price-based screen (two under the reserve-only screen) despite repeated qualifying exits. Second, this and the serial-group counts themselves depend materially on how teams are constructed and on the length of the observation window (Section~\ref{sec:robust}). We therefore report the primary results together with their sensitivity, and we do not treat any single construction as a validated ownership classification.

The contributions are fivefold. (i) A period-separated measurement of repeated team-associated exits with explicit denominators and censoring. (ii) A group-level cash-flow accounting that includes all launches of a group, whole-window flows, unsold inventory valued on the bonding curve and an empirical fee adjustment. (iii) A sensitivity analysis of team construction, window length and thin-market filtering, with bootstrap intervals and a label-permutation benchmark. (iv) An agent-based simulation of a team exit on a bonding curve, used for a systematic parameter study. (v) Released code, derived data and an audit sample for human labelling.

\section{Relation to previous research}
\textbf{Manipulation and pump-and-dump.} Coordinated pump-and-dump schemes in cryptocurrency markets have been measured and modelled in finance and computer science \cite{dhawan2023,gandal2018,lamorgia2020,lamorgia2023,clough2023,hu2023}, and the older literature on trade-based price manipulation \cite{allen1992,aggarwal2006,kyle2008} is the background for why we keep an observational label separate from a legal or intent-based one. Our episodes combine a sharp collapse with concentrated selling by a wallet set, which is a pattern of behaviour rather than a finding of manipulation.

\textbf{Rug pulls and scam tokens.} Scam and rug-pull tokens on decentralized exchanges have been studied through token features, transaction patterns and dataset construction \cite{xia2021,cernera2023,mazorra2022}. Huynh et al. study serial scammers and scam clusters on decentralized exchanges using transaction relationships and cluster-aware economic analysis \cite{huynh2025}; their emphasis on the group as an accounting unit motivates keeping internal transfers separate from external economic gains. Chen et al. distinguish several Solana rug-pull mechanisms and evaluate detection against a manually verified benchmark \cite{chen2026}, and a Solana rug-pull dataset has been released \cite{alhaidari2025}. These distinctions matter here: a price collapse with associated selling does not by itself identify authority abuse or liquidity withdrawal, and our checks validate calculations against the supplied records, not a ground-truth fraud label.

\textbf{Pump.fun and meme coins.} Szwajcok et al. investigate meme-coin production and coordinated behaviour on Pump.fun \cite{szwajcok2026}, and a prospective measurement study identifies groups of early buyers and fast-wallet teams on the same launchpad and tests whether following their earliest signal pays \cite{gafarov2026}; other work studies Solana meme-coin trends, market manipulation in the meme-coin ecosystem, early prediction of fraudulent launches and behavioural trace datasets \cite{mancino2025,mongardini2025,li2026rug,hu2026melt}. Measurement error in collector-generated graduation labels on Pump.fun \cite{kamat2026b} is a reminder that the migration status of a token needs its own validation, which is why we treat our ambiguity guard as a heuristic. Our narrower contribution is a period-separated measurement design that connects repeated launch-associated exits, finite-curve accounting and the inclusion of unsuccessful or unlabelled launches in selected groups' recorded cash flow, together with an explicit sensitivity analysis of the team construction. We do not claim to discover coordinated launch behaviour for the first time.

\textbf{Snipers and early buyers.} Sniper bots and extractable-value studies \cite{cernera2023www,cernera2025,qin2022} show that early buyers with privileged timing are common. We therefore require both early group buying and group selling in the collapse window, and we do not read a high group sale share as intent.

\textbf{Address clustering.} Heuristic address clustering has long been used on blockchains and is known to merge unrelated users or miss related ones \cite{reid2011,androulaki2013,harrigan2016,victor2020,beres2021}; identity multiplication is a general problem of open systems \cite{douceur2002}. Our funding-based groups inherit these limitations, and Section~\ref{sec:robust} reports how serial-group counts change under stricter constructions.

\textbf{Automated market makers.} The constant-product relation behind the finite-sale identity is standard in the analysis of automated market makers \cite{angeris2019,xu2023}. The current official Pump program documentation describes virtual and real reserves and the possibility of migration \cite{pumpdocs}; it supports the distinction between a curve state and immediately withdrawable money, but it does not establish historical fees, migration thresholds or contract versions for the 2024--2025 observations. Our theoretical model is consequently normalized and stylized; it is not a historical Pump program emulator.

\section{Economic framework}
\subsection{Finite sale proceeds and marked prices}
Consider a normalized constant-product curve with virtual SOL reserve $x$, virtual token reserve $y$ and invariant $k=xy$. Ignoring fees, a net SOL purchase of $u$ acquires
\begin{equation}
q=y-\frac{k}{x+u}.
\end{equation}
At a later state $(x,y)$, selling $q$ tokens yields the gross proceeds
\begin{equation}\label{eq:sale}
S(q;x,y)=x-\frac{xy}{y+q}=\frac{xq}{y+q}.
\end{equation}
The marginal price is $p=x/y$. Valuing the whole position at that price gives $pq=xq/y$, which exceeds the finite sale proceeds whenever $q>0$. In fact,
\begin{equation}
\frac{pq}{S(q;x,y)}=1+\frac{q}{y}.
\end{equation}
Thus, a high peak price does not imply that all holders can sell their positions at that price. The discrepancy grows with position size relative to the remaining virtual token reserve. A correct liquidation model updates the curve state after each sale and respects available real liquidity. It does not credit every wallet with the same peak quotation.

Virtual reserve is also not the same as funded real SOL reserve. In the simulation of Section~\ref{sec:sim} the virtual SOL reserve starts at $x_0=30$ SOL, the real reserve is $R=x-x_0$ up to fees retained on sales, and every token held by a participant was first bought. This keeps the real reserve nonnegative when the inventories are sold back, and it does not authorize withdrawing the initial virtual SOL. Mechanisms involving minted inventory, integer rounding, protocol graduation or a second pool would need additional states and constraints.

\subsection{Accounting at the group boundary}
Let $G$ denote the set of wallets assigned to a candidate group, $B_G$ its recorded SOL spent on buys and $S_G$ its recorded SOL received on sells. The directly calculated empirical measure is
\begin{equation}\label{eq:cash}
F_G=S_G-B_G.
\end{equation}
We call $F_G$ \emph{recorded trading cash flow}. It is not automatically realized profit. A seller might have acquired tokens before the dataset begins, received an unobserved token transfer, paid costs outside the trade table, traded elsewhere or retained unsold inventory. Conversely, a creator might receive issuance-related income absent from trade rows. Recovering these items requires additional evidence.

Internal SOL transfers have a different role. If wallet $a\in G$ sends SOL to wallet $b\in G$, the transfer changes individual balances but cancels when group balances are consolidated. Adding it to sales receipts would double-count movement of the same capital. External funding is not trading profit either: it may be capital supplied, borrowed money, a service payment or an unrelated transfer. Our transfer ledgers describe recorded flows across and within the group boundary separately; they are never added to equation~\eqref{eq:cash} as profit.

For a fully observed run with closed token inventories, write $\Pi_G$ and $\Pi_O$ for the changes in the group's and the other traders' cash, $\Phi$ for fees paid out of the system and $R_{\mathrm{end}}$ for the real reserve left in the pool. Conservation gives
\begin{equation}\label{eq:wealth}
\Pi_G+\Pi_O+\Phi+R_{\mathrm{end}}=0.
\end{equation}
The identity follows by summing all cash accounts, the real reserve and the fee sink before and after every transfer or trade. Internal transfers disappear from the sum. A buy moves cash into the reserve and the fee sink; a sale reverses part of that movement, and a fee kept in the reserve on sales remains in $R_{\mathrm{end}}$. A group's positive result in this closed model is financed by other traders' losses and reduced by fees. Repeated internal trading alone cannot create aggregate wealth.

Equation~\eqref{eq:wealth} is a theorem of the specified closed model. It is \emph{not} an empirically verified balance identity for the observed data. Missing inventory and venue information prevents substituting observed $F_G$ for $\Pi_G$ and claiming that the residual measures victim losses. We retain this separation throughout.

\section{Agent-based simulation of a team exit on a bonding curve}\label{sec:sim}
This section builds the computational experiments on an agent-based simulation of a team that launches a token on a Pump.fun-style bonding curve, funds a set of sybil wallets, buys early, trades against arriving other traders and finally sells. We verify the implementation against the original simulation trade by trade and use it to study how the team's result depends on the parameters. The simulation is a stylized mechanism model: its parameters are not estimated from the data of the empirical sections, and its outcomes must not be read as predictions of real profit.

\subsection{Model}
\textbf{Bonding curve.} The curve has virtual reserves $x$ (SOL) and $y$ (tokens) with invariant $k=xy$, starting at $x_0=30$ SOL and $y_0=10^{9}$ tokens (the constants of the original simulation; the public description of the protocol uses a slightly different token reserve). A buy of $s$ SOL pays a fee $\varphi_b s$ out of the system, the remainder $u=(1-\varphi_b)s$ enters the real reserve $R$ and the virtual reserve, $x\leftarrow x+u$, $y\leftarrow k/x$, and the buyer receives the tokens $q=y_{\text{old}}-y_{\text{new}}$. A sale of $q$ tokens sets $y\leftarrow y+q$, $x_{\text{new}}=k/y$, with gross proceeds $x-x_{\text{new}}$ and net proceeds $(1-\varphi_s)(x-x_{\text{new}})$, the finite-sale rule of equation~\eqref{eq:sale}. In the original simulation the sale fee stays in the real reserve; the experiments also run the variant in which it leaves the system. The recorded trade price is the SOL paid or received per token. In the protocol the curve is complete, and the token graduates, when the real reserve reaches about 85 SOL (about 793 of the 1,073 million virtual tokens sold); the real reserve of a curve therefore never exceeds 85 SOL, and all stopping reserves in the experiments are at most 85 SOL.

\textbf{Agents and timeline.} The creator holds $250$ SOL. It sends a random amount, uniform on $[1,3]$ SOL, to each of $N$ sybil wallets while keeping $20$ SOL, buys tokens for $b_0=2$ SOL, and $B$ randomly chosen sybil wallets spend all their SOL at launch. Then, second by second, between 2 and 5 transactions arrive. Each is made by an outside trader with probability $\pi_t=\min(c,t/\tau)$ ($0$ during the first second) and otherwise by a uniformly drawn team wallet. An outside transaction is by a new wallet with probability $\nu=0.3$ (endowment uniform on $[0.2,3]$ SOL) and otherwise by a uniformly drawn existing other trader. A trader holding tokens sells with probability $\rho_o$ (other traders) or $\rho_s$ (team); a sale is the whole holding or, with probability $0.3$, a uniform fraction of at least $0.2$; otherwise the trader buys with a uniform fraction (at least $0.2$) of its SOL. The run stops when the real reserve reaches $R_{\text{stop}}$ or after $T$ seconds. Then all remaining positions are sold in a forced sequence: team wallets and the creator first, other traders afterwards (the order of the original program), with the sybil wallets' cash swept to the creator.

\textbf{Outcomes and accounting.} The team result is $\Pi_G=$ final creator SOL minus its endowment, the other traders' result $\Pi_O$ is the sum of their final minus initial SOL, $\Phi$ the fees paid out and $R_{\text{end}}$ the real reserve left after all sales. Summing all cash accounts gives the closed-system identity \eqref{eq:wealth} in the form $\Pi_G+\Pi_O+\Phi+R_{\text{end}}=0$, which holds to at most $10^{-12}$ SOL in every run. Further outputs are the duration, the reserve at the stop, the peak price relative to the launch price, the team's share of circulating tokens at the stop, and whether the empirical episode screen of Section~\ref{sec:screen} flags the run.

\begin{table}[htbp]\centering\small\caption{Parameters of the simulation: baseline values (those of the supplied program, with 30 sybil wallets) and values used in the experiments.}\label{tab:simparams}
\resizebox{\linewidth}{!}{\begin{tabular}{lrl}\toprule Parameter & Baseline & Experiments \\ \midrule
Initial virtual SOL $x_0$ & 30 & 10--60 \\
Initial virtual tokens $y_0$ & $10^9$ & fixed \\
Buy and sell fee $\varphi_b=\varphi_s$ & 0.01 & 0--0.03 \\
Sell fee stays in reserve & yes & yes, no \\
Creator endowment $E$ (SOL) & 250 & fixed \\
Creator launch buy $b_0$ (SOL) & 2 & 0--20 \\
Sybil wallets $N$ & 30 & 6--110 \\
Early buyers $B$ & 5 & 0--20 \\
Team sell probability $\rho_s$ & 0.2 & 0--1 \\
Other-trader sell probability $\rho_o$ & 0.8 & 0--1 \\
Cap on the share of other traders $c$ & 0.3 & 0--0.7 \\
Share ramp $\tau$ (s) & 50 & 10--200 \\
New-other-trader probability $\nu$ & 0.3 & 0.1--0.8 \\
Other-trader wallet SOL & 0.2--3 & (0.1,1)--(2,10) \\
Stop reserve $R_{\text{stop}}$ (SOL) & 80 & 20--85 \\
Maximal duration $T$ (s) & 780 & 120--1500 \\
Transactions per second & 2--5 & max 3--12 \\
Final sale order & team first & team first, others first, random \\
\bottomrule\end{tabular}}\par\smallskip\begin{minipage}{.98\linewidth}\small Fixed in all experiments: sybil funding uniform on [1,3] SOL, creator keeps 20 SOL, partial-sale probability 0.3 with fraction in [0.2,1], buy fraction in [0.2,1], minimal trade 0.001 SOL.\end{minipage}\end{table}

\subsection{Implementation and verification}
The implementation separates the parameters, the curve with explicit fee and reserve accounting, the simulation, the episode screen and the experiment machinery. It uses one seeded random generator per run and the same sequence of draws as the original, so that both can be compared trade by trade. Three differences to the original were corrected or made explicit: a statement that raised an error when the reserve ended at a rounding-level negative value (it affected 10 of 84 comparison runs, all with zero fees); the funding of the team wallets can no longer give a negative amount; and the final sale order and the fee treatment are parameters. The forced sales are given a time step so that second-level measures can be applied.

Verification has four parts. (i) With identical seeds the implementation reproduces the original trade sequences in all 74 comparison runs in which the original does not fail (seven parameter settings, twelve seeds each), to a relative tolerance of $10^{-12}$. (ii) The conservation identity holds in all runs of all experiments (largest absolute residual $10^{-12}$ SOL), for both fee treatments and all sale orders. (iii) Unit tests check the curve round trip without fees, that a finite sale is worth less than its marginal valuation, determinism, the stopping rules and parameter validation. (iv) Every experiment point is replicated with deterministic seeds, so that all figures can be regenerated. A run takes about three milliseconds, which allows the 154\,400 runs reported here.

\subsection{Baseline behaviour}
\begin{figure}[htbp]\centering\includegraphics[width=.85\linewidth]{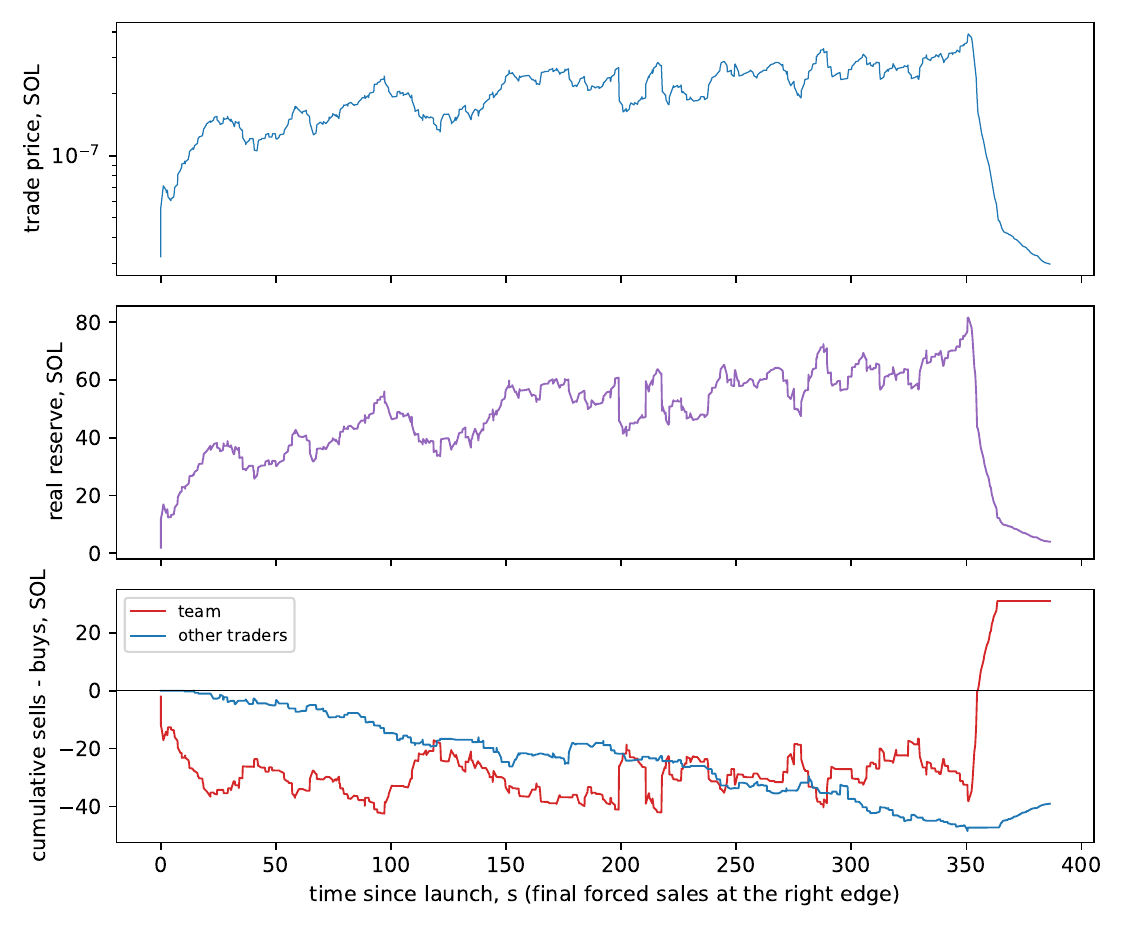}\caption{One baseline run: trade price, real reserve and cumulative sells minus buys of the team and of other traders. The vertical drop at the right is the forced final sequence.}\label{fig:simexample}\end{figure}

Figure~\ref{fig:simexample} shows one baseline run and Table~\ref{tab:simbase} the distribution over 10\,000 runs. The team's launch purchases lift the price, other traders arrive and buy, team wallets trade among themselves and with other traders, and the reserve grows until it reaches the stop; the forced team sales then push the price and reserve down within seconds, and other traders, who sell last, receive the rest. In the baseline every run ends with a positive team result (mean 29.6 SOL, return 0.48 on the SOL committed), financed by other traders' mean loss of 39.8 SOL and fees of 5.1 SOL. 100\% of the runs stop at the reserve cap after on average 261 seconds.

\begin{table}[htbp]\centering\small\caption{Baseline simulation (10\,000 runs).}\label{tab:simbase}
\begin{tabular}{lrrrr}\toprule Outcome & Mean & Median & 5\% & 95\% \\ \midrule
Team profit (SOL) & 29.6 & 29.5 & 21.1 & 38.6 \\
Team return on committed SOL & 0.481 & 0.476 & 0.330 & 0.645 \\
Other traders' net result (SOL) & -39.8 & -39.8 & -50.5 & -29.3 \\
Duration before the final sales (s) & 261 & 259 & 168 & 360 \\
Reserve at stop (SOL) & 81.8 & 81.2 & 80.1 & 85.3 \\
Peak price / launch price & 11.98 & 11.91 & 11.41 & 12.81 \\
Team share of circulating tokens at stop & 0.685 & 0.685 & 0.604 & 0.766 \\
Fees paid out (SOL) & 5.12 & 5.10 & 3.40 & 6.95 \\
Other-trader wallets & 80 & 80 & 55 & 107 \\
\bottomrule\end{tabular}\par\smallskip\begin{minipage}{.96\linewidth}\small Share of runs with positive team profit: 100.0\%; stopped by the reserve cap: 100.0\%; flagged by the empirical episode screen without the migration guard: 100.0\%, with the guard: 0.0\%.\end{minipage}\end{table}

\subsection{One-at-a-time sensitivity}
\begin{figure}[htbp]\centering\includegraphics[width=\linewidth]{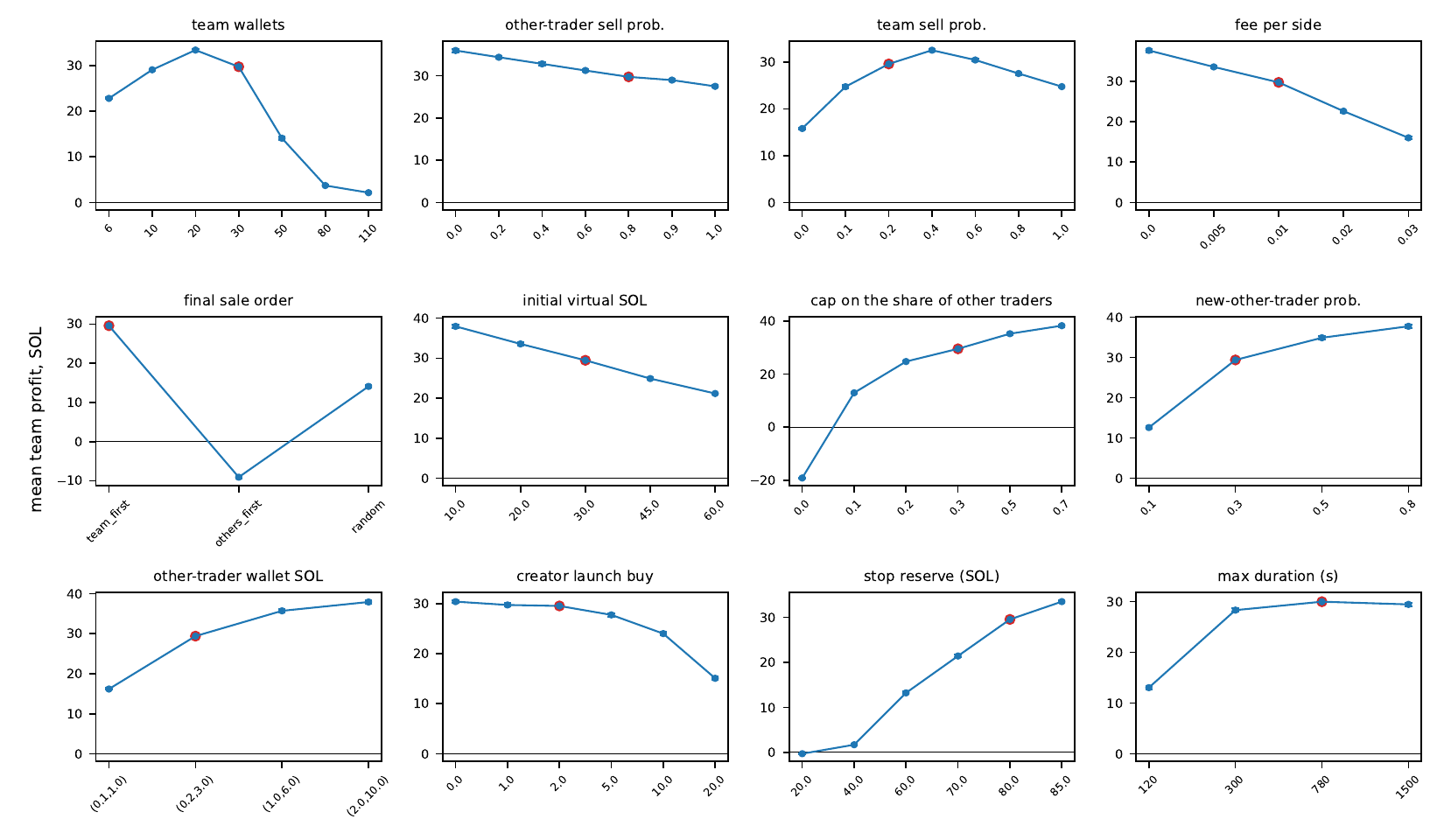}\caption{Mean team profit when one parameter is varied (red: baseline value; bars: 95\% intervals).}\label{fig:simoat}\end{figure}

Each of 16 parameters is varied alone over the values of Table~\ref{tab:simparams}, with 1\,000 replications per value; Figure~\ref{fig:simoat} shows the mean team profit with 95\% intervals, Appendix Figure~\ref{fig:simoatoth} the other traders' result, Appendix Table~\ref{tab:simoat} the extremes, and Appendix Table~\ref{tab:simoatfull} every point. The number of early buyers, the treatment of the sale fee, the speed at which the share of other traders ramps up and the transaction rate move the mean profit by at most 2.9 SOL and are not plotted.

\textbf{Size and behaviour of the team.} Team size matters non-monotonically: the mean profit is 22.8 SOL with 6 wallets, peaks at 33.4 (at 20 wallets) and falls to 2.2 with 110. Section~\ref{sec:team} examines the mechanism. The number of early buyers has practically no effect (29.8 to 29.8 SOL), possibly because early buyers are team wallets that would trade anyway in the main loop. The team's own selling propensity has an interior optimum (15.8 SOL if team wallets never sell before the final sequence, 32.5 (at 0.4) at the best value, 24.7 if they always sell). The creator's launch purchase lowers the profit for large values (30.4 at 0 SOL, 15.1 at 20 SOL).

\textbf{Other traders.} The model needs other traders: with no share of other traders the team loses 19.2 SOL on average, with a cap of 0.1 it earns 12.9, and with 0.7 38.2. More new wallets, larger endowments of other traders and fewer other-trader sales all raise the team profit (new-wallet probability 0.1 to 0.8: 12.6 to 37.8; wallet size of other traders from (0.1,1) to (2,10) SOL: 16.2 to 37.9; other-trader sale probability 0 to 1: 36.0 to 27.5). The speed at which the share of other traders ramps up (10 to 200 seconds) and the transaction rate make almost no difference.

\textbf{Curve, fees and stopping rule.} Fees reduce profit by about 7.2 SOL per percentage point of fee on each side (37.6 at no fee, 16.0 at 3\%); keeping the sale fee in the reserve rather than paying it out changes the mean by 2.9 SOL. A larger initial virtual SOL reserve (a stiffer curve) lowers profit (38.0 at 10, 21.2 at 60 SOL). The stopping reserve is, with the share of other traders, the strongest single driver: the mean profit rises from $-$0.3 SOL at a 20 SOL stop to 33.5 at 85 SOL (graduation), consistent with more money of other traders being in the pool when the team sells; the maximal duration matters only when it is so short that the run ends by time (120 s: 13.0 SOL).

\textbf{Order of the final sales.} The largest structural effect is the order of the forced final sequence. With the original order (team first) the mean profit is 29.6 SOL; if other traders sell first the team loses 9.1 SOL; with a random order the profit is 14.1 SOL. Much of the baseline profit is therefore the advantage of exiting before the other traders, not of the preceding trading, and the model's headline number depends on this design choice.

\subsection{Team size in detail}\label{sec:team}
\begin{figure}[htbp]\centering\includegraphics[width=.95\linewidth]{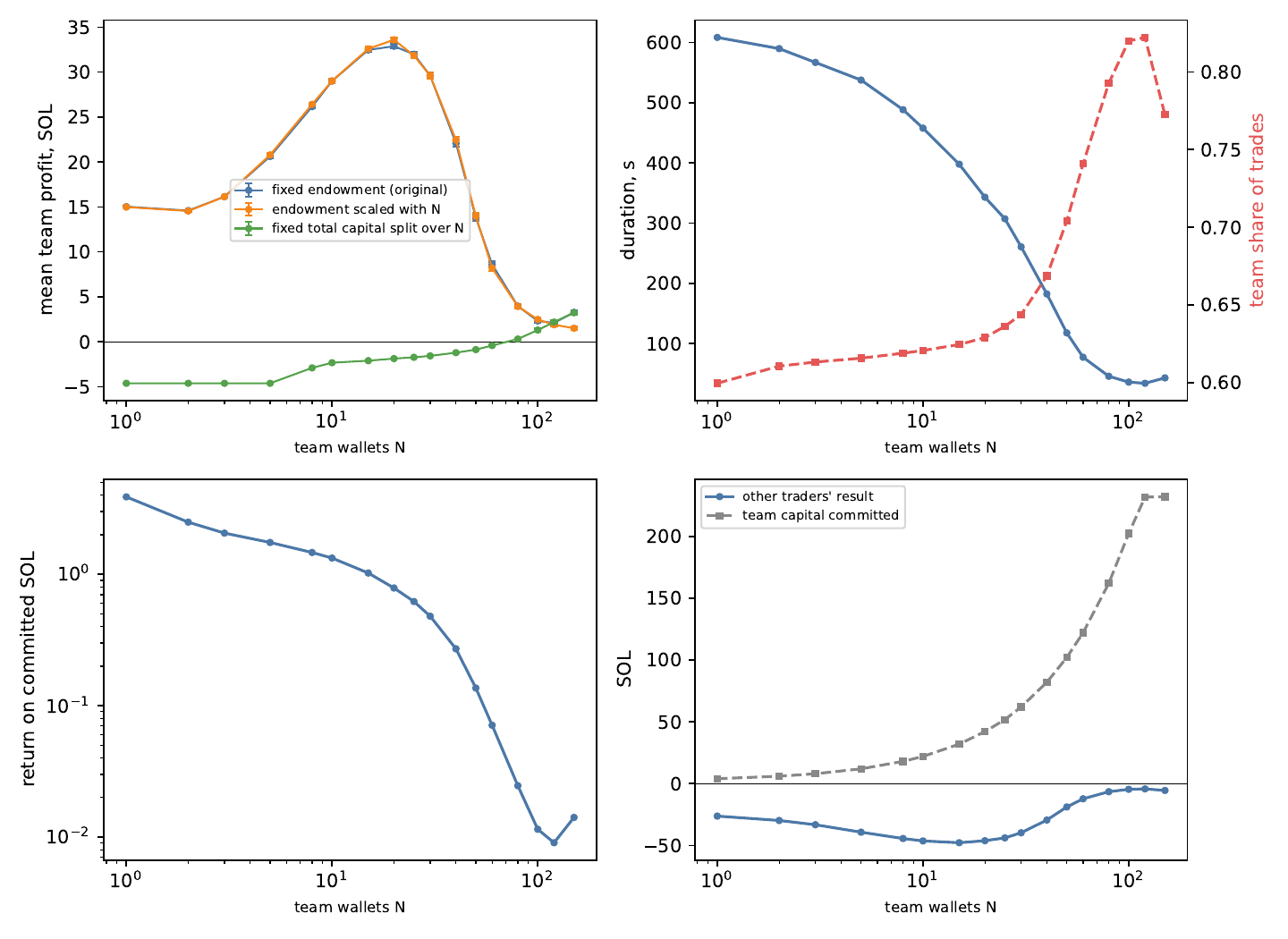}\caption{Team size. Top left: mean team profit in the three designs (95\% intervals). Top right: duration and team share of trades (fixed endowment). Bottom left: return on committed SOL. Bottom right: other traders' result and committed team capital.}\label{fig:simteam}\end{figure}

Team size is the parameter whose effect is least obvious (the mean profit peaks at 20 wallets and falls to about 2 SOL at 110), so we examine it separately: 17 sizes from 1 to 150 with 1,000 runs each, in three designs that separate the number of wallets from the capital they hold. In the \emph{fixed endowment} design the creator has the original 250 SOL, so that with many wallets the later ones are funded less or not at all. In the \emph{scaled endowment} design the endowment grows with $N$ and every wallet receives its full random amount. In the \emph{fixed total capital} design the same 230 SOL are split evenly over $N$ wallets. Two further grids cross $N$ with the stopping reserve and the cap on the share of other traders (300 runs per cell). Figure~\ref{fig:simteam} and Appendix Table~\ref{tab:simteam} show the results.

\textbf{The decline is not caused by the creator running out of money.} With the fixed endowment the mean profit rises from 15.0 SOL for a single wallet to a maximum of 32.9 at 20 wallets and falls to 2.1 at 120 wallets, where it is positive in only 78\% of the runs. Scaling the endowment so that every wallet is fully funded gives practically the same curve (the two designs differ by less than 1 SOL for every size up to 120 wallets and by 1.8 SOL at 150, where the fixed endowment funds only about 115 wallets), so the exhaustion of the endowment is not the explanation of the decline.

\textbf{The team fills the curve itself.} As $N$ grows the committed capital rises from 4 to 232 SOL, the team's share of all trades from 0.60 to 0.82, and the run ends much sooner, after 608 seconds for one wallet and 36 seconds for 100 wallets. Because other traders arrive with a probability that grows with time, a large team reaches the stopping reserve before many of them have come, and other traders' loss shrinks from 47.8 SOL (15 wallets) to 4.3 SOL (120 wallets); there is little money of other traders left to take. The return on the SOL committed falls steadily from 3.9 for one wallet to 0.01 for 100 wallets: small teams use their capital far more effectively, and the maximum of the total profit is a compromise between capital and return.

\textbf{What matters is capital relative to the stopping reserve.} In the fixed-total-capital design the same capital is deployed at launch by at most five early buyers; with five or fewer wallets all of it enters the curve at once, the stop is reached immediately and the team loses only fees ($-$4.6 SOL). Profit then rises with the number of wallets, because a smaller share of the capital is spent at launch, and becomes positive only above about 80 wallets (3.2 SOL at 150). Across stopping reserves of 40, 60, 70, 80 and 85 SOL the best team size is 10, 10, 20, 20 and 20 wallets, which corresponds to a committed capital of 0.37 to 0.60 times the stopping reserve (Appendix Figure~\ref{fig:simteamgrid}, middle). A larger share of other traders raises the profit at every size and moves the best size to 20--30 wallets; with few other traders (cap 0.1) no size earns more than 13 SOL and teams of 80 or more lose.

These statements concern wallets that cooperate on one token under this stopping rule and arrival process. They should not be transferred to the group sizes of the empirical sections, where a group is inferred from creators and funding links, usually has one launch per creator wallet, and its wallets are not known to trade the same token.

\subsection{Interactions}
\begin{figure}[htbp]\centering\includegraphics[width=.95\linewidth]{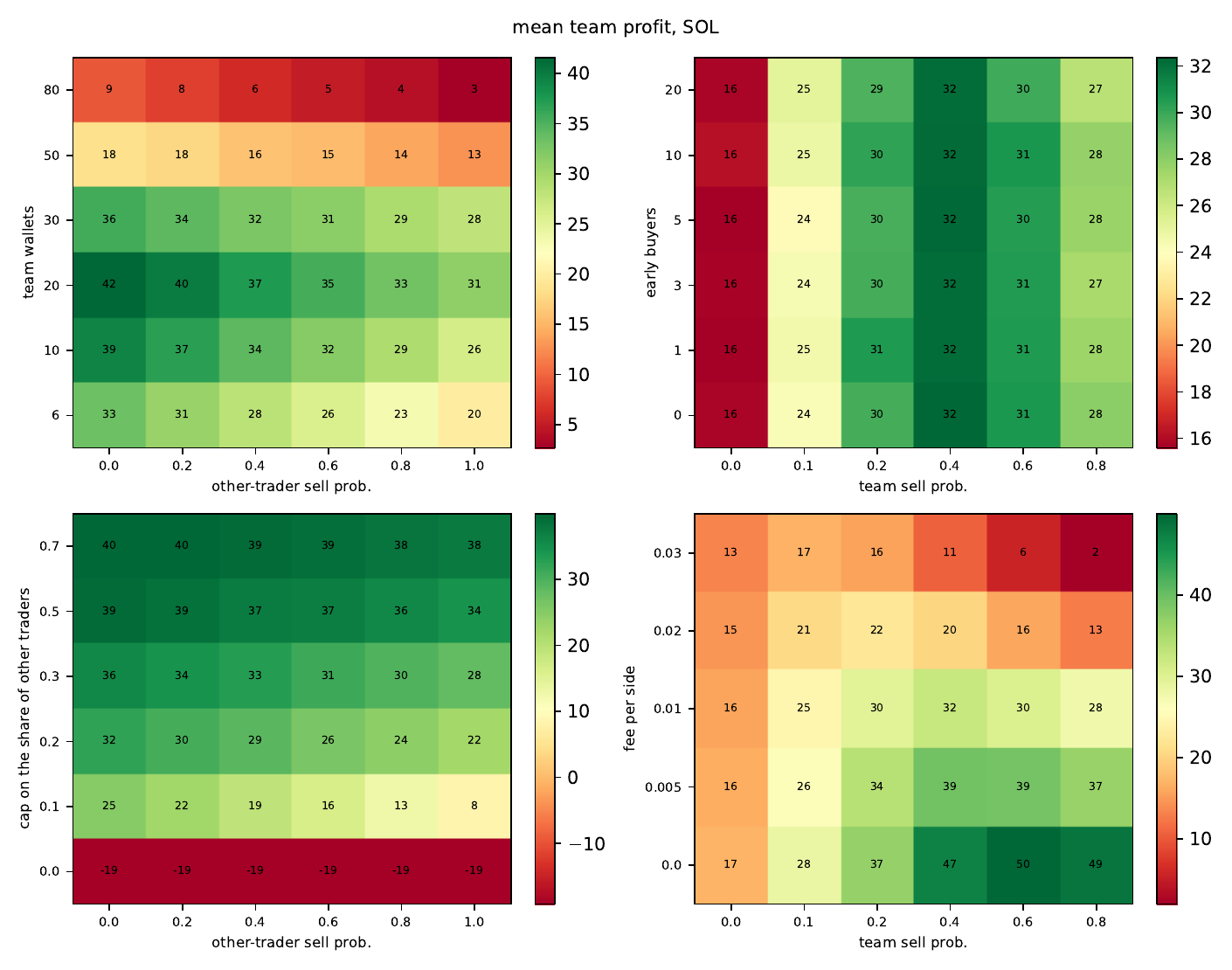}\caption{Mean team profit (SOL) in four two-way grids; numbers are cell means.}\label{fig:simgrid}\end{figure}

Figure~\ref{fig:simgrid} crosses pairs of parameters (300 replications per cell). The team result is positive in all cells of three grids; only the other trader-share grid has cells with a mean loss (those with no other traders, 17\% of its cells), so selling propensities, team size, early buying and fees move the size of the profit but not its sign, in contrast to the market-structure parameters. The best cell of the team-size grid has 41.6 SOL (20 wallets, other traders never selling) and the worst 2.7 (80 wallets, other traders always selling); fees of 3\% with a team sell probability of 0.8 leave only 1.9 SOL.

\subsection{Global sensitivity}
\begin{figure}[htbp]\centering\includegraphics[width=\linewidth]{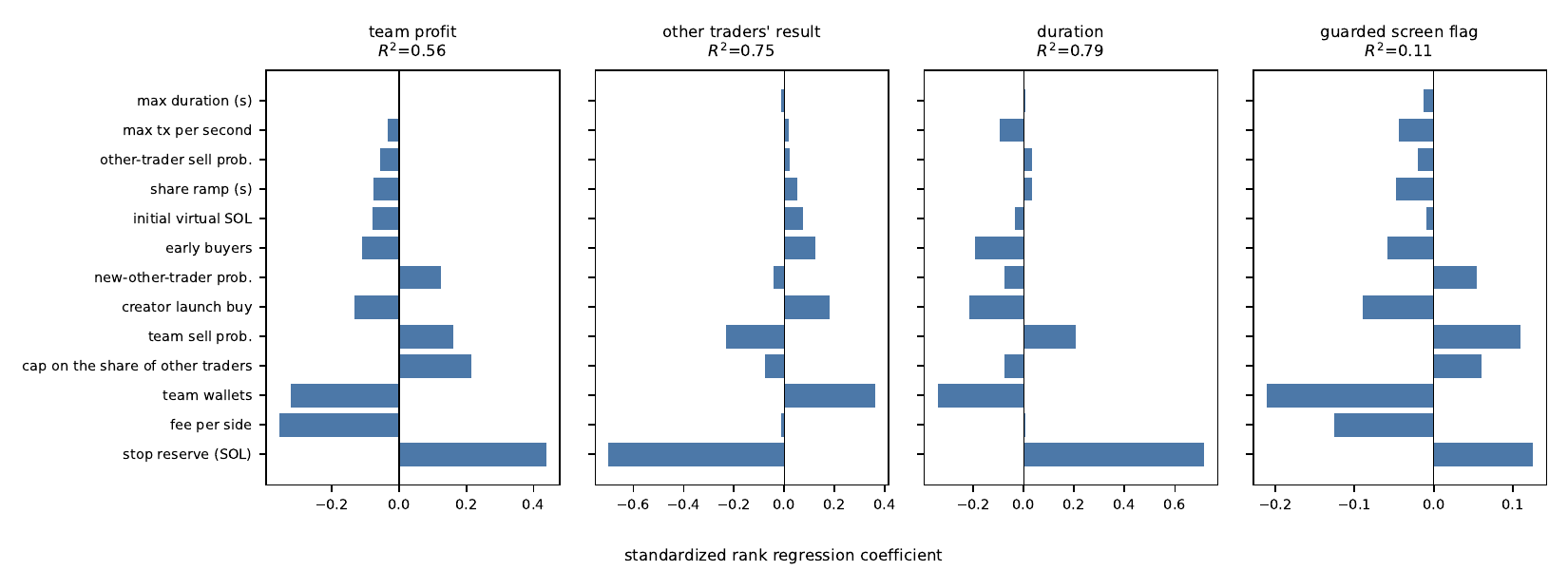}\caption{Standardized rank regression coefficients in the Latin-hypercube design for four outcomes.}\label{fig:simsrrc}\end{figure}

To rank the parameters jointly, a Latin-hypercube design of 20\,000 points varies 13 parameters over the ranges of Appendix Table~\ref{tab:simsrrc} at once. The team profit is positive in 48\% of the design points; a rank regression explains 0.56 of the variance of its ranks. The standardized coefficients (Figure~\ref{fig:simsrrc}) rank the stopping reserve first (+0.44), followed by the cap on the share of other traders (+0.22), the fee ($-$0.36) and team size ($-$0.32); the team's selling propensity and the arrival of new other traders have smaller positive effects, and the initial virtual reserve a negative one.

\subsection{Link to the empirical episode screen}\label{sec:screen}
The reserve-only screen of Section~\ref{sec:reserve} (an implied price at least twice the launch price, an implied-price fall of at least 80\% and a reserve fall of at least 50\% within ten minutes, team net sales of at least 1 SOL and at least half of the sales in the window) is applied to the simulated trades exactly as to the data; it uses only the real reserve and SOL amounts. Without the migration guard, 100\% of the baseline runs are flagged: the simulated exit is rug-pull-like by construction. With the guard, which treats a recorded reserve of at least 70 SOL before the drop as ambiguous, none is flagged, because the baseline stops at 80 SOL; this shows how much of the empirical count depends on that heuristic (Section~\ref{sec:robust}). Over the Latin-hypercube design 40\% of the runs are flagged without and 20\% with the guard; the guarded flag is more frequent for low stop reserves, small teams, a small initial virtual reserve and a higher team selling propensity (Appendix Table~\ref{tab:simsrrc}).

Table~\ref{tab:simscreen} compares descriptors of the flagged baseline runs with the empirical medians. The simulated peaks are higher (13.5 against 7.6 times the launch price implied by the reserve), built more slowly (6.8 against 2.5 minutes) and the reserve falls less deeply by the time of the first qualifying price drop (0.78 against 0.99), while the team's share of sales in the window is similar (0.89 against 0.96) and the drop delay is of the same order (12 against 22 seconds). The model therefore reproduces the qualitative pattern the screen looks for, but it has not been calibrated to the data and should not be used to infer prevalence.
\begin{table}[htbp]\centering\small\caption{Episodes found by the empirical screen: baseline simulation (guard off) against the pooled empirical medians of the reserve-only screen.}\label{tab:simscreen}
\begin{tabular}{lrr}\toprule Descriptor & Simulation median & Empirical median \\ \midrule
Peak / launch price (implied by the reserve) & 13.47 & 7.59 \\
Drop delay (s) & 12.00 & 22.00 \\
Team share of sales in window & 0.89 & 0.96 \\
Peak age (min) & 6.80 & 2.52 \\
Reserve fall (lower bound) & 0.78 & 0.99 \\
Implied price fall & 0.81 & 0.85 \\
\bottomrule\end{tabular}\par\smallskip\begin{minipage}{.9\linewidth}\small Empirical: reserve-only screen\, nine windows; simulation: 10\,000 flagged runs.\end{minipage}\end{table}

\subsection{Limits of the original model, and an extended model}\label{sec:ext}
The original model is a mechanism model and, as such, internally consistent, but four of its features decide its results more than the parameters we vary. (i) It is conditioned on success: every token attracts other traders and reaches the stop reserve, whereas in the data about two thirds of all launches never reach 5 SOL of real reserve and only 1\% graduate. (ii) Everything is forced: after the stop, all positions are sold in a fixed order, so other traders realize their losses and the order of the sales alone can change the sign of the team's result. (iii) The other traders are noise traders: their arrival does not depend on the price path, their selling does not depend on the drawdown, their sizes do not have a heavy tail, and nobody snipes the launch. (iv) The team neither observes nor decides: it trades at random and exits by a reserve threshold, and the model has no cost of repeating the exercise on many tokens. In addition, the curve constants differ from the public description (virtual token reserve of $10^9$ instead of $1.073\times10^9$), fees are partly retained in the reserve, and stop reserves above the graduation level of 85 SOL are impossible on the real curve (the experiments above are limited to 85 SOL for that reason). Token prices are not used anywhere in the extended model or in its screen: outcomes are measured in SOL, and the screen of Section~\ref{sec:reserve} uses the real reserve.

The extended model keeps the structure (curve, team wallets, other traders) and changes these points. A token draws a latent \emph{attention} level $A$ from a lognormal distribution; other traders arrive as a Poisson process with rate $\lambda_t=\lambda_0 A\min(1,t/\tau)e^{-t/\tau_d}\min(1+\kappa r_t,6)$, where $r_t$ is the recent rise of the price implied by the reserve (momentum), so most tokens attract almost nobody. Other traders have heavy-tailed endowments (truncated lognormal); another trader holding tokens sells with probability $\min(1,h_0+h_1d_t)$, where $d_t$ is the fall from the running price peak (panic selling); with probability $q$ a token is also bought in its first seconds by snipers who sell early. The team (creator and $N$ sybil wallets) buys at launch, trades in the background at a Poisson rate and exits by a rule: reserve threshold, price multiple, drawdown from the peak or a fixed time, with a maximal holding time of five minutes. Nothing is forced: the run ends at graduation (85 SOL), after a post-exit period or at the time limit, other traders keep their tokens, and every outcome is reported as cash (SOL received minus SOL paid, fees and a launch cost of 0.02 SOL included) and as cash plus the value of the remaining tokens on the final curve state, either each holder selling alone (optimistic) or all holders selling together (pooled, pessimistic). Fees of 1\% leave the system on both sides, and the same closed-system identity holds for cash (largest residual of the order of $10^{-12}$ SOL).

\subsection{Calibration}\label{sec:calibration}
\begin{figure}[htbp]\centering\includegraphics[width=.55\linewidth]{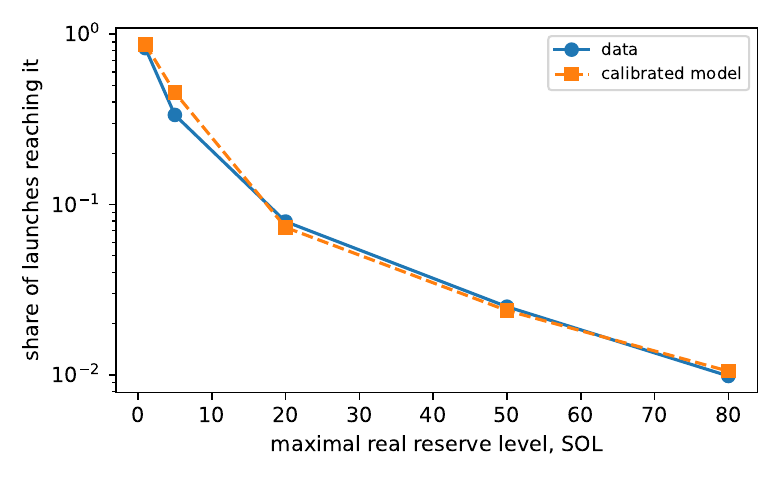}\caption{Share of launches reaching a given maximal real reserve: data (nine windows pooled) and the calibrated organic tokens.}\label{fig:reach}\end{figure}

The model has too many parameters to fix by hand, and the side of the other traders and the team side are informed by different data. We calibrate in two stages by approximate Bayesian computation (random search from a wide prior, then a refinement round around the best draws; Appendix Table~\ref{tab:calparams}). In stage A, \emph{organic} tokens (one wallet, no team trading, no exit) fit the side of the other traders to the unconditional distribution of the maximal real reserve and of the number of trades per launch in the nine windows (Figure~\ref{fig:reach}). In stage B, team tokens fit the team side to the median descriptors of the episodes found by the reserve-only screen: peak age, drop delay, team share of sales, implied rise, reserve fall and implied price fall. Table~\ref{tab:calfit} reports the fit. The tail of the reach distribution is matched closely (share of launches reaching 20, 50 and 80 SOL: 7.3\%, 2.4\% and 1.0\% against 7.9\%, 2.5\% and 1.0\%), the share reaching 5 SOL is too high (45\% against 34\%), and the episode descriptors are within about 20\% of the data except the implied rise (5.6 against 7.6). Only marginal moments are matched; no moment of the team's cash flow, of the team's trading after launch or of its exits before the peak was used. This matters below. The calibration selects a large team (51 wallets with up to 2.0 SOL each) that trades in the background at 0.85 transactions per second and exits when the reserve reaches 41 SOL, so that it deploys about 68 SOL per token.

\begin{table}[htbp]\centering\small\caption{Fit of the calibrated extended model to the empirical moments it was fitted to.}\label{tab:calfit}
\begin{tabular}{lrr}\toprule Moment & Model & Data \\ \midrule
P(max reserve $\geq$ 1 SOL) & 0.870 & 0.827 \\
P(max reserve $\geq$ 5 SOL) & 0.454 & 0.335 \\
P(max reserve $\geq$ 20 SOL) & 0.073 & 0.079 \\
P(max reserve $\geq$ 50 SOL) & 0.024 & 0.025 \\
P(max reserve $\geq$ 80 SOL) & 0.010 & 0.010 \\
P($\geq$ 10 trades) & 0.605 & 0.693 \\
P($\geq$ 50 trades) & 0.217 & 0.194 \\
Median trades per launch & 14 & 14 \\
Episode peak / launch price implied by the reserve (median) & 5.56 & 7.59 \\
Episode drop delay, s (median) & 19.00 & 22.00 \\
Episode team share of sales (median) & 1.00 & 0.96 \\
Episode peak age, min (median) & 2.09 & 2.52 \\
Episode reserve fall (median) & 0.99 & 0.99 \\
Episode implied price fall (median) & 0.81 & 0.85 \\
\bottomrule\end{tabular}\par\smallskip\begin{minipage}{.85\linewidth}\small Reach and trade moments: unconditional over all launches with a complete first hour, nine windows pooled; episode moments: episodes found by the screen with the migration guard on. Model episodes in the best draws: 1596 of 2500 team launches.\end{minipage}\end{table}

\subsection{Results of the extended model}\label{sec:extresults}
\begin{figure}[htbp]\centering\includegraphics[width=.95\linewidth]{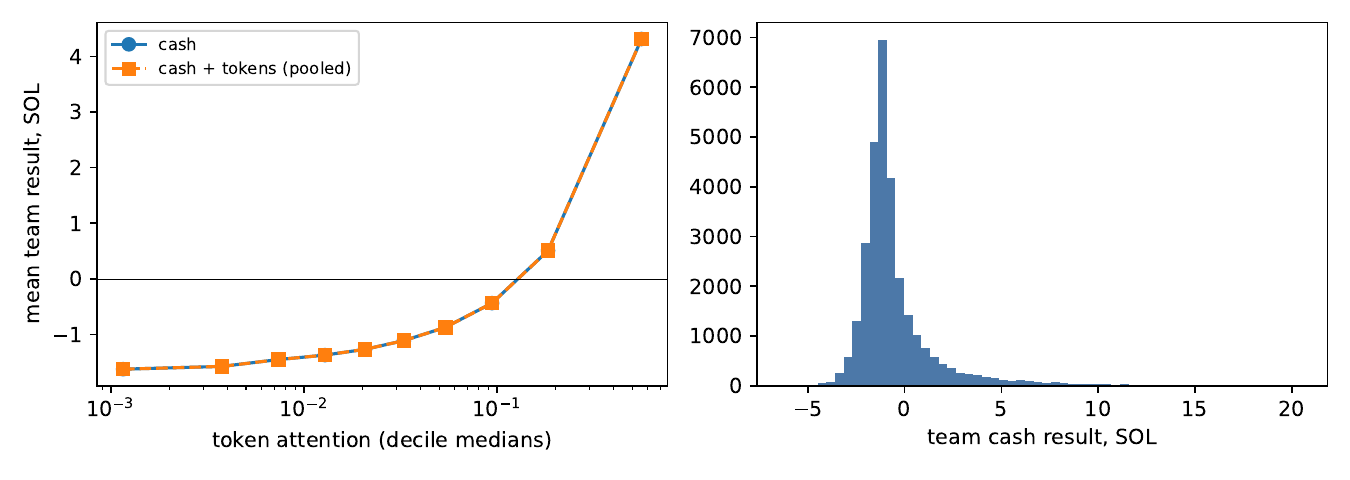}\caption{Calibrated extended model. Left: mean team result by attention decile (cash, and cash plus tokens on a pooled sale). Right: distribution of the team's cash result.}\label{fig:x2base}\end{figure}

\textbf{One launch.} Table~\ref{tab:x2base} and Figure~\ref{fig:x2base} show one team launch in the calibrated model, over 30\,000 runs. The team's mean cash result is $-$0.49 SOL per launch (median $-$1.04) and it is positive in 22\% of the launches; other traders lose 2.29 SOL in cash, but only 0.90 SOL once their remaining tokens are valued, because most of their loss is unrealized. The result is a lottery on attention: the mean team cash rises from $-$1.62 SOL in the lowest to 4.31 SOL in the highest attention decile, and the share of profitable launches from 0.2\% to 96\%. The screen flags 69\% of the team launches, which is not a prevalence: in this model every team token reaches the exit reserve with the team's own money.

\begin{table}[htbp]\centering\small\caption{Calibrated extended model: one team launch (30\,000 runs).}\label{tab:x2base}
\begin{tabular}{lrrrrr}\toprule Outcome (SOL) & Mean & Median & 5\% & 95\% & Share $>$0 \\ \midrule
Team cash result & $-$0.5 & $-$1.0 & $-$2.5 & 3.7 & 22\% \\
Team cash + tokens (individual sale) & $-$0.5 & $-$1.0 & $-$2.5 & 3.7 & 22\% \\
Team cash + tokens (pooled sale) & $-$0.5 & $-$1.0 & $-$2.5 & 3.7 & 22\% \\
Other traders' cash result & $-$2.3 & $-$0.5 & $-$10.3 & 0.8 & 23\% \\
Other traders' cash + tokens (pooled sale) & $-$0.9 & $-$0.2 & $-$5.2 & 0.9 & 30\% \\
Peak real reserve (SOL) & 41.4 & 41.1 & 40.7 & 42.6 & \\
\bottomrule\end{tabular}\par\smallskip\begin{minipage}{.9\linewidth}\small Cash: SOL received minus SOL paid, fees and launch cost included. Tokens still held at the end are valued at the final curve state, either each holder selling alone (optimistic) or all holders selling together (pooled, pessimistic).\end{minipage}\end{table}

\textbf{Exit rules.} Appendix Table~\ref{tab:x2exit} compares exit rules. Holding without any exit leaves the team with unsold tokens and a cash result of $-$42.02 SOL. Among the exits, the rules that leave early do best: exiting when the price has doubled gives 0.48 SOL (43\% of the launches positive; the screen flags none of them), against $-$0.55 for the calibrated reserve rule at 40 SOL and $-$0.20 at 80 SOL, which waits for the other traders but pays for the team's trading in the meantime and leaves the other traders 3.54 SOL worse off. A reserve exit at 10 SOL is the best single setting in the sweep of Appendix Table~\ref{tab:x2oat} (0.80 SOL, 56\% positive). A large exit leaves a visible collapse; an early one does not.

\begin{figure}[htbp]\centering\includegraphics[width=.95\linewidth]{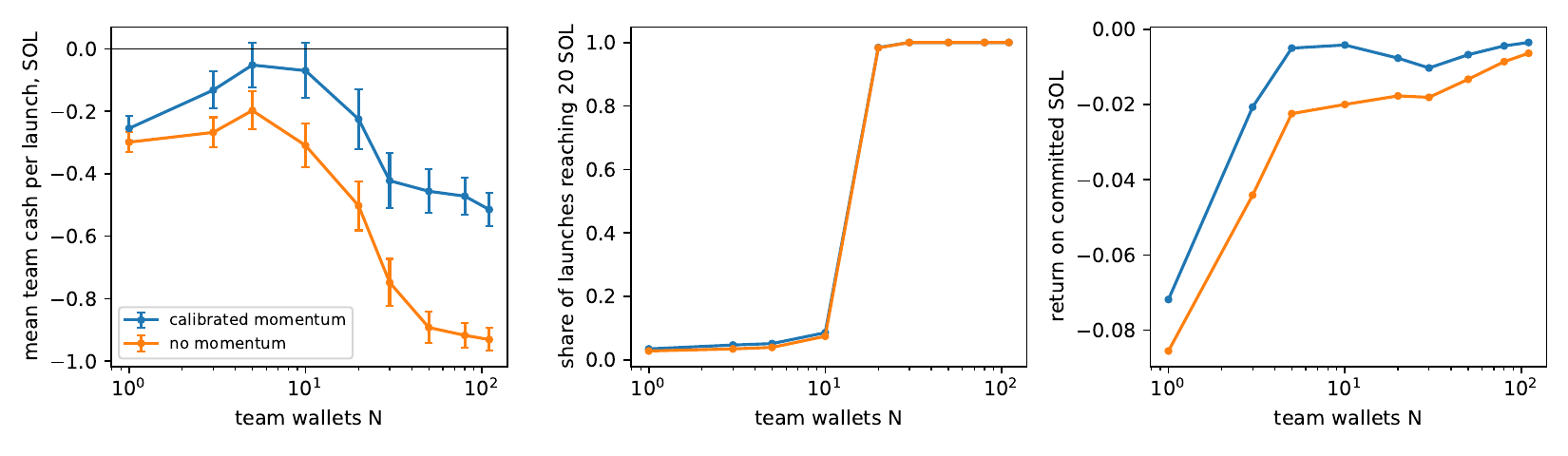}\caption{Team size in the calibrated extended model, with the calibrated momentum and without it (endowment scaled with the number of wallets).}\label{fig:x2team}\end{figure}

\textbf{Team size.} The interior optimum of the original model disappears. Under the calibrated side of the other traders the mean team cash is between $-$0.05 and $-$0.51 SOL for every team size from 1 to 110 wallets (Figure~\ref{fig:x2team}), highest with five wallets and falling slowly as the team's own capital grows (from 9 to 145 SOL), and the share of launches reaching 20 SOL is close to one for 20 or more wallets and small for fewer, because the team's own money fills the curve. Momentum in the other traders' arrival changes the picture little. The optimum of the original model therefore depended on its fixed arrival process and on the team's capital relative to the stop reserve; it should not be read as a property of real teams.

\begin{figure}[htbp]\centering\includegraphics[width=.95\linewidth]{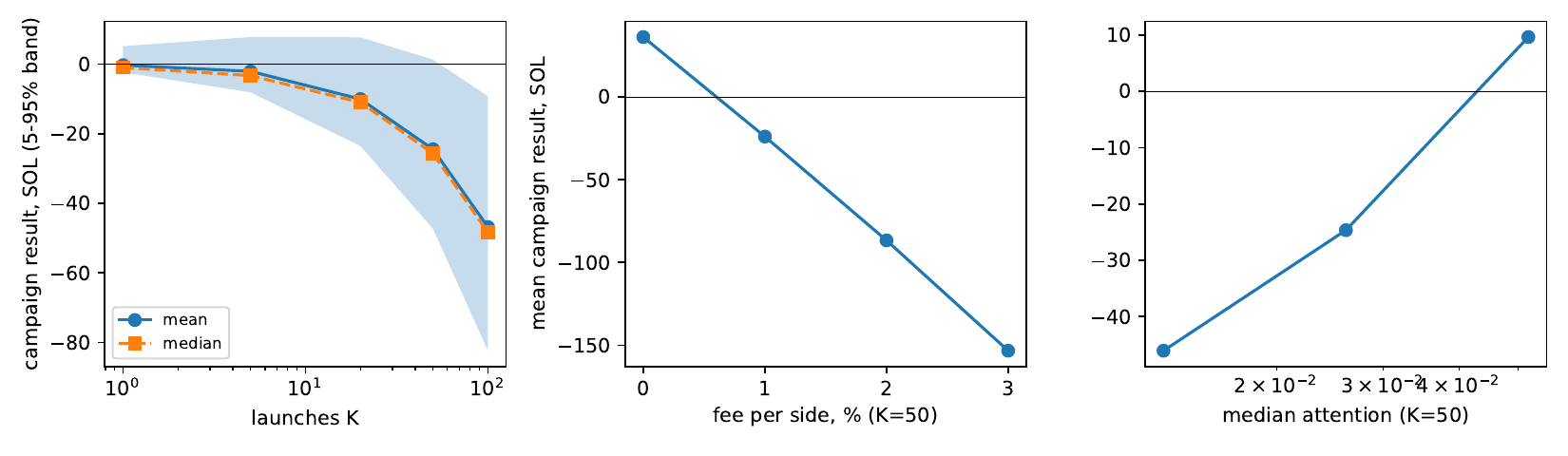}\caption{Campaigns of $K$ launches: distribution of the result by $K$, and the dependence on the fee and on the median attention (50 launches).}\label{fig:x2camp}\end{figure}

\textbf{Repeated launches.} A team that repeats the strategy $K$ times starts every launch from its current cash. Appendix Table~\ref{tab:x2camp} and Figure~\ref{fig:x2camp} show that the mean result is $-$0.29 SOL for one launch, $-$10.05 for 20 and $-$46.81 for 100, and the share of campaigns with a positive result falls from 24\% to 2\%; nobody is ruined, because the losses per launch are small. The result depends on the fee and on the attention the tokens draw far more than on anything else: with 50 launches, a fee of 0 gives 36.2 SOL (all campaigns positive), 1\% gives $-$24.0 SOL (7\% positive) and 2\% gives $-$86.7 SOL; doubling the median attention turns the mean to 9.6 SOL (67\% positive) and halving it gives $-$46.0 SOL. Under the effective cost of about 2\% per side measured in Section~\ref{sec:fees}, a team that trades as the calibrated team does loses on average; its profit comes from a minority of launches with high attention. This is consistent in direction with the negative group-level flows of several windows, but the model was not fitted to them.

\begin{figure}[htbp]\centering\includegraphics[width=\linewidth]{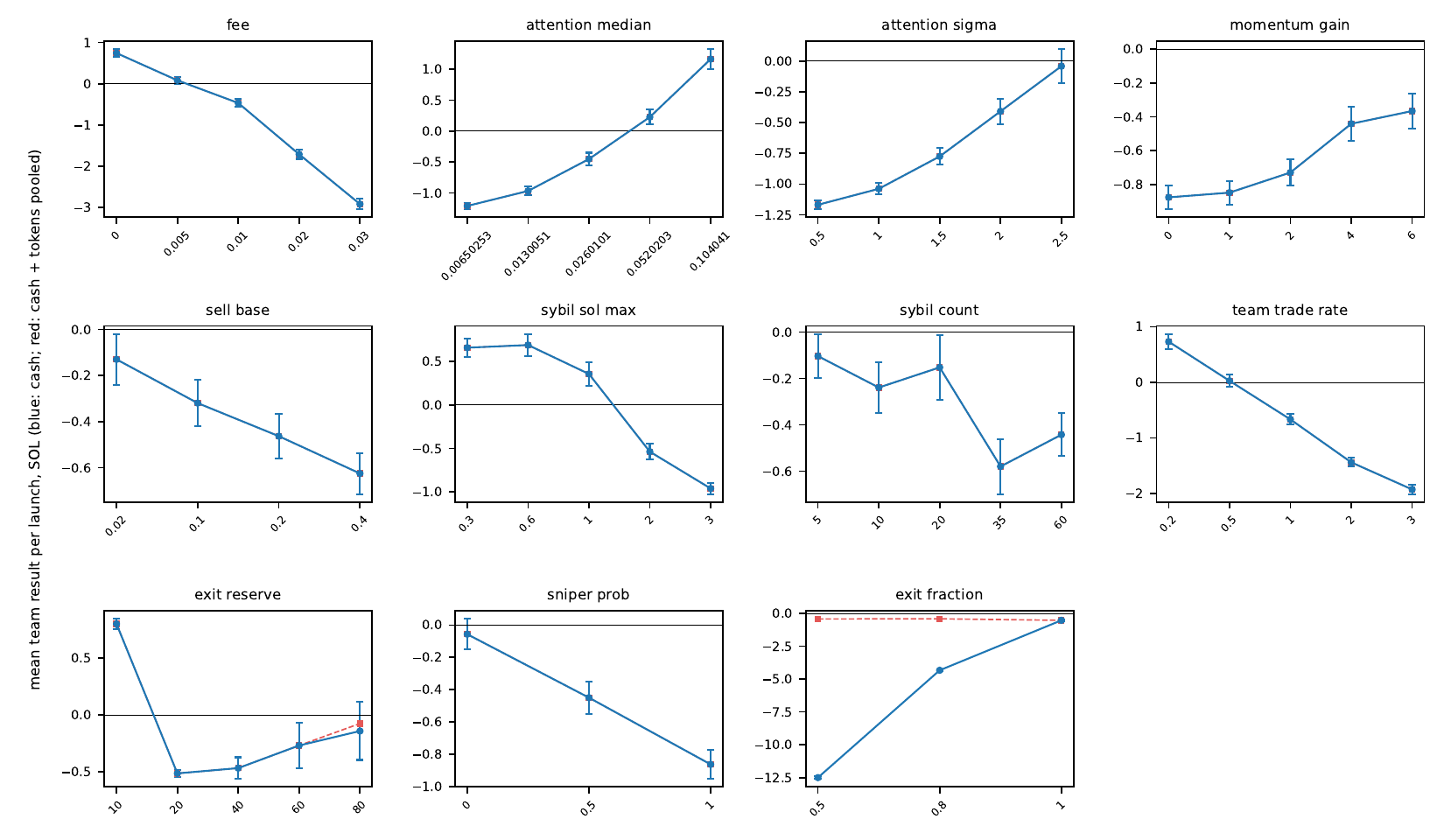}\caption{Extended model: mean team result per launch when one parameter is varied around the calibration (blue: cash; red: cash plus tokens, pooled sale).}\label{fig:x2oat}\end{figure}

\textbf{Sensitivity.} Figure~\ref{fig:x2oat} and Appendix Table~\ref{tab:x2oat} vary one parameter at a time around the calibration, and Appendix Figure~\ref{fig:x2srrc} ranks 12 parameters jointly in a Latin-hypercube design of 15\,000 points (team cash positive in 46\% of the points; rank regression $R^2=0.40$). The median attention comes first (standardized coefficient +0.49), followed by the fee (-0.26), the team's background trading rate (-0.21) and the SOL per wallet (-0.20); other traders' selling propensity, momentum, the launch cost and the graduation level matter little. Two sweeps are cliffs rather than slopes: a team that sells only half of its position at the exit ends with unsold tokens and a cash result of $-$12.5 SOL, and a team whose capital does not reach the exit reserve waits for the time limit. The launch cost, the graduation level, the post-exit period, the number of early buyers and the panic selling of other traders change the mean team result by at most 0.3 SOL and are not plotted.

\subsection{Teams that react to the other traders' inflow}\label{sec:policy}
The idea that a team should stop pumping once other traders buy early, and may sell into their demand before the reserve peaks, can be tested in the model because the model allows counterfactual policies. The team observes the other traders' net SOL inflow and the reserve. Policies: stop all team trading when the other traders' inflow over 30 seconds exceeds a threshold; abandon the token at 60 seconds when the other traders' cumulative net inflow is below a threshold (cut the losses); exit when the windowed inflow falls to a fraction of its maximum; exit when the reserve falls from its running maximum; and combinations. All policies keep the calibrated reserve exit and a holding limit of five minutes. Table~\ref{tab:x2policy} and Appendix Figure~\ref{fig:x2pol} give the results.

\begin{table}[htbp]\centering\small\caption{Observation-based team policies in the calibrated extended model (mean SOL per launch, 8\,000 launches per policy).}\label{tab:x2policy}
\resizebox{\linewidth}{!}{\begin{tabular}{lrrrrrrrr}\toprule Policy & Team cash & Cash $>$0 & Low att. & Mid att. & High att. & Other traders' cash & Team buys after 60 s & Screen flag \\ \midrule
static (calibrated rule, 300 s limit) & $-$0.49 & 21\% & $-$1.52 & $-$1.18 & 1.25 & $-$2.34 & 24.8 & 69.1\% \\
static, no hold limit & $-$0.58 & 21\% & $-$1.57 & $-$1.23 & 1.08 & $-$2.16 & 26.4 & 69.8\% \\
no extra team trading & 0.87 & 52\% & $-$0.11 & 0.09 & 2.67 & $-$3.02 & 0.0 & 0.0\% \\
stop pumping when inflow >= 2 & $-$0.22 & 22\% & $-$1.45 & $-$1.10 & 1.93 & $-$2.56 & 25.9 & 67.3\% \\
stop pumping when inflow >= 5 & $-$0.40 & 22\% & $-$1.52 & $-$1.14 & 1.50 & $-$2.44 & 25.2 & 69.5\% \\
abandon at 60 s if other traders < 1 & $-$0.32 & 15\% & $-$1.02 & $-$0.88 & 0.92 & $-$1.89 & 2.4 & 9.7\% \\
abandon at 60 s if other traders < 3 & $-$0.32 & 17\% & $-$1.02 & $-$0.83 & 0.90 & $-$2.03 & 0.4 & 1.6\% \\
fade exit 0.4 & $-$0.58 & 21\% & $-$1.51 & $-$1.16 & 0.91 & $-$2.21 & 23.8 & 65.8\% \\
drawdown exit 0.15 & $-$0.20 & 29\% & $-$0.66 & $-$0.49 & 0.55 & $-$1.90 & 5.7 & 23.1\% \\
stop pumping + abandon & $-$0.12 & 17\% & $-$0.98 & $-$0.79 & 1.38 & $-$2.27 & 1.1 & 4.1\% \\
stop pumping + abandon + fade + drawdown & $-$0.12 & 29\% & $-$0.49 & $-$0.38 & 0.49 & $-$1.84 & 0.2 & 1.1\% \\
no extra trading + abandon + drawdown & 0.28 & 44\% & 0.08 & 0.11 & 0.64 & $-$1.94 & 0.0 & 0.0\% \\
\bottomrule\end{tabular}}\par\smallskip\begin{minipage}{.98\linewidth}\small All policies keep the calibrated reserve exit rule and, except the second row, a holding limit of 300 s. Low, mid and high attention are terciles of the token's attention level. Inflow thresholds are in SOL over a 30 s window (stop pumping, fade) or cumulative (abandon).\end{minipage}\end{table}

Four results. First, the main cost is the team's continual trading, not a lack of reaction: without extra team trading the mean result is 0.87 SOL per launch (52\% positive) against $-$0.49 SOL (21\%), a difference of 1.36 SOL, which is the fees and the slippage of about 25 SOL of team purchases after the first minute. Second, stopping the pumping only when the inflow is strong recovers little ($-$0.22 SOL with a threshold of 2 SOL), because most of the extra trading happens in tokens that attract no other traders. Third, cutting the losses early helps the tokens with low and medium attention (mean result $-$1.02 and $-$0.83 SOL against $-$1.52 and $-$1.18) but costs part of the upside of the high-attention tokens (0.90 against 1.25 SOL); the combination of stopping, abandoning and exiting on the reserve drawdown gives $-$0.12 SOL, better than the static team but worse than not trading. Fourth, the exit on fading inflow almost never triggers (in under 1\% of the launches), because inflow in the model is noisy and small. A team that does not trade after launch, abandons weak tokens and exits on a drawdown gets 0.28 SOL (44\% positive).

\begin{figure}[htbp]\centering\includegraphics[width=\linewidth]{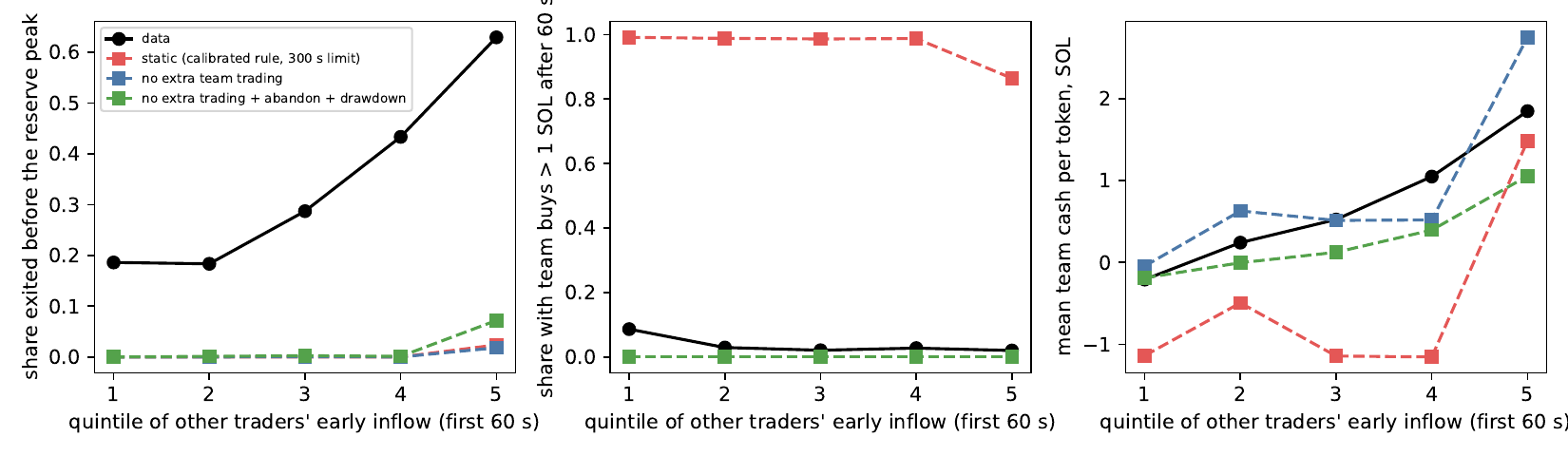}\caption{Quintiles of the other traders' early inflow: data (black) and simulated populations. Left: share of positions mostly sold before the reserve peak. Middle: share of tokens with team purchases above 1 SOL after 60 s. Right: mean team cash per token.}\label{fig:x2pat}\end{figure}

The data of Section~\ref{sec:inflow} can be compared with these populations (Figure~\ref{fig:x2pat}). In the data a team buys more than 1 SOL after the first minute in only a few per cent of the tokens, whereas the calibrated static team does so in almost all; and the team's net result rises with the early inflow of other traders in a way that the model without extra team trading reproduces closely and the static team does not. In that sense the data resemble a team that does not keep trading, and the calibration of stage B, which used only episode descriptors, did not identify this. What no model population reproduces is the share of exits before the reserve peak: 19\% to 63\% across the inflow quintiles in the data against at most a few per cent in the model, with every policy. The simulated teams hold their position until the reserve is high, while a third of the real positions are mostly sold before the peak, and the mechanism that makes real teams sell early (information about the token, a plan to leave a token with a small position, or knowledge of other buyers) is not in the model.

\subsection{Team types, the early-inflow pattern and group-level results}\label{sec:stagec}
The policy experiments and the data point the same way: real teams rarely keep buying after launch, and the calibrated pump-and-collapse team is not typical. We therefore add a second team type, a \emph{quiet seller}: a small team (about 4 wallets, a launch purchase of about 0.50 SOL) that trades in the background at only 0.018 transactions per second and sells a share of 10\% of its holdings per second while the other traders' inflow over 30 seconds is at least 2.2 SOL (after 36 seconds, and in any case within five minutes). The team of a token is pump-and-collapse with probability $\pi$ and a quiet seller otherwise. Stage C of the calibration fits $\pi$ and the parameters of the quiet seller to the early-inflow moments of Section~\ref{sec:inflow} (share of positions mostly sold before the reserve peak, share of tokens with team purchases after the first minute, mean team cash and share positive, and the episode rate, by quintile of the early inflow where available), keeping stages A and B.

\begin{figure}[htbp]\centering\includegraphics[width=\linewidth]{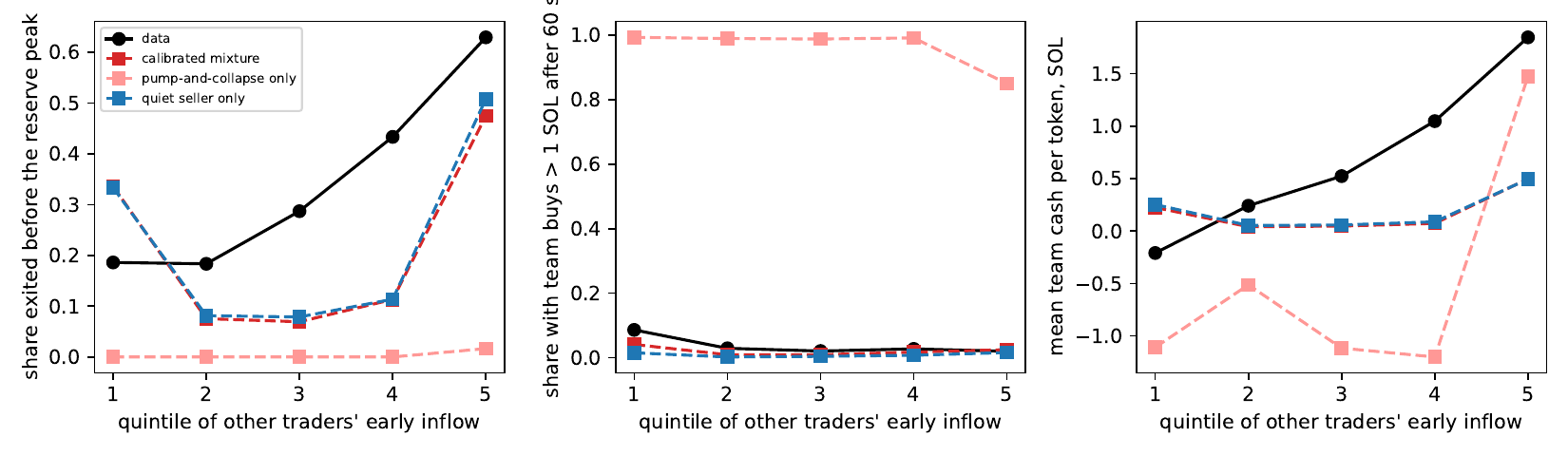}\caption{Quintiles of the other traders' early inflow: data (black) and simulated tokens of three team populations. Left: share of positions mostly sold before the reserve peak. Middle: share of tokens with team purchases above 1 SOL after 60 s. Right: mean team cash per token.}\label{fig:stagec}\end{figure}

The fit is partial. The best mixture has $\pi=0.7\%$, so almost all teams are quiet sellers. It reproduces the low share of tokens with further team buying (2.0\% against 3.7\% in the data) and a positive mean team result, but not the gradient across the inflow quintiles: the simulated share of exits before the peak is U-shaped (34\%, 8\%, 7\%, 11\%, 47\% across the quintiles) while it rises monotonically in the data (19\%, 18\%, 29\%, 43\%, 63\%), the overall share of such exits is 21\% against 34\%, the mean team cash per token is 0.18 against 0.69 SOL, and the team result rises with the inflow only in the top quintile (Figure~\ref{fig:stagec}). The distance of the best draw corresponds to a mean squared deviation of about 5 in units of the tolerances used in the fit. The quiet seller explains the low pumping and the positive average, not the graduated early exits that the data show.

\textbf{Group-level results.} A group in the data repeats the exercise on many tokens, so the model can be compared with the groups' results, which were not used in the calibration (Table~\ref{tab:x2groups}). For the repeat groups with at least ten positioned tokens, the mean team cash per token is 0.69 SOL, 79\% of the tokens are positive, and 90\% of the groups have a positive total. The pump-and-collapse team alone is far from this: it loses on average ($-$0.46 SOL per token), has 22\% positive tokens and only 16\% of the groups with a positive total. The quiet seller and the mixture are much closer in sign (0.18 SOL per token, 57\% positive tokens, 99\% of the groups positive) but not in size: the median group total is 3.2 SOL in the model against 10.4 in the data. For the serial groups, which were selected by their collapses, the data lie between the two types (64\% positive tokens, 71\% of the groups positive); a mixture with 10--20\% of pump-and-collapse tokens gives 51--46\% positive tokens and 38--66\% positive groups, close in shares and concentration but not in size (mean team cash per token 0.07 SOL against 1.63 in the data, median group total 1.2 against 31.5 SOL).

\begin{table}[htbp]\centering\small\caption{Group-level results of the data and of simulated groups (positioned tokens: reserve peak of at least 5 SOL and a team position).}\label{tab:x2groups}
\resizebox{\linewidth}{!}{\begin{tabular}{lrrrrrr}\toprule Population & Groups & Mean team cash per token (SOL) & Tokens with cash $>$0 & Top-decile share of positive results & Groups with total $>$0 & Median group total (SOL) \\ \midrule
Data: repeat groups with $\geq$10 positioned tokens & 1657 & 0.69 & 79\% & 62\% & 90\% & 10.4 \\
Model: calibrated mixture & 600 & 0.18 & 57\% & 49\% & 98\% & 3.1 \\
Model: pump-and-collapse only & 600 & $-$0.46 & 22\% & 86\% & 16\% & $-$9.7 \\
Model: quiet seller only & 600 & 0.18 & 57\% & 47\% & 99\% & 3.2 \\
\midrule
Data: serial groups (price screen) & 62 & 1.63 & 64\% & 70\% & 71\% & 31.5 \\
Model: share of pump-and-collapse tokens 0.0 &  & 0.18 & 57\% & 47\% & 98\% & 2.7 \\
Model: share of pump-and-collapse tokens 0.1 &  & 0.07 & 51\% & 64\% & 66\% & 1.2 \\
Model: share of pump-and-collapse tokens 0.2 &  & $-$0.03 & 46\% & 72\% & 38\% & $-$1.6 \\
Model: share of pump-and-collapse tokens 0.5 &  & $-$0.27 & 34\% & 84\% & 19\% & $-$6.9 \\
Model: share of pump-and-collapse tokens 1.0 &  & $-$0.49 & 22\% & 86\% & 17\% & $-$9.4 \\
\bottomrule\end{tabular}}\par\smallskip\begin{minipage}{.98\linewidth}\small Top-decile share: share of all positive token results that comes from the best 10\% of the tokens. Data: first-hour team cash per token; model: cash of the run. Simulated groups have as many positioned tokens as the data groups they are matched to.\end{minipage}\end{table}

\textbf{A testable implication.} If other traders' attention is a lottery, the team's positive results are concentrated in few tokens. In the data the best 10\% of the tokens give 62\% of all positive results (70\% for the serial groups, and a median of 28\% inside the repeat groups). The simulated populations bracket this: 47\% for the quiet seller and 86\% for the pump-and-collapse team. The data are consistent with a lottery in that profit is concentrated, but the concentration alone does not separate the team types. Taken together, the model now agrees with the data on the sign structure of the group results, on the low rate of further team buying and on the concentration, and disagrees on the size of the profits, on the gradient of early exits across the inflow quintiles, and on the average profit of the serial groups, which is an order of magnitude larger than the calibrated attention distribution produces; we have not tested whether the attention drawn by the tokens of serial groups differs from that of all launches.

\subsection{Analytic results for a team exit}\label{sec:analytic}
The simulations show how the team's result depends on the parameters. A part of this dependence can be derived exactly. Let $x=30+R$ be the virtual SOL reserve for a real reserve $R$, let $k$ be the invariant of the curve, and let $f$ be the fee per side. Measure a token quantity as $\nu=q/k$. A purchase that raises the real reserve from $R_0$ to $R_1$ (net of the fee) yields $\nu=1/x_0-1/x_1$, and a sale of $\nu$ moves the virtual reserve from $x$ to $1/(1/x+\nu)$.

\paragraph{Exit identity.} Suppose the team has bought $\nu$ tokens, and other traders then trade in any way, so that the real reserve at the exit is $R_e$ ($x_e=30+R_e$). Token supply is conserved, so the tokens held by other traders are fixed by $R_e$ and $\nu$ alone. When the team sells all of its tokens, the reserve falls by
\begin{equation}
g(\nu,R_e)=x_e-\frac{1}{1/x_e+\nu}=\frac{\nu x_e^{2}}{1+\nu x_e},
\label{eq:exitgross}
\end{equation}
and the team's profit is
\begin{equation}
\pi=(1-f)\,g(\nu,R_e)-\frac{R_b}{1-f},\qquad \nu=\frac{1}{30}-\frac{1}{30+R_b},
\label{eq:exitprofit}
\end{equation}
where $R_b$ is the real reserve raised by the team's purchases. The result does not depend on the order or the size of the other traders' trades, only on the reserve at which the team sells. We verified Equation~\eqref{eq:exitprofit} on 3\,976 runs of the extended model in which the team buys at the launch and sells everything after 240 seconds while other traders trade freely; the largest absolute difference between the simulated and the computed team result is $2\times10^{-13}$ SOL.

\paragraph{Zero-sum form.} Without fees, $\pi=D-R_f$, where $D=R_e-R_b$ is the net amount that other traders spent after the team's purchases and $R_f$ is the reserve left after the team's sale. The team gains exactly what other traders lose, measured at the curve value of their remaining tokens.

\paragraph{Break-even.} Setting $\pi=0$ in Equation~\eqref{eq:exitprofit} gives the reserve at which the exit pays for the fees and for the team's own price impact,
\begin{equation}
x^{*}=\frac{R_b+\sqrt{R_b^{2}+4(1-f)^{2}\,30\,x_b}}{2(1-f)^{2}},\qquad x_b=30+R_b .
\end{equation}
With $f=1\%$ and team capital $R_b$ between 1 and 60 SOL, the other traders must add between 0.3 and 1.4 SOL to the reserve; with $f=2\%$ and $R_b=20$ SOL they must add 1.3 SOL. The exit itself is therefore cheap, and the negative average results of the calibrated team (Section~\ref{sec:extresults}) must come from something else.

\paragraph{Costs of continued trading.} In the extended model the calibrated team without its background trading earns $+0.87$ SOL per launch on average, and with it $-0.54$ SOL (3\,000 launches each). The extra team volume of 104 SOL per launch, multiplied by the fee, explains $-1.04$ of the difference of $-1.41$ SOL, about three quarters; we have not decomposed the remainder, which is likely due to the team's own price impact when it raises the reserve before the exit.

\begin{figure}[htbp]\centering\includegraphics[width=.95\linewidth]{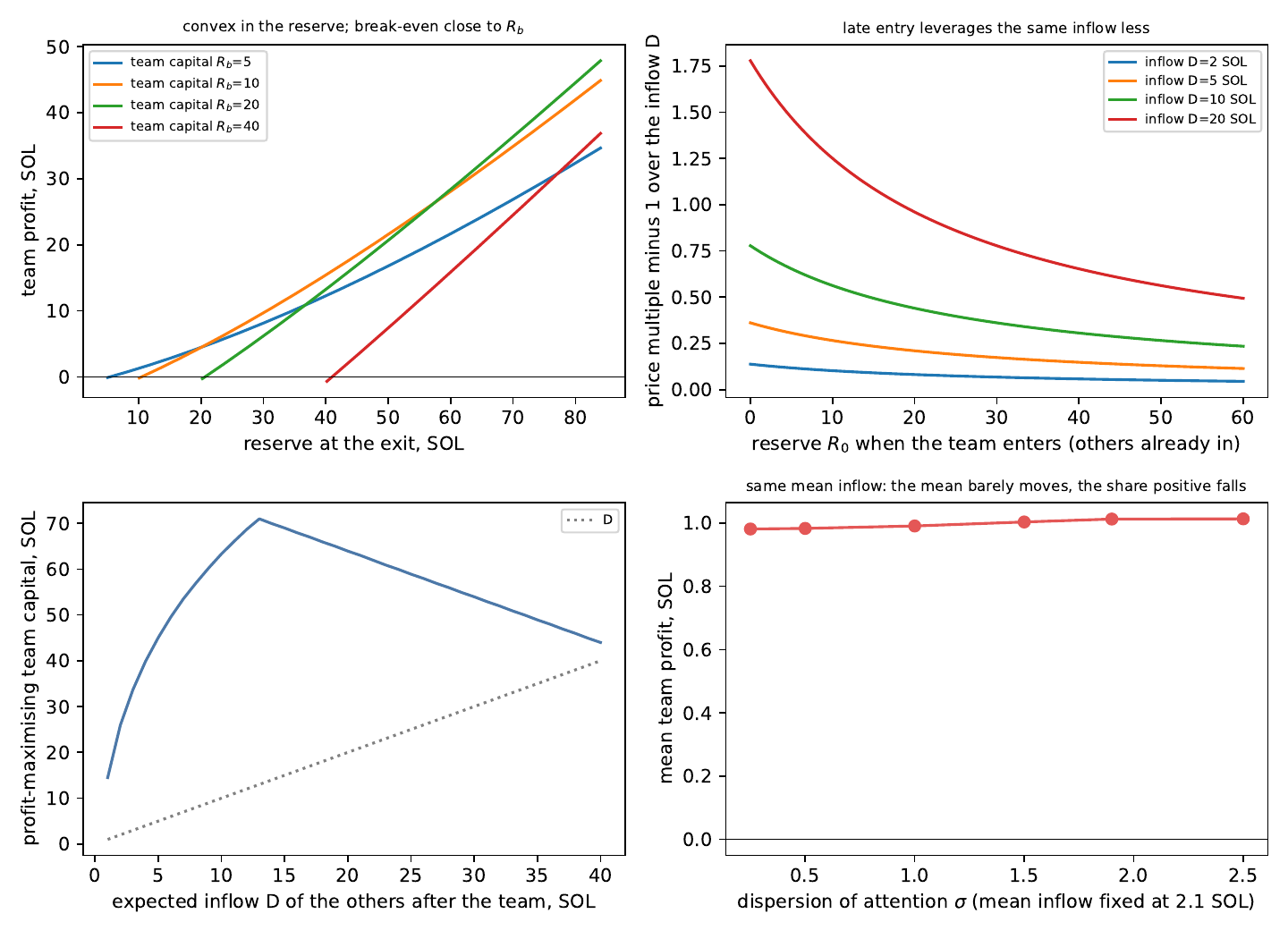}
\caption{Analytic results ($f=1\%$). Upper left: team profit against the reserve at the exit for four team capitals; upper right: price multiple minus one over an inflow $D$ against the reserve $R_0$ at which the team enters; lower left: profit-maximizing team capital against the expected inflow; lower right: mean team profit against the dispersion of attention at a fixed mean inflow.}\label{fig:analytic}\end{figure}

\paragraph{Convexity and the leverage of a late entry.} The profit is convex in the exit reserve, because $\partial^{2}g/\partial x_e^{2}=2\nu/(1+\nu x_e)^{3}>0$. Each additional SOL from other traders raises the profit more than the previous one, as for an option. For a small position, the ratio of the sale price to the purchase price is $(x_e/x_b)^{2}$, so an inflow $D$ into a token whose reserve is already $R_0$ raises the price by the factor $\bigl((30+R_0+D)/(30+R_0)\bigr)^{2}$. For $D=10$ SOL the price rises by 78\% when $R_0=0$, by 44\% when $R_0=20$ and by 31\% when $R_0=40$ (Figure~\ref{fig:analytic}, upper right). A team that enters a token which other traders have already bought has less to gain from the same inflow, which is the mechanism behind the lower team activity after a large early inflow (Section~\ref{sec:inflow}).

\paragraph{Team size.} If the team expects an inflow $D$ after its purchases, the profit-maximizing capital grows with $D$: in the calculation with $f=1\%$ it is 26 SOL for $D=2$, 45 SOL for $D=5$ and 63 SOL for $D=10$. Beyond that, the graduation level restricts the position ($R_b+D\le84$). An interior optimum exists only for small $D$, because the team's own sale moves the price against it (Figure~\ref{fig:analytic}, lower left).

\paragraph{Random inflow.} In the extended model with the calibrated attention distribution, the rise of the reserve caused by other traders increases with the token's attention $A$ roughly as $D+0.1\approx 8.7\,A^{0.69}$ with a logarithmic scatter of 0.82 (a fit to the simulation). Inserting this relation and the lognormal attention of the calibration (median 0.026, $\sigma=1.89$) into Equation~\eqref{eq:exitprofit} gives a mean team result of $+1.03$ SOL and a share of 46\% of positive launches for the team that buys at the launch and sells after 240 seconds; the simulation gives $+1.02$ SOL and 43\%. Because the relation between $D$ and $A$ was fitted to the same simulation, this is a consistency check and not an independent prediction. Keeping the mean inflow fixed at 2.1 SOL and raising the dispersion of attention from $\sigma=0.25$ to 2.5 hardly changes the mean profit ($0.98$ to $1.01$ SOL) but lowers the share of profitable launches from 74\% to 36\% (Figure~\ref{fig:analytic}, lower right). Convexity is thus present but weak in the range of inflows of the calibrated model; dispersion mainly concentrates the profit in few tokens.

\paragraph{A stopping rule.} Approximate the exit reserve by a diffusion with drift $\mu$ and variance $\sigma^{2}$ per unit of time. A team that is not constrained by time compares the sale now with waiting an instant. Waiting is worthwhile as long as
\begin{equation}
\mu\;\ge\;-\frac{\sigma^{2}}{x_e\,(1+\nu x_e)\,(2+\nu x_e)} .
\end{equation}
The right-hand side is small, so in practice the team should sell when the net flow of other traders turns negative. This agrees with the observation-based policies of Section~\ref{sec:policy}, but we have not tested the rule itself in the simulation, and the reaction of other traders to the team's sale is not included in it.

\paragraph{Limits.} Equation~\eqref{eq:exitprofit} assumes that the team sells its whole position at once and that its tokens were bought when it was the only trader, or that the other traders' purchases are included in $\nu$; for purchases mixed with other traders' purchases the average cost must be computed from the actual path. The link between the inflow and the attention is empirical, and the distribution of the inflow is not derived from the arrival process of the model. The fee of 1\% per side is the parameter of the simulation; the transaction sample of Section~\ref{sec:fees} suggests about 2\%.

\subsection{Limitations and use as a base model}
The original model supports comparative statements about a mechanism, not estimates. The extended model is calibrated, but only partly. Stages A and B match marginal moments (the distribution of the maximal reserve and of trades per launch, the median descriptors of the episodes of the reserve screen); the pump-and-collapse team they select deploys about 68 SOL per token, trades continually and exits late, which the early-inflow data contradict. Stage C adds a quiet seller and matches the low rate of further team buying and the sign structure of the group results, but not the gradient of early exits across the inflow quintiles, and it underestimates the size of the profits of serial groups by an order of magnitude. Other traders are still of one type, react to the price path but not to the team, and the attention of a token is independent of the team; there is no competition between tokens, no creator rewards and no heterogeneity of teams. Results such as the loss of the static team, the importance of the fee and the cost of continual team trading are therefore conditional statements about this specification, and the numbers must not be quoted as estimates of real profit.

The simulation is meant as a base for further experiments: the parameters are explicit, an experiment is a list of parameter changes with a replication count, the calibration is a separate step, and the team's decision rules read observable state (reserve, the inflow of other traders, time), so a new strategy requires a new rule and a design. The natural next steps are to add the early-inflow pattern of Section~\ref{sec:inflow} to the calibration (share of tokens with team buying after the first minute, share of exits before the peak, team result by inflow quintile), which would identify the team's behaviour; to let the team type vary between tokens (quiet exit, pump and collapse, abandonment); to calibrate the campaign level to the group cash flows of the windows; and to make the other traders react to the team's visible selling.

\section{Data and independent period construction}
\subsection{Source universe and observation windows}
The data come directly from the Solana blockchain: the records of token launches and trades on Pump.fun and of SOL transfers between wallets, for ten data windows. Nine windows contain recorded launches and are usable. The February 2024 window contains four trades and no token launches; it is reported as unusable for this design, rather than assigned zero incidence or zero profit.

Across usable windows there are 424\,216 recorded launches and 31\,724\,252 trade rows after joining trades to those launches and restricting records to creation or later within the respective period. The complete trade tables contain additional tokens not launched in the recorded window. Those rows are not silently treated as new launches. Table~\ref{tab:windows} reports the actual recorded window endpoints in UTC; names of archives are labels, not substitutes for those endpoints.

\begin{table}[htbp]\centering\small
\caption{Recorded UTC windows and newly extracted launch cohorts. Each row is analyzed separately.}\label{tab:windows}
\begin{tabular}{lrrrrr}\toprule
Period & Start (UTC) & End (UTC) & Hours & Launches & Cohort trades \\ \midrule
2024-04 & 04-10 19:57 & 04-11 19:38 & 23.7 & 11\,916 & 280\,944 \\
2024-07 & 07-10 19:59 & 07-12 03:25 & 31.4 & 14\,282 & 1\,314\,540 \\
2024-10 & 10-10 20:00 & 10-12 05:20 & 33.3 & 22\,398 & 2\,264\,190 \\
2025-02 & 02-10 20:00 & 02-12 02:58 & 31.0 & 73\,394 & 4\,853\,879 \\
2025-03 & 03-10 20:00 & 03-11 19:40 & 23.7 & 25\,856 & 2\,339\,412 \\
2025-05 & 05-10 20:00 & 05-13 03:24 & 55.4 & 57\,939 & 4\,693\,425 \\
2025-06 & 06-21 20:00 & 06-25 03:16 & 79.3 & 84\,821 & 7\,036\,574 \\
2025-09 & 09-12 19:59 & 09-15 19:18 & 71.3 & 81\,233 & 5\,273\,041 \\
2025-12 & 12-12 20:00 & 12-15 19:36 & 71.6 & 52\,377 & 3\,668\,247 \\
\bottomrule\end{tabular}
\end{table}

These are purposefully selected interesting windows, not a random sample of calendar time. Their observed duration varies from approximately 23.7 to 79.3 hours. Two of the windows, June and September 2025, were chosen at periods of relatively low and high SOL prices. We use these labels descriptively. We add a daily SOL/USD covariate (Section~\ref{sec:solusd}) but have no market-wide attention index or representative panel of high and low regimes. Consequently, a contrast between the two labeled windows cannot identify a price-of-SOL effect on token demand or serial activity.

Token metadata provide a recorded creator address, creation timestamp, mint and bonding-curve identifier. Trade rows provide transaction type, trader address, SOL amount, token amount, reported price, reserve fields, transaction signature, block identifier and time. SOL-transfer rows provide source, destination, amount, signature and time. No direct token-transfer table or explicit verified migration field is present in these datasets. The recorded real reserve is bounded by the graduation level of the curve: in the first five windows its maximum is 85.01 to 85.11 SOL (112 to 1,004 tokens per window reach 84 SOL or more), whereas in the last four windows 5, 1, 100 and 85 tokens carry values above 86.5 SOL, up to about 48,500 SOL, which cannot belong to the curve and presumably describe the pool after migration. Such values exceed the 70 SOL threshold of the ambiguity guard, so these tokens never enter an episode, and none of the tokens of the primary serial groups ends with a recorded reserve above 86.5 SOL. Tokens near graduation are valued on the curve state in the inventory marking of Section~\ref{sec:robust}, which is approximate for tokens that have migrated. Original numeric columns include single-precision values; our outputs cannot restore precision lost before ingestion.

\subsection{Censoring and calendar time}
A launch enters the primary event and cash-flow comparisons only when a complete first hour fits inside its own recorded window. This is a boundary rule, not a guarantee that every real transaction was collected during that hour. A token without a valid early price remains in the complete-hour launch denominator, but cannot produce a price-based episode. We separately report the number eligible for the price-based random-allocation control.

We aggregate valid trade prices to observed seconds. Missing seconds are not forward-filled, and the absence of another trade is not a zero price. Calendar transaction intensity divides counts by elapsed calendar seconds, including idle seconds; it is distinct from transactions per active second. This addresses the reference script's tendency to omit quiet intervals from an activity denominator.

When several records share a second, the record identifier provides a reproducible ordering for extraction, but does not establish instruction order. The event definition uses second-level price aggregation and conservative reserve ranges. It therefore does not claim transaction-by-transaction causal ordering within a second. Shared signatures and SOL transfers can also reflect the same economic instruction; we keep transfer and trade ledgers separate rather than summing them into an apparent gain.

\section{Inferring candidate teams}
\subsection{Creator and funding components}
Every recorded creator starts as a node. We find its earliest eligible SOL funding observed \emph{within its own period}, at least 0.2 SOL, before its first launch. Funding in the same second as the launch is excluded from grouping. This is not a reconstruction of a wallet's lifetime first funding, and a missing predecessor is not evidence that a wallet self-funded.

High-activity addresses are excluded as funding bridges if their incoming plus outgoing transfer-endpoint incidence exceeds 1,000 within that period. A self-transfer contributes two endpoint appearances. The threshold is configurable and does not remove those addresses' token launches from the launch universe. The exclusion reduces obvious service-induced linking but does not establish that every remaining funder is private.

We connect two creators when one directly funds the other under the rule above, or when an eligible common first funder funds between two and fifty creators. Connected components define the creator/funding groups. The cap applies to each common funder; transitive chains can still produce a larger component. Component size, therefore, must be inspected rather than assumed to be bounded by fifty creators.

\subsection{Conservative trading-wallet extension}
The primary accounting scope contains the recorded creators, uniquely assigned qualifying shared first funders, and additional funded repeat early buyers. An additional buyer must purchase at least two different tokens of the same group within the first 60 seconds of those tokens, and receive eligible funding from a core assigned wallet strictly before its first qualifying early purchase. High-activity bridge addresses and buyers linked to multiple groups are excluded from this extension. Each retained wallet is assigned to exactly one group in a period.

Neither token returns, collapse labels nor the sign of a group's cash flow determines grouping. However, some links require multiple launches, so the complete group is a \emph{retrospective} within-period construction. It is not presented as a real-time detector available before every included token. Direct funding and repeat early buying are supporting relations, not proof of common beneficial ownership. A remaining common funder might be a service, a group might contain independent participants, and an operator may use undetected wallets.

A creator-wallet-only sensitivity retains these same group memberships but counts trades only by recorded creator wallets in each group. It tests the effect of \emph{accounting scope}. It does not rebuild the graph into isolated creators and is not evidence that the funding clustering is independently validated. Re-estimating clusters under alternative graph rules is a separate future robustness task.

\section{Episode definition and statistical design}
\subsection{Price and reserve measurements}
For price estimation, a trade must be a recorded buy or sell with finite positive SOL and token amounts, at least 0.001 SOL and at least 100 tokens, and a positive reported price. The product of reported price and token amount must match SOL within the greater of 0.00001 SOL and 5\% of the recorded SOL amount. The second-level price is the sum of valid SOL amounts divided by the sum of valid token amounts. It is an observed-trade VWAP rather than a last dust-trade quote or a ratio derived from reserve differences.

A higher-volume sensitivity independently recalculates these candles using at least 0.01 SOL and 1,000 tokens. It can alter the baseline, local peaks and crossing times, so its labels need not be a strict subset of the primary labels. It is not merely a deletion of already selected primary episodes.

The baseline is the first valid candle within 60 seconds of creation. A local peak must be at least twice this baseline and occur by minute 50. For each observed price, we inspect a trailing ten-minute window and retain the local maximum. Candidate records are the first crossings of each of the 70\%, 80\% and 90\% drawdown levels for each such peak. Selection therefore ranges over this explicitly enumerated crossing set; it is not an unconstrained search over every possible endpoint pair.

Reserve values are treated as \emph{recorded reserve measurements}. If a second contains several values, let $R_{p,\min}$ denote the minimum at the peak second and $R_{d,\max}$ the maximum at the drop second. We define a conservative lower bound on the recorded fall as
\begin{equation}
D_R^{L}=1-\frac{R_{d,\max}}{R_{p,\min}}.
\end{equation}
Using the opposite extrema would make a fall easier to declare. This range treatment reduces ambiguity from unknown within-second order. It does not validate the recorded reserve field against protocol state or turn that field into observed liquidity-provider withdrawal.

\subsection{Primary retrospective screen}
A candidate qualifies when all the following conditions hold:
\begin{enumerate}
\item the launch has a complete first hour and a valid early baseline; its local peak reaches at least twice that baseline;
\item price falls at least 80\% from the local peak within ten minutes, with the peak no later than minute 50;
\item the peak-second minimum recorded reserve is at least 5 SOL and the conservative recorded reserve fall is at least 50\%;
\item in the window from the later of creation and 60 seconds before the peak through the drop second, assigned group sales minus purchases are at least 1 SOL and the group's share of all recorded sales is at least 50\%;
\item assigned group wallets record at least 0.01 SOL of buying within the token's first 60 seconds;
\item the protocol-or-measurement ambiguity guard described below is not triggered.
\end{enumerate}
Only the earliest qualifying candidate is retained per mint. A \emph{serial group} has at least three distinct qualifying tokens in that period. Simultaneous or overlapping token lifetimes are allowed. ``Serial'' means repeated launch-associated episodes, not necessarily nonoverlapping, fully closed capital cycles.

The ambiguity guard excludes candidates if a recorded reserve of at least 70 SOL has occurred by the drop, or if the minimum drop-second reserve is within $10^{-7}$ SOL of 0.00123192. These are explicit heuristics intended to flag potentially ambiguous protocol or measurement behavior. They are \emph{not} verified migration criteria. We also vary the guard on and off. In particular, a reserve above 70 is not declared a historical migration threshold, and the fixed small value is not declared a proven protocol signature.

\subsection{Group cash flow and outcome selection}
For every selected serial group, we sum equation~\eqref{eq:cash} over \emph{all its own-token launches having a complete first hour}, including launches that do not satisfy the episode screen. Trades by its wallets on unrelated groups' tokens are not included. The accounting horizon is the first hour of each included token, not a single portfolio hour or the group's lifetime. Overlapping first hours can expose the same capital to several tokens; we do not infer independent reinvestment cycles from them.

Cash-flow sums retain recorded buy/sell rows with finite nonnegative SOL and token amounts, including trades too small to define a price candle. The stricter price-quality thresholds affect the price series and episode construction, not the inclusion of otherwise finite trade amounts in the financial ledger. Thus, excluding a dust quote does not silently delete its recorded cash movement.

Positive episode-window net selling is imposed by selection. It would be circular to use the selected windows' positive sum as proof of a profitable strategy. Including other launches and a wider, fixed observation horizon reduces this particular selection problem. It still does not identify lifetime profit, total wallet wealth or a causal return to conducting a scheme.

An illustrative additional friction rate $f$ is applied to recorded turnover as
\begin{equation}
F_G(f)=(1-f)S_G-(1+f)B_G.
\end{equation}
Rates of 0.5\%, 1\% and 2\% per side are \emph{incremental hypothetical deductions}. We have not established whether historical trade amounts already include specific protocol charges. These scenarios are therefore not estimates of actual fee-adjusted profit, and should not be interpreted as historical fee reconstruction.

\subsection{Sensitivity and allocation comparison}
Section~\ref{sec:robust} supplements the comparisons below with equal-length windows, alternative team constructions, a label-permutation benchmark that keeps group launch counts, and bootstrap intervals. Uncertainty is reported with group-cluster bootstrap intervals \cite{efron1979,cameron2008}, permutation $p$-values computed as $(1+k)/(1+B)$ so that they are never zero \cite{phipson2010}, and Wilson intervals for the proportions of the audit sample \cite{wilson1927}. 

Within each usable period we vary price drawdown in $\{0.7,0.8,0.9\}$, reserve drawdown in $\{0.3,0.5,0.7\}$, group sale share in $\{0.3,0.5,0.7\}$, minimum group net sales in $\{0.5,1,2\}$ SOL, serial episode count in $\{2,3,5\}$ and ambiguity guard status. This gives 486 detector settings per window. All remain retrospective descriptions; we do not choose the setting that maximizes apparent gain and report it as a prospective strategy.

The allocation comparison preserves every eligible token's group and creation-hour stratum and the observed number of episode labels in each hour. Within a stratum, it randomly reallocates those labels without replacement across full-hour tokens with a valid early baseline. For each of 1,000 replications it counts groups receiving at least three labels. This asks whether observed repeat concentration resembles a simple allocation conditional on launch opportunity and hour. It is \emph{not a calibrated fraud or skill hypothesis test}: the original labels themselves depend on group trading, so exchangeability under a causal null has not been established. Quantiles describe this comparator's distribution, not confidence intervals for population prevalence.

We additionally restrict token exposure to launches whose complete first hour ends within the first twelve hours of the window. This fixes a shorter exposure interval but \emph{holds full-period group membership fixed}. It is a partial duration sensitivity, not a complete equal-length rerun of funding-graph inference. Unequal windows and retrospective information remain material limitations.

\section{Results across independently analyzed periods}
\subsection{Incidence and repeat concentration}
Table~\ref{tab:incidence} reports the primary screen. There are 679 distinct token episodes across the nine separately analyzed windows. Of these, 307 belong to the 62 serial group--period records. A count across windows is a count of analysis records, not of unique real-world organizations: identical or related participants could appear in more than one window, and we have deliberately not matched them.

\begin{table}[htbp]\centering\small
\caption{Primary episode incidence and serial groups. Rates use all complete-hour launches.}\label{tab:incidence}
\begin{tabular}{lrrrrrr}\toprule
Period & Full hour & Price eligible & Episodes & Per 1,000 & Groups & Serial ep. \\ \midrule
2024-04 & 11\,527 & 11\,527 & 38 & 3.30 & 5 & 28 \\
2024-07 & 13\,920 & 13\,916 & 43 & 3.09 & 6 & 21 \\
2024-10 & 21\,955 & 21\,955 & 36 & 1.64 & 1 & 9 \\
2025-02 & 71\,262 & 70\,285 & 328 & 4.60 & 32 & 167 \\
2025-03 & 24\,013 & 23\,798 & 39 & 1.62 & 4 & 18 \\
2025-05 & 57\,327 & 57\,187 & 66 & 1.15 & 6 & 26 \\
2025-06 & 83\,907 & 83\,898 & 48 & 0.57 & 3 & 18 \\
2025-09 & 78\,940 & 78\,630 & 42 & 0.53 & 1 & 3 \\
2025-12 & 51\,422 & 51\,274 & 39 & 0.76 & 4 & 17 \\
\bottomrule\end{tabular}
\par\smallskip\begin{minipage}{.98\linewidth}\small A serial group has at least three qualifying tokens in its own period. Price-eligible counts are also the allocation-control universe; missing valid early prices are not filled.\end{minipage}
\end{table}

The February 2025 window has the largest observed episode count and the highest incidence per complete-hour launch. The June and September 2025 windows have the lowest normalized incidence, despite being substantially longer than several early windows. Duration alone therefore does not explain the ranking. It can still affect serial-group counts and graph completeness, particularly when a threshold requires three episodes. A single October 2024 group has nine qualifying episodes, illustrating how a low group count can coexist with concentrated repeated behavior.

\begin{table}[htbp]\centering\small
\caption{Conditional episode timing and selling; late-price availability for episodes in serial groups.}\label{tab:behavior}
\begin{tabular}{lrrrrrr}\toprule
Period & Peak age, min & Drop, sec & Peak/base & Sale share & Late / serial & Late/peak \\ \midrule
2024-04 & 3.96 & 84 & 2.85 & 99.6\% & 9/28 & 0.135 \\
2024-07 & 3.85 & 56 & 3.90 & 98.3\% & 8/21 & 0.127 \\
2024-10 & 2.01 & 36 & 3.73 & 98.4\% & 7/9 & 0.138 \\
2025-02 & 2.31 & 13 & 3.85 & 93.7\% & 59/167 & 0.122 \\
2025-03 & 3.00 & 32 & 3.32 & 97.1\% & 8/18 & 0.154 \\
2025-05 & 3.75 & 16 & 3.54 & 74.0\% & 7/26 & 0.153 \\
2025-06 & 6.98 & 32 & 3.61 & 83.9\% & 8/18 & 0.136 \\
2025-09 & 1.62 & 8 & 3.84 & 84.4\% & 3/3 & 0.138 \\
2025-12 & 3.28 & 48 & 3.86 & 71.9\% & 10/17 & 0.150 \\
\bottomrule\end{tabular}
\par\smallskip\begin{minipage}{.98\linewidth}\small First four numeric columns are medians across all primary episodes. Late/serial gives the count with valid prices 5--10 minutes after the drop (within the first hour); late/peak is the median ratio among only those observed serial episodes. Missing later trades are not assigned a zero price.\end{minipage}
\end{table}

Table~\ref{tab:behavior} describes event timing and selling concentration conditional on passing the screen. Peak multiples are measured from valid observed prices. The sample's high drawdowns and positive event-window net selling partly reflect the definition; these conditional medians are not unconditional characteristics of every launch. Later-price observations are frequently absent. Their absence does not establish a permanently worthless token, and the conditional late-price statistics do not fill those missing outcomes.

\begin{figure}[htbp]\centering
\includegraphics[width=\linewidth]{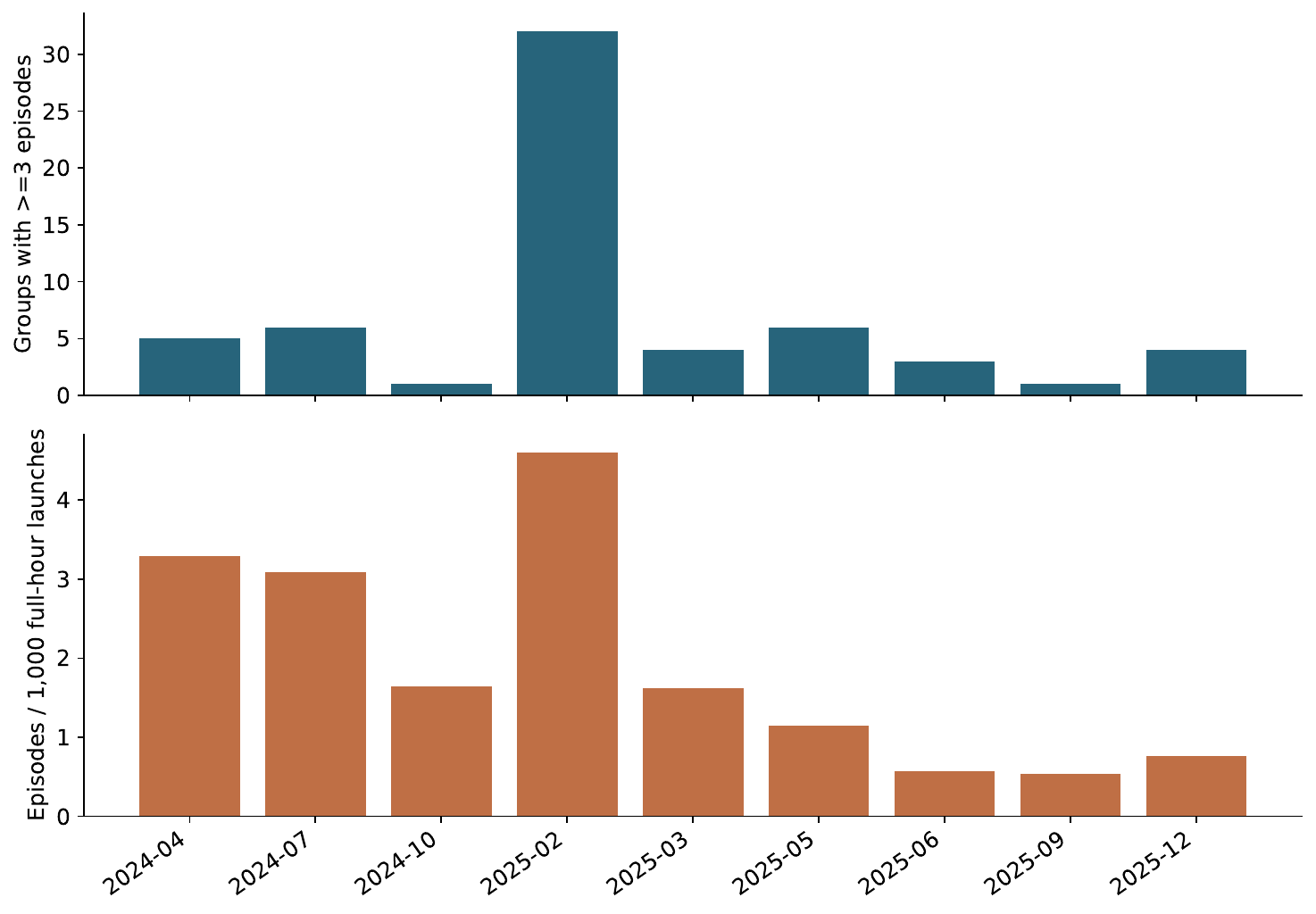}
\caption{Serial-group counts and token-level episode incidence. Each bar uses a separately reconstructed period. The denominator in the lower panel includes every recorded launch with a complete first hour, including launches without a usable early price.}\label{fig:frequency}
\end{figure}

\subsection{Recorded group outcomes including other launches}
Table~\ref{tab:cash} consolidates selected serial groups' first-hour own-token trades. These are sums across groups \emph{within the named period}; there is no pooled financial total. May, June and September 2025 are negative. In June and September every selected serial group has negative recorded first-hour cash flow. March has positive aggregate flow but a negative median group result, showing the importance of concentration. Section~\ref{sec:robust} tests how stable these signs are, including a repeat of the screen that uses only the bonding-curve reserve (Section~\ref{sec:reserve}), under which the May sign reverses.

\begin{table}[htbp]\centering\small
\caption{Recorded cash flow of primary serial groups on their own complete-hour launches (SOL).}\label{tab:cash}
\begin{tabular}{lrrrrrr}\toprule
Period & All launches & Buys & Sells & Net & Median net & Positive \\ \midrule
2024-04 & 37 & 1\,752.72 & 1\,808.74 & 56.02 & 5.04 & 100\% \\
2024-07 & 38 & 1\,644.05 & 1\,793.00 & 148.95 & 38.10 & 83\% \\
2024-10 & 10 & 510.01 & 543.28 & 33.27 & 33.27 & 100\% \\
2025-02 & 3\,705 & 28\,524.47 & 32\,460.82 & 3\,936.35 & 96.22 & 75\% \\
2025-03 & 93 & 2\,994.42 & 3\,620.09 & 625.68 & -33.62 & 50\% \\
2025-05 & 317 & 5\,945.85 & 5\,754.17 & -191.69 & -27.06 & 50\% \\
2025-06 & 227 & 5\,085.71 & 4\,808.63 & -277.09 & -120.19 & 0\% \\
2025-09 & 5 & 316.84 & 152.52 & -164.32 & -164.32 & 0\% \\
2025-12 & 97 & 1\,482.99 & 2\,092.03 & 609.04 & 122.19 & 100\% \\
\bottomrule\end{tabular}
\par\smallskip\begin{minipage}{.98\linewidth}\small All launches counts every recorded launch of the selected groups, including censored launches; money columns use complete-hour launches only. Positive is the fraction of selected groups with positive recorded cash flow. None of these quantities identifies true profit.\end{minipage}
\end{table}

For April, adding a hypothetical 2\% deduction to each side changes the aggregate from 56.02 SOL to approximately $-15.21$ SOL. The corresponding changes are scenario arithmetic on turnover, not discoveries of historical costs. The very large February aggregate also needs to be read with its exposure: selected groups have thousands of launches, including many nonqualifying tokens. Gross trading, signed cash flow and episode count measure different aspects of the same activity.

\paragraph{How to read the negative flows.} The recorded flow is the SOL received from sales minus the SOL paid for purchases of the groups' own tokens in the first hour. It leaves out the tokens still held, income from other sources and everything the groups did outside the window, so a negative value does not mean that a group lost money overall. It is also a small difference between large flows: in May the net flow is $-$192 SOL on a turnover of 11\,700 SOL ($-$1.6\%), in June $-$277 SOL on a turnover of 9\,894 SOL ($-$2.8\%). A cost of about two per cent per side (Section~\ref{sec:fees}) is as large as or larger than these nets: it deepens the negative values and turns small positive ones, such as April, negative, whereas it leaves the periods with a large positive flow positive. The sign of a small net, in either direction, should therefore be taken as undetermined. 

\begin{figure}[htbp]\centering
\includegraphics[width=\linewidth]{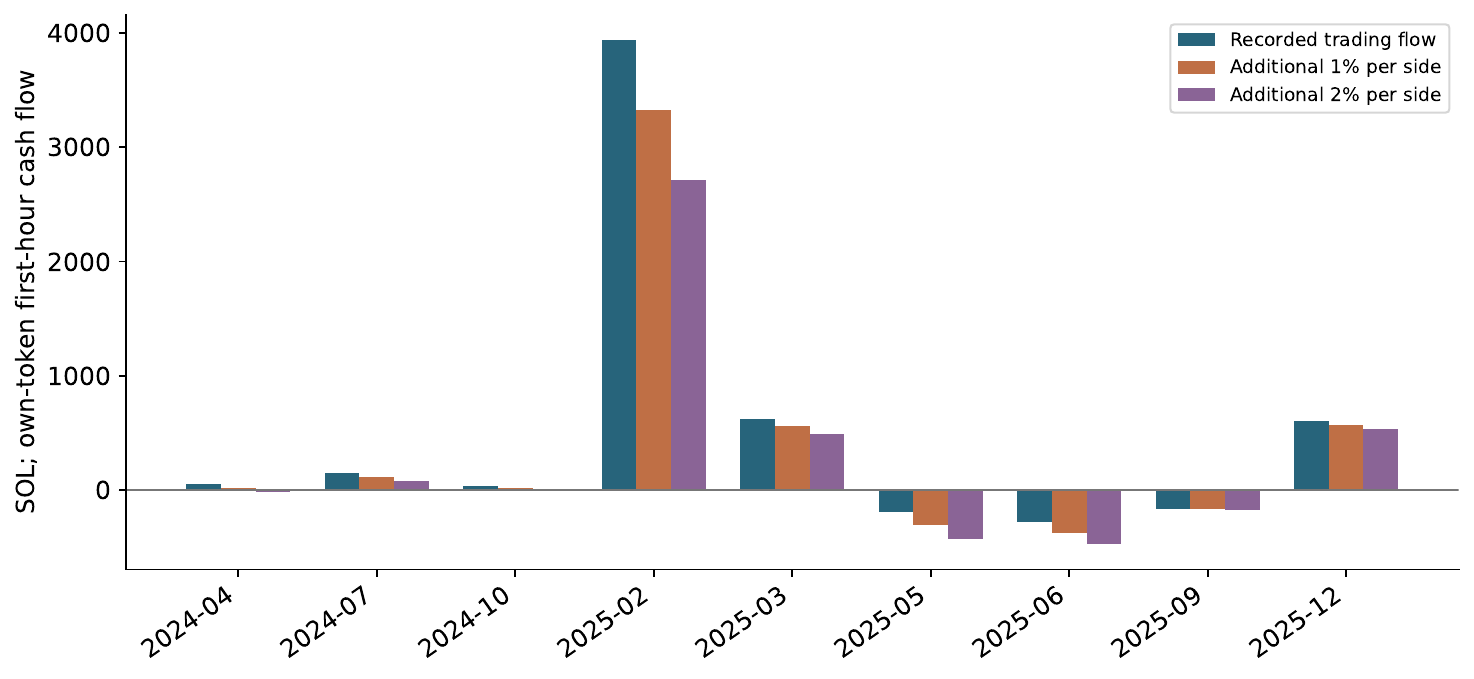}
\caption{Recorded first-hour own-token cash flow of the serial groups selected within each period, together with hypothetical additional per-side friction. Positive values are not identified realized profit; negative values are not identified lifetime losses.}\label{fig:cashflow}
\end{figure}

\begin{figure}[htbp]\centering
\includegraphics[width=\linewidth]{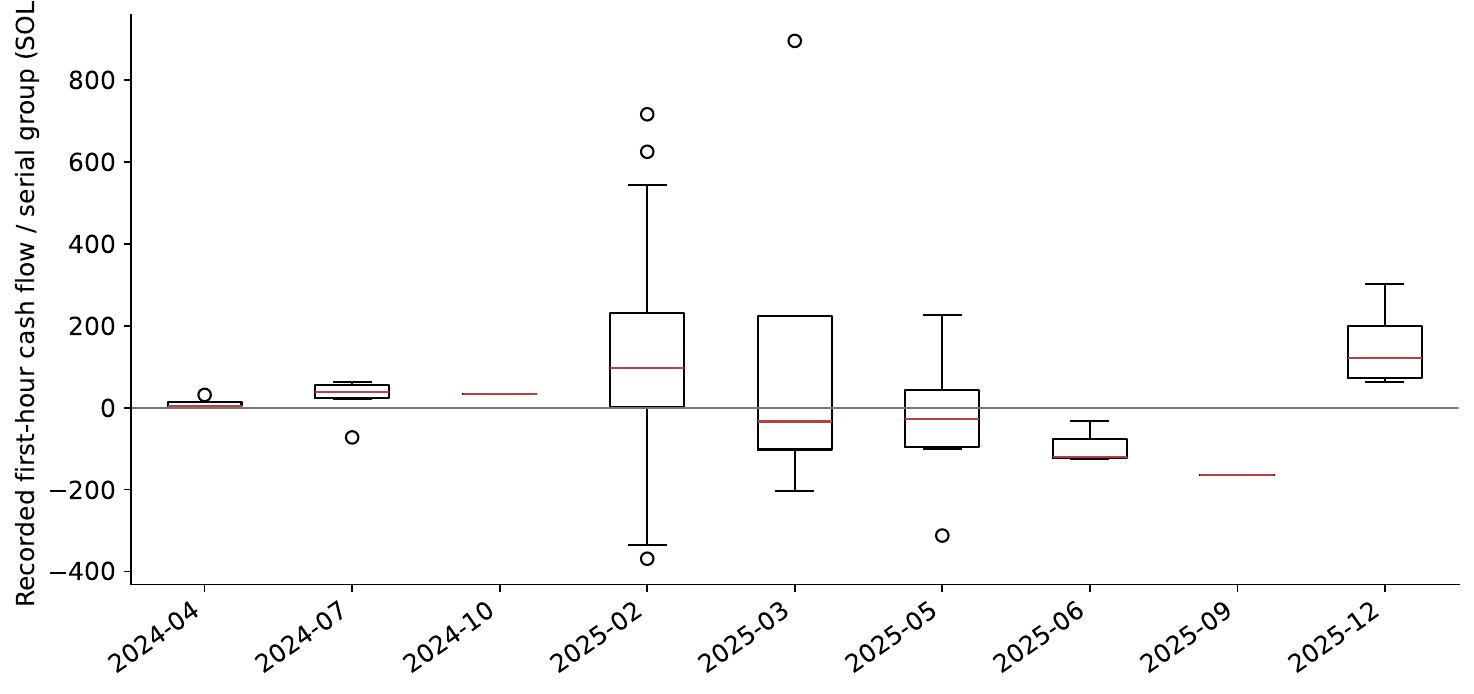}
\caption{Distribution of recorded first-hour own-token cash flow across primary serial groups. A period with a positive aggregate need not have a positive median. Individual groups are observations inside their own periods; the plot does not join their trading histories. Boxes use the observed group distribution, not sampling uncertainty.}\label{fig:groupdistribution}
\end{figure}

Role attribution can change the sign even when the set of selected groups is fixed. Appendix Table~\ref{tab:roles} partitions cash flow for the \emph{same primary serial groups} into creator-wallet and additional linked-wallet portions, as independently recomputed in SQL. This partition is distinct from the creator-scope detector, which reselects episodes and possibly groups. In the June cases, creator-wallet selling contributes positive cash flow, while the linked noncreator portion is sufficiently negative to reverse the consolidated result. Omitting related buying wallets can therefore produce a misleadingly favorable picture.

\subsection{Sensitivity, ambiguous collapses and concentration benchmark}
\begin{table}[htbp]\centering\small
\caption{Episode and serial-group sensitivity; allocation comparator.}\label{tab:robustness}
\begin{tabular}{lrrrrrrr}\toprule
Period & E & E-vol & E-cr & G & G-vol & G-cr & Alloc. med./95\% \\ \midrule
2024-04 & 38 & 36 & 36 & 5 & 4 & 5 & 0 / 1 \\
2024-07 & 43 & 41 & 41 & 6 & 5 & 6 & 0 / 1 \\
2024-10 & 36 & 36 & 32 & 1 & 2 & 1 & 0 / 1 \\
2025-02 & 328 & 319 & 263 & 32 & 30 & 24 & 7 / 10 \\
2025-03 & 39 & 39 & 32 & 4 & 4 & 3 & 0 / 1 \\
2025-05 & 66 & 63 & 37 & 6 & 6 & 2 & 0 / 1 \\
2025-06 & 48 & 44 & 23 & 3 & 3 & 1 & 0 / 1 \\
2025-09 & 42 & 40 & 42 & 1 & 1 & 1 & 0 / 0 \\
2025-12 & 39 & 39 & 32 & 4 & 4 & 2 & 0 / 0 \\
\bottomrule\end{tabular}
\par\smallskip\begin{minipage}{.98\linewidth}\small vol: candles recalculated using trades of at least 0.01 SOL and 1,000 tokens. cr: creator-wallet trade scope within the same groups. Allocation numbers are descriptive comparator quantiles, not inferential significance thresholds.\end{minipage}
\end{table}

The higher-volume price reconstruction produces somewhat fewer episodes in most windows, while creator-only trading scope changes some periods much more strongly. May falls from 66 primary episodes to 37 under creator scope; June falls from 48 to 23. Such differences indicate dependence on the accounting boundary, rather than a universally stable fraud count. The grid includes settings with the ambiguity guard disabled.

\begin{figure}[htbp]\centering
\includegraphics[width=\linewidth]{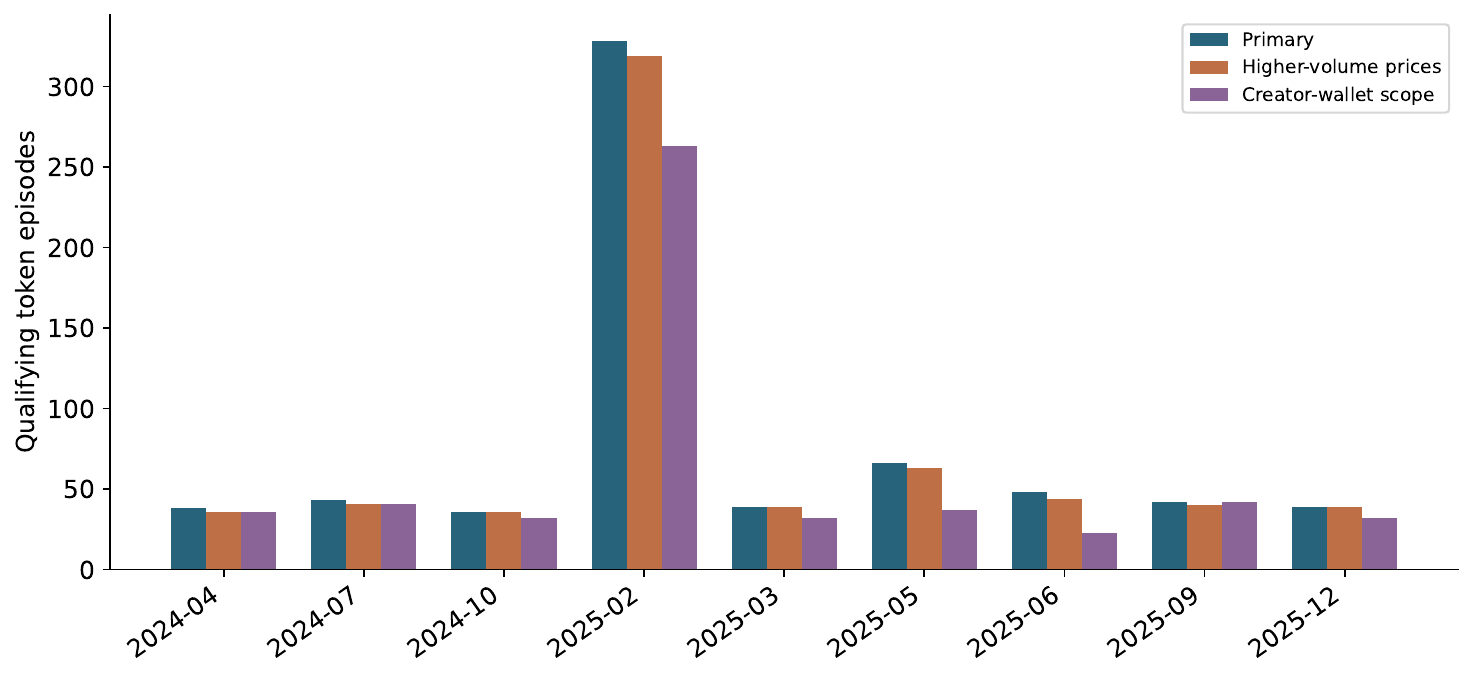}
\caption{Token-episode counts under independently recalculated higher-volume prices and creator-wallet trading scope. Creator scope retains the primary inferred group memberships. It does not constitute an alternative isolated-creator clustering.}\label{fig:scope}
\end{figure}

The observed serial counts usually exceed the descriptive random-allocation median and upper quantile. February has 32 groups against a comparator median of seven and a 95th percentile of ten. October and September have only one observed group, so their interpretation should remain modest even when the comparator's typical count is zero. We do not translate exceedance frequencies into a fraud probability or a predictive significance claim.

\begin{figure}[htbp]\centering
\includegraphics[width=\linewidth]{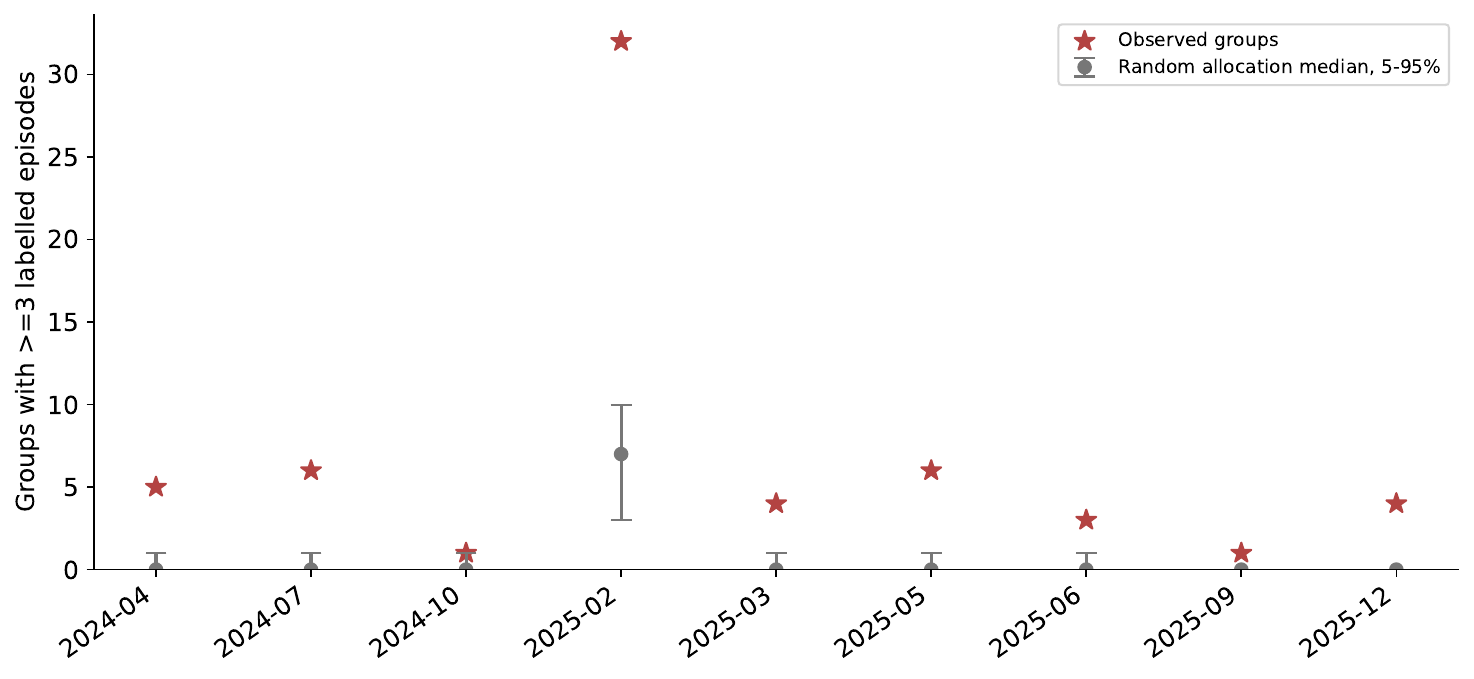}
\caption{Observed repeat concentration and hour-stratified allocation comparison. Error bars are the comparator's 5th--95th percentiles over 1,000 allocations. They are neither sampling confidence intervals nor a validated null distribution for intentional misconduct.}\label{fig:allocation}
\end{figure}

\begin{table}[htbp]\centering\small
\caption{Ranges across the detector grid within each period. Each guard setting has 243 parameter combinations.}\label{tab:grid}
\begin{tabular}{lrrrr}\toprule
Period & Guard: episodes & Guard: groups & No guard: episodes & No guard: groups \\ \midrule
2024-04 & 1--52 & 0--7 & 1--53 & 0--7 \\
2024-07 & 8--101 & 0--22 & 22--120 & 1--25 \\
2024-10 & 2--76 & 0--9 & 5--82 & 0--11 \\
2025-02 & 45--624 & 1--73 & 118--738 & 8--81 \\
2025-03 & 9--91 & 1--15 & 22--107 & 1--17 \\
2025-05 & 3--167 & 0--25 & 16--190 & 0--30 \\
2025-06 & 3--156 & 0--16 & 6--165 & 0--19 \\
2025-09 & 0--141 & 0--4 & 4--160 & 0--5 \\
2025-12 & 0--115 & 0--19 & 3--122 & 0--19 \\
\bottomrule\end{tabular}
\par\smallskip\begin{minipage}{.98\linewidth}\small Ranges combine different threshold settings and do not represent confidence intervals. The minimum series count itself varies between two, three and five.\end{minipage}
\end{table}

The twelve-hour exposure check changes the number of groups meeting the repeat threshold, especially for longer windows. It does not preserve the full-period ranking mechanically. As graph membership still uses the full window, this result should be read as evidence that the observation horizon matters, not as complete adjustment for duration or prospective information.

\subsection{Examples of episodes}\label{sec:examples}
The episodes found by the screen are heterogeneous. We sort the 679 primary episodes by simple rules on observable features, checked in this order so that the types are mutually exclusive: \emph{thin market} (fewer than 20 recorded trades before the drop); \emph{creator dump} (the creator wallet makes at least 80\% of the team's sales in the window and at most two team wallets sell); \emph{multi-wallet exit} (five or more team wallets sell in the window); \emph{gradual decline} (the price falls over at least two minutes); and \emph{other fast exit}. Table~\ref{tab:epitypes} gives the counts. Most episodes are creator dumps (64\%), but only 34\% of them belong to serial groups, whereas 75\% of the multi-wallet exits do; thin-market episodes are about a seventh of all, which is why the trade-count filter of Section~\ref{sec:robust} removes many April episodes.

\begin{table}[htbp]\centering\small\caption{Types of primary episodes (rules in the text; medians).}\label{tab:epitypes}
\resizebox{\linewidth}{!}{\begin{tabular}{lrrrrrrrr}\toprule Type & Episodes & Share & In serial groups & Team sales (SOL) & Selling wallets & Peak to drop (s) & Trades before drop & Peak / launch price \\ \midrule
Creator dump & 436 & 64\% & 34\% & 47 & 1 & 26 & 52 & 3.7 \\
Multi-wallet exit & 118 & 17\% & 75\% & 26 & 10 & 7 & 188 & 4.6 \\
Thin market & 90 & 13\% & 58\% & 52 & 1 & 32 & 14 & 2.8 \\
Other fast exit & 28 & 4\% & 57\% & 33 & 3 & 16 & 100 & 5.2 \\
Gradual decline & 7 & 1\% & 29\% & 20 & 2 & 289 & 98 & 6.2 \\
\bottomrule\end{tabular}}\par\smallskip\begin{minipage}{.98\linewidth}\small The migration guard removes 165 further candidate episodes (median recorded peak reserve 75 SOL, median 1 selling wallet); they are shown in the last panel of Figure~\ref{fig:gallery} but not counted here.\end{minipage}\end{table}

\begin{figure}[htbp]\centering\includegraphics[width=\linewidth]{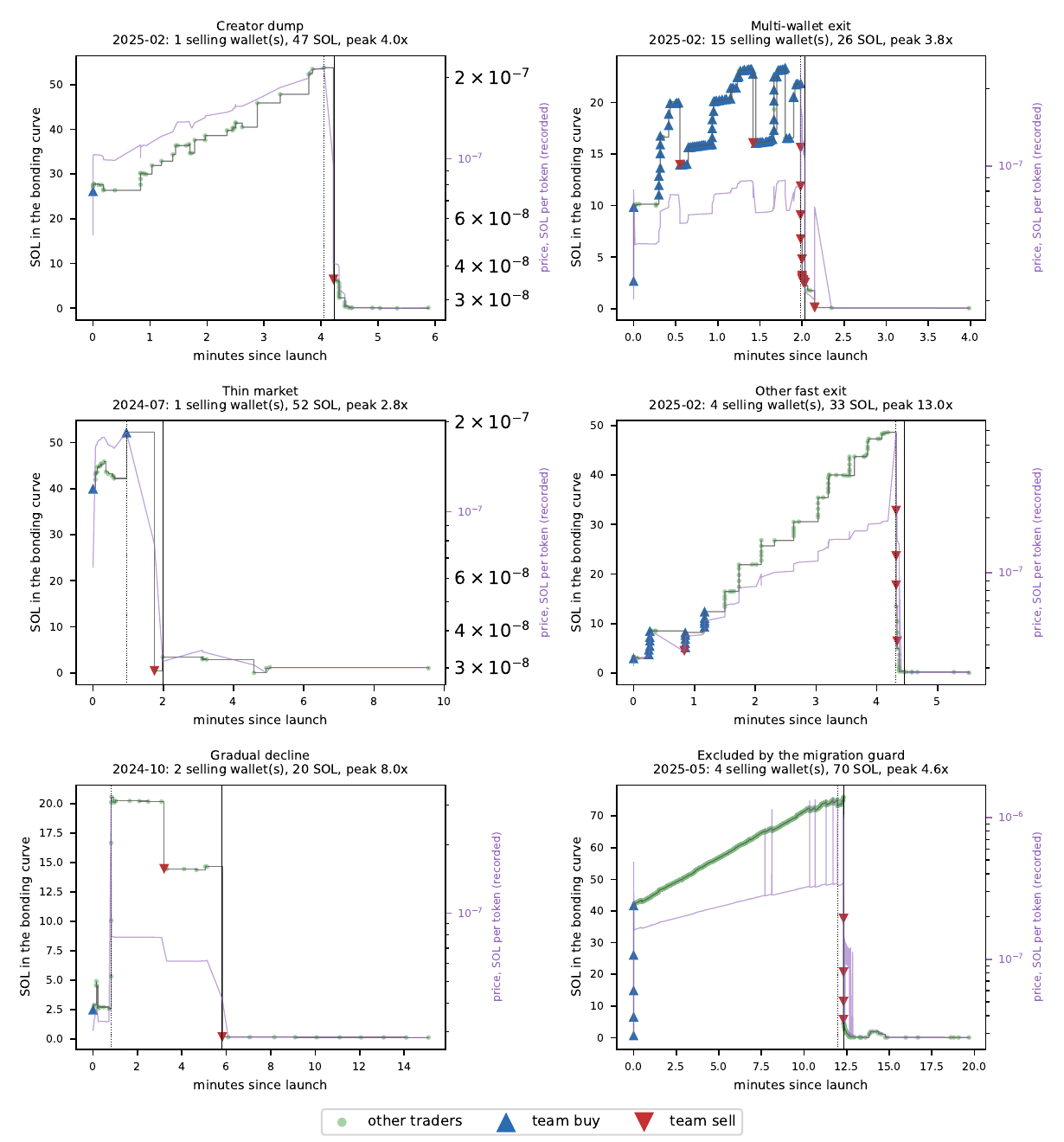}\caption{Representative episode of each type (closest to the type median in team sales). Blue triangles: team buys; red triangles: team sells; dots: other traders; vertical axis: SOL in the bonding curve after the trade; thin line (right axis): recorded price; dotted vertical line: peak; solid line: first qualifying drop.}\label{fig:gallery}\end{figure}

Figure~\ref{fig:gallery} shows, for every type, the episode whose team sales are closest to the median of the type, so that no example is picked by eye. Team purchases and sales are marked, other traders' trades are dots and every trade is drawn at the SOL in the bonding curve after it (left axis); the thin line on the right axis is the recorded price. In the creator dump the price rises on purchases of other traders and one wallet sells at the peak. In the multi-wallet exit many team wallets buy during the rise and then sell within seconds. The thin-market episode consists of a handful of trades, the gradual decline of a few steps over minutes, and the example excluded by the migration guard rises for twelve minutes to a recorded reserve of about 75 SOL before falling, a pattern that cannot be told from a normal graduation without a migration label.

\begin{figure}[p]\centering\includegraphics[width=.97\linewidth]{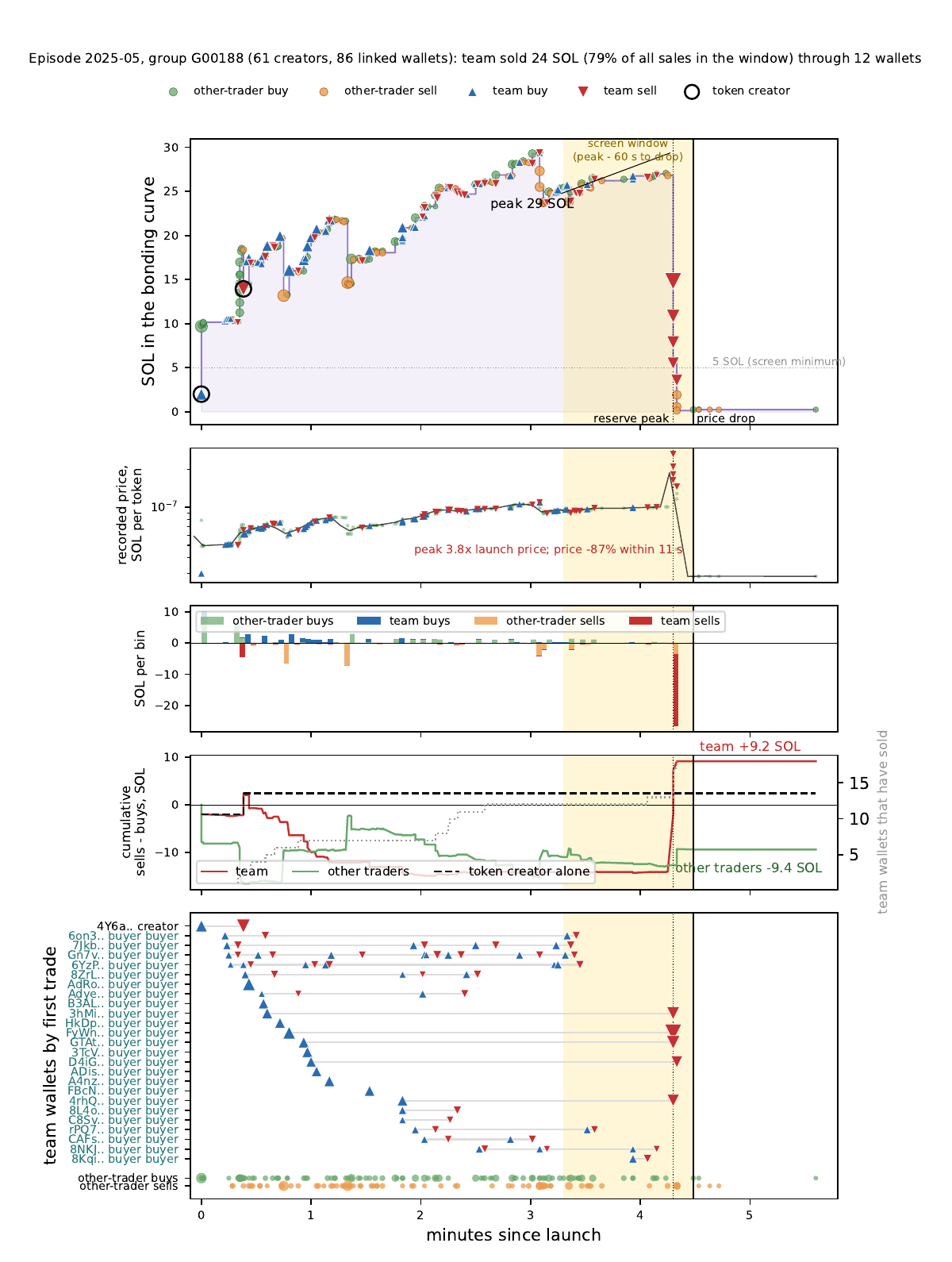}\caption{One episode in detail. Top to bottom: SOL in the bonding curve after each trade (marker area grows with the SOL traded; black ring: token creator; shaded: the screen's window from 60 s before the peak to the drop); the recorded price with its 5-second volume-weighted average (secondary, because the recorded price may be unreliable); SOL traded per time bin by team and other traders (up: buys, down: sells); cumulative sells minus buys of the team, of the other traders and of the token creator, with the number of team wallets that have sold (right axis); and a timeline of the team wallets in order of their first trade (other traders' buys and sells in the last two rows).}\label{fig:detail}\end{figure}

Figure~\ref{fig:detail} shows one episode in detail. It is chosen by an explicit rule and not at random: a serial group, at least six selling team wallets, other traders accounting for at least half of the trades before the drop and for at least 30 SOL of purchases from at least 50 wallets, and the highest score among the candidates. It is therefore a clear and well-observed case, not a representative one. The token belongs to group G00188 in 2025-05, whose 61 creators and 86 linked wallets were connected by funding and repeat early buying. The price rises 3.8-fold over about four minutes while 65 other-trader wallets buy 40 SOL and team wallets trade on both sides; the cumulative team flow stays negative, at about $-$14 SOL. Then 12 team wallets sell 24 SOL within seconds (79\% of all sales in the window), the price falls 87\% in 11 seconds and the real reserve falls from about 29 SOL to nearly zero. The team ends this token about 9 SOL ahead and other traders about 9 SOL behind. The lower panel shows that most of the selling wallets had traded earlier in the same token, so that the coincidence of price collapse, reserve loss and concentrated team selling is visible wallet by wallet. It does not show common ownership: the wallets are linked only through the funding and early-buying evidence described in the section on candidate teams, and the pattern is equally compatible with coordinated trading by independent participants.

\subsection{Early inflow of other traders and team behaviour}\label{sec:inflow}
If many other traders buy a token early, a team gains nothing by pushing the price up with its own money: any further team purchase costs fees and slippage, and the team's profit comes from the other traders' SOL. Teams that watched the inflow could therefore stop buying and sell before the reserve peaks, instead of collapsing the token afterwards. We test this with the repeat groups of the primary construction (groups with three or more launches in the window), using their tokens with a complete first hour, a recorded reserve peak of at least 5 SOL and a team position (team purchases of at least 0.1 SOL): 82\,333 tokens of 12\,768 groups. Other traders are all wallets outside the token's group. The early inflow is the other traders' purchase volume in SOL in the first 60 seconds; the reserve peak and all flows use SOL amounts and the real reserve only. A team has \emph{exited} when it has sold more than half of its largest holding before the reserve peak (token amounts are used only for this position).

Further team buying after the first minute is rare: more than 1 SOL of team purchases after 60 seconds occurs in 3.7\% of the tokens, and in 8.6\% of those with the lowest early inflow against about 2.4\% of the others (Table~\ref{tab:inflow}). Exits before the peak are common: 34\% of the positions are mostly sold before the reserve peak. The exit share rises from 19\% to 63\% across the quintiles of the early inflow, the median first team sale moves from +6 to $-$15 seconds relative to the peak, and the team's net first-hour cash per token rises from $-$0.21 to 1.85 SOL (positive in 53\% and 91\% of the tokens). The screen sees almost none of this: only 0.5\% of the tokens are episodes of the reserve screen, 0.06\% of the tokens whose position was mostly sold before the peak against 0.79\% of the others. Quiet exits are common and invisible to an episode screen.

\begin{table}[htbp]\centering\small\caption{Early inflow of other traders and team behaviour (tokens of repeat groups, quintiles of the other traders' purchases in the first 60 s).}\label{tab:inflow}
\resizebox{\linewidth}{!}{\begin{tabular}{lrrrrrrrr}\toprule Quintile & Early inflow (SOL, median) & Tokens & Team buys $>$1 SOL after 60 s & Exited before peak & First team sale vs peak (s) & Team net cash (SOL) & Net $>$0 & Screen episode \\ \midrule
Q1 & 1.0 & 16\,467 & 8.6\% & 19\% & 6 & $-$0.21 & 53\% & 1.4\% \\
Q2 & 4.7 & 16\,466 & 2.9\% & 18\% & 3 & 0.24 & 76\% & 0.3\% \\
Q3 & 8.2 & 16\,467 & 2.1\% & 29\% & 2 & 0.52 & 84\% & 0.2\% \\
Q4 & 14.5 & 16\,466 & 2.7\% & 43\% & 0 & 1.05 & 89\% & 0.3\% \\
Q5 & 35.4 & 16\,467 & 2.0\% & 63\% & $-$15 & 1.85 & 91\% & 0.6\% \\
\bottomrule\end{tabular}}\par\smallskip\begin{minipage}{.98\linewidth}\small Exited: more than half of the team's largest holding sold before the reserve peak. First team sale vs peak: median seconds after (positive) or before (negative) the reserve peak. Team net cash: sales minus purchases of the team in the first hour. Screen episode: episode of the reserve-only screen.\end{minipage}\end{table}

The pattern across quintiles overstates the role of decisions. Tokens with strong early inflow peak later, which leaves more time to sell before the peak. Within groups and controlling for the timing of the peak, each e-fold increase (2.7 times) of the early inflow raises the exit probability by 0.09, compared with 0.16 without the control, and among tokens whose reserve peaks after five minutes the effect is 0.008 (standard error 0.005) (Table~\ref{tab:inflowrobust}): indistinguishable from zero, although the raw gradient remains (48\% to 80\% exited). The association of the early inflow with the team's result (0.68 SOL per log unit within groups, 0.57 for late peaks) is robust, but it measures the other traders' money brought to the token, not a team decision. Finally, tokens in which a team kept buying despite a high early inflow end worse (0.63 against 1.47 SOL), but within groups the coefficient of extra buying on the result is +0.21 (standard error 0.26), so the difference reflects which tokens need pumping and not an estimated cost of pumping. The data are consistent with teams that mostly do not keep buying and often sell early, but they do not show that teams react to the inflow.

\begin{table}[htbp]\centering\small\caption{Within-group regressions on the log early inflow, with and without control for the timing of the reserve peak (coefficient and cluster-robust standard error).}\label{tab:inflowrobust}
\resizebox{\linewidth}{!}{\begin{tabular}{llrrr}\toprule Sample & Outcome & No control & Control: log peak time & Tokens \\ \midrule
all & Exited before peak & +0.163 (0.004) & +0.092 (0.003) & 82\,333 \\
all & Team net cash (SOL) & +0.741 (0.040) & +0.683 (0.046) & 82\,333 \\
peak at least 300 s & Exited before peak & +0.007 (0.005) & +0.008 (0.005) & 10\,553 \\
peak at least 300 s & Team net cash (SOL) & +0.591 (0.098) & +0.574 (0.098) & 10\,553 \\
peak reserve at least 20 SOL & Exited before peak & +0.022 (0.005) & +0.067 (0.005) & 16\,703 \\
peak reserve at least 20 SOL & Team net cash (SOL) & +0.403 (0.215) & +0.500 (0.209) & 16\,703 \\
\bottomrule\end{tabular}}\par\smallskip\begin{minipage}{.9\linewidth}\small Group fixed effects, standard errors clustered by group; both specifications control for the log of the team's purchases in the first 60 s.\end{minipage}\end{table}

\section{Robustness to accounting, team construction and window length}\label{sec:robust}

\subsection{Reliance on the reserve: a price-free screen}\label{sec:reserve}
Token prices in the primary screen are computed from trade amounts (SOL divided by tokens), and a price can be wrong for a trade even when the SOL amount and the curve state are right. On a constant-product curve the marginal price is a function of the real reserve, $p=(x_0+R)^2/k$ with $x_0=30$ SOL, so the reserve carries the same information without the token amounts. As a check, the trade price lies within 10\% of the reserve-implied price for 96\% to 100\% of the trades of at least 0.01 SOL (in the sampled tokens of each window), but more than 25\% away for up to 3\% of them; the price extremes on which a screen depends are drawn from these tails. Two features of the price screen also depend on how a price is defined. The rise is measured from the average price of the first trade, which for a large launch purchase understates the price of the resulting curve state. And an 80\% fall of the implied price is only possible if the peak reserve exceeded about 37 SOL, yet 20\% of the primary episodes have a recorded peak reserve below that (47\% in May and 52\% in June 2025).

We therefore repeat the screen using only the real reserve and SOL amounts. The rise is measured against the empty curve (an implied price at least twice the launch price, i.e.\ a peak reserve of at least 12.4 SOL); the collapse is an implied-price fall of at least 80\% and a reserve fall of at least 50\% within ten minutes; peak and drop use the conservative bounds of the reserve within a second; the team conditions use SOL amounts only; and the migration guard (recorded reserve of at least 70 SOL before the drop) is kept. Table~\ref{tab:reserveedit} compares the two screens.

\begin{table}[htbp]\centering\small\caption{Price-based and reserve-only screens (migration guard on).}\label{tab:reserveedit}
\resizebox{\linewidth}{!}{\begin{tabular}{lrrrrrrrrr}\toprule Period & Price episodes & Reserve episodes & Both & Price only & Reserve only & Serial groups (price/reserve) & Flow, price screen (SOL) & Flow, reserve screen (SOL) & 95\% interval \\ \midrule
2024-04 & 38 & 44 & 38 & 0 & 6 & 5/5 & 56 & 56 & [18, 104] \\
2024-07 & 43 & 57 & 42 & 1 & 15 & 6/7 & 149 & 706 & [$-$3, 1\,840] \\
2024-10 & 36 & 46 & 32 & 4 & 14 & 1/1 & 33 & 33 & -- \\
2025-02 & 328 & 358 & 278 & 50 & 80 & 32/30 & 5\,353 & 5\,593 & [3\,377, 8\,029] \\
2025-03 & 39 & 38 & 31 & 8 & 7 & 4/4 & 923 & 1\,103 & [115, 2\,783] \\
2025-05 & 66 & 59 & 40 & 26 & 19 & 6/6 & $-$167 & 648 & [263, 1\,057] \\
2025-06 & 48 & 38 & 28 & 20 & 10 & 3/4 & $-$130 & $-$80 & [$-$363, 204] \\
2025-09 & 42 & 44 & 41 & 1 & 3 & 1/1 & $-$160 & $-$160 & -- \\
2025-12 & 39 & 40 & 38 & 1 & 2 & 4/4 & 609 & 609 & [281, 981] \\
\bottomrule\end{tabular}}\par\smallskip\begin{minipage}{.98\linewidth}\small Flow: own-token cash flow of the serial groups of the respective screen over the whole window plus unsold inventory valued on the curve, same complete-hour launches and the same group construction; the interval is a group bootstrap for the reserve screen (omitted for fewer than three groups).\end{minipage}\end{table}

The reserve screen finds 724 episodes against 679; 84\% of the price-based episodes are also found by it. The price-only episodes are concentrated in May (26 of 66), June (20 of 48) and February 2025 (50), which is where the price criterion is least supported by the reserve; we did not analyse the reserve-only episodes further. Serial-group counts change little. The cash flow of the serial groups changes in sign in one window: in May 2025 it is $-$167 SOL under the price screen and 648 SOL under the reserve screen, so the negative May result of the primary analysis depends on the price definition. In June it remains negative but smaller ($-$80 SOL, interval $-$363 to 204), and September rests on the same single group. July differs because the reserve screen finds a seventh serial group (706 SOL against 149 SOL). The negative results of the price-based analysis for May and June (Section~\ref{sec:robust}, Table~\ref{tab:flows2}) should therefore be read as results of that screen, not of the data in general. The remainder of this section keeps the price-based screen as the primary definition for continuity; the reserve screen rests on fields that do not depend on token amounts and should be preferred in future work.
\subsection{Cash flow beyond the first hour}
The first-hour measure may understate the groups' outcome if their tokens are sold later. Table~\ref{tab:flows2} compares it with whole-window flow on the same complete-hour launches and with whole-window flow plus the fee-free curve value of tokens still held at the end of the window (Section~\ref{sec:economics_link}). In the three negative periods the sign is unchanged: May -191.7, -172.9, -167.4 SOL; June -277.1, -217.8, -129.6 SOL; September -164.3, -164.3, -159.6 SOL. Several positive periods increase when later sales are included (February 3\,936.4 to 5\,352.9 SOL). Unsold inventory valued on the curve is small compared with turnover. The sign is therefore not a first-hour truncation artifact. The group-cluster bootstrap, however, gives intervals that include zero in May and June, and only one group is involved in September.
For the fee question, the reserve fields show that recorded SOL equals the change in the curve's real reserve for both sides: the median ratio of reserve change to recorded SOL equals 1.0000 to four decimals for buys and for sells in every window (range 1.0000 to 1.0000). Recorded amounts thus exclude any protocol fee paid outside the curve, so the recorded flows overstate net outcomes by the fees; the effective cost is measured on a transaction sample in Section~\ref{sec:fees}.
\begin{table}[htbp]\centering\small
\caption{Primary serial groups' own-token cash flow: first hour, whole recorded window, and whole window plus unsold inventory valued on the final curve state (SOL).}\label{tab:flows2}
\resizebox{\linewidth}{!}{\begin{tabular}{lrrrrrrrrr}\toprule
Period & Groups & 1 h & Window & Window + inventory & 95\% interval & Inventory value & Open/tokens & Median & Pos. \\ \midrule
2024-04 & 5 & 56.0 & 56.0 & 56.0 & [18, 109] & 0.0 & 0/37 & 5.0 & 100\% \\
2024-07 & 6 & 148.9 & 148.9 & 149.1 & [-101, 321] & 0.2 & 4/38 & 38.1 & 83\% \\
2024-10 & 1 & 33.3 & 33.3 & 33.3 & -- & 0.0 & 0/10 & 33.3 & 100\% \\
2025-02 & 32 & 3\,936.4 & 5\,172.4 & 5\,352.9 & [2\,990, 7\,887] & 180.6 & 455/3\,635 & 115.1 & 84\% \\
2025-03 & 4 & 625.7 & 849.8 & 923.4 & [-133, 2\,690] & 73.7 & 35/93 & 47.4 & 75\% \\
2025-05 & 6 & -191.7 & -172.9 & -167.4 & [-892, 615] & 5.5 & 251/309 & -26.3 & 50\% \\
2025-06 & 3 & -277.1 & -217.8 & -129.6 & [-359, 156] & 88.2 & 223/227 & -62.1 & 33\% \\
2025-09 & 1 & -164.3 & -164.3 & -159.6 & -- & 4.7 & 2/5 & -159.6 & 0\% \\
2025-12 & 4 & 609.0 & 609.0 & 609.2 & [281, 981] & 0.2 & 14/97 & 122.3 & 100\% \\
\bottomrule\end{tabular}}
\par\smallskip\begin{minipage}{.98\linewidth}\small All three measures use the same complete-hour launches of the same groups. Inventory value is the fee-free curve sale value of tokens the group still holds at window end, capped at real reserve (a lower bound on realisable value after later trading is unknown). The interval is a group-cluster bootstrap over serial groups (2,000 draws); it is omitted for fewer than three groups. Neither column is identified realized profit.\end{minipage}
\end{table}

\begin{figure}[htbp]\centering\includegraphics[width=.9\linewidth]{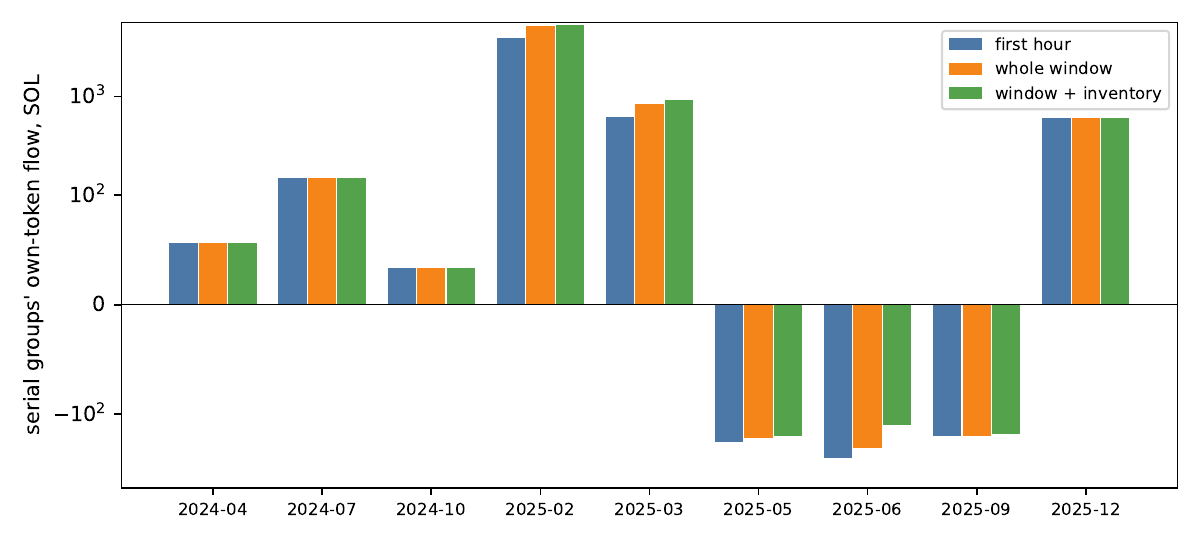}\caption{Serial groups' own-token flow under the three accounting measures (symmetric log axis).}\label{fig:flowmeasures}\end{figure}
\subsection{Team construction and equal-length windows}
\begin{figure}[htbp]\centering\includegraphics[width=.9\linewidth]{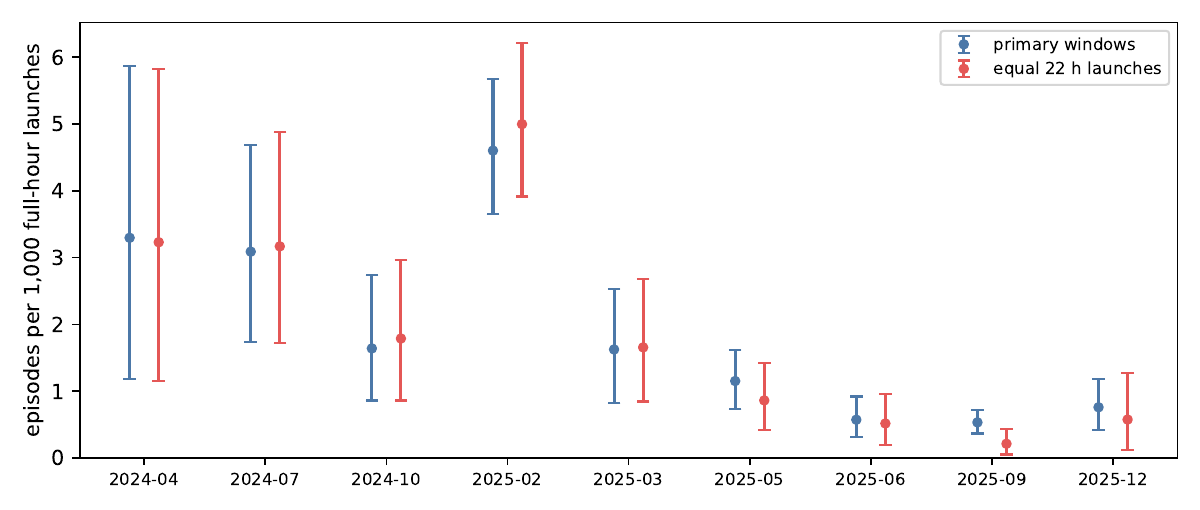}\caption{Episode incidence with group-cluster bootstrap intervals, primary windows and equal 22-hour launch windows.}\label{fig:ratesci}\end{figure}

Table~\ref{tab:variants2} reconstructs groups under stricter rules. Removing the funded repeat early buyers (creators and shared funders only) leaves April and October unchanged but cuts serial groups in May from 6 to 2, June from 3 to 1 and February from 32 to 24, and it changes the sign of the May and June flows (window plus inventory: May 258.5, June 52.1 SOL). The ten-fold tighter limit of five creators per funder changes little, and the per-hour bridge rule behaves like the no-buyer variant because both drop the same extension. The primary negative result for May and June therefore depends on a behavioural wallet extension whose validity is untested. September's group is created from creators and funding alone and remains negative.
Windows differ in length from 23.7 to 79.3 hours, and a threshold of three episodes is easier to reach in a longer window. Restricting to launches in the first 22 hours of each window and rebuilding groups inside it gives 22 serial groups in February (primary 32), 2 in July (6), 1 in May (6), 1 in June (3), 0 in September (1) and 1 in December (4). Episode incidence per launch changes much less in most windows (Figure~\ref{fig:ratesci}; February 4.60 to 5.00, July 3.09 to 3.17) but falls in September (0.53 to 0.21), so incidence is more comparable across windows than serial-group counts, though not uniformly. Equal-length rebuilding uses the transfer-based first funders already extracted for each creator's first launch and therefore applies to launches inside the shorter window.
\begin{table}[htbp]\centering\small
\caption{Episodes / serial groups under alternative team constructions.}\label{tab:variants2}
\resizebox{\linewidth}{!}{\begin{tabular}{lrrrrrrrr}\toprule
Period & v1 & no buyers & cap 5 & bridge/h & eq22 & eq22 no b. & eq22 cap5 & eq22 br/2 \\ \midrule
2024-04 & 38/5 & 36/5 & 38/5 & 36/5 & 36/5 & 34/5 & 36/5 & 34/5 \\
2024-07 & 43/6 & 42/6 & 43/6 & 42/6 & 29/2 & 29/2 & 29/2 & 29/2 \\
2024-10 & 36/1 & 34/1 & 34/1 & 34/1 & 25/1 & 23/1 & 23/1 & 23/1 \\
2025-02 & 328/32 & 263/24 & 328/30 & 263/24 & 239/22 & 196/17 & 239/20 & 196/17 \\
2025-03 & 39/4 & 32/3 & 39/4 & 32/3 & 38/4 & 31/3 & 38/4 & 31/3 \\
2025-05 & 66/6 & 37/2 & 66/6 & 37/2 & 20/1 & 11/0 & 20/1 & 11/0 \\
2025-06 & 48/3 & 23/1 & 47/3 & 23/1 & 11/1 & 6/0 & 11/1 & 6/0 \\
2025-09 & 42/1 & 42/1 & 42/1 & 42/1 & 4/0 & 4/0 & 4/0 & 4/0 \\
2025-12 & 39/4 & 36/3 & 36/3 & 40/3 & 9/1 & 9/1 & 9/1 & 9/1 \\
\bottomrule\end{tabular}}
\par\smallskip\begin{minipage}{.98\linewidth}\small v1: primary construction. no buyers: creators and shared funders only. cap 5: at most five creators per shared funder (primary 50). bridge/h: bridge exclusion at 42 transfer endpoints per window hour (equals the primary 1,000 in the shortest window) and no buyer extension. eq22: only launches in the first 22 hours of each window, groups rebuilt inside that equal-length window. br/2: bridge threshold 21 per hour. Bridge variants omit the repeat-buyer extension because those transfer links were extracted under the primary rule.\end{minipage}
\end{table}

\subsection{Uncertainty, thin markets and the concentration benchmark}
Table~\ref{tab:rates2} reports group-cluster bootstrap intervals for incidence, e.g. 4.60 per 1,000 in February (95\% interval 3.65 to 5.68) against 0.53 in September (0.37 to 0.72); the ordering of the extreme periods is therefore not a sampling accident, although the selection of windows is not random. The revised concentration benchmark shuffles group labels among price-eligible launches within each creation hour and keeps the observed episode set, so group launch counts are preserved. In every window the observed number of serial groups exceeds the permuted median, with permutation $p$ at most 0.051 (the weakest cases are windows with a single serial group). This shows that episodes are concentrated in groups, not that groups act fraudulently.
\begin{table}[htbp]\centering\small
\caption{Episode incidence with cluster-bootstrap intervals and the label-permutation comparison.}\label{tab:rates2}
\resizebox{\linewidth}{!}{\begin{tabular}{lrrrrrrrr}\toprule
Period & Per 1,000 & 95\% int. & Eq22 per 1,000 & 95\% int. & Serial & Permuted: median [5\%, 95\%] & $p$ serial & $p$ max \\ \midrule
2024-04 & 3.30 & [1.18, 5.87] & 3.23 & [1.15, 5.83] & 5 & 0 [0, 1] & 0.001 & 0.001 \\
2024-07 & 3.09 & [1.73, 4.69] & 3.17 & [1.71, 4.88] & 6 & 0 [0, 1] & 0.001 & 0.002 \\
2024-10 & 1.64 & [0.86, 2.74] & 1.79 & [0.86, 2.97] & 1 & 0 [0, 0] & 0.051 & 0.001 \\
2025-02 & 4.60 & [3.65, 5.68] & 5.00 & [3.91, 6.21] & 32 & 7 [3, 10] & 0.001 & 0.001 \\
2025-03 & 1.62 & [0.82, 2.54] & 1.66 & [0.84, 2.69] & 4 & 0 [0, 1] & 0.001 & 0.001 \\
2025-05 & 1.15 & [0.73, 1.62] & 0.86 & [0.42, 1.42] & 6 & 0 [0, 1] & 0.001 & 0.001 \\
2025-06 & 0.57 & [0.31, 0.92] & 0.52 & [0.19, 0.95] & 3 & 0 [0, 1] & 0.001 & 0.001 \\
2025-09 & 0.53 & [0.37, 0.72] & 0.21 & [0.05, 0.44] & 1 & 0 [0, 0] & 0.011 & 0.011 \\
2025-12 & 0.76 & [0.42, 1.18] & 0.57 & [0.12, 1.27] & 4 & 0 [0, 0] & 0.001 & 0.001 \\
\bottomrule\end{tabular}}
\par\smallskip\begin{minipage}{.98\linewidth}\small Intervals resample groups (with their launches and episodes) 2,000 times. In the permutation comparison group labels are shuffled among price-eligible launches within each creation hour (1,000 draws) while the observed episode set is held fixed; $p$ is the share of draws, plus one, with at least as many serial groups (or as large a maximum group). It tests concentration of episodes by group label, not fraud.\end{minipage}
\end{table}

Many episodes occur in very thin markets. Requiring at least 20 recorded trades before the drop reduces April episodes from 38 to 14 and its serial groups from 5 to 1, and February from 328 to 297; the June and December counts hardly change (Table~\ref{tab:guards2}). Using a full-curve guard at 84 SOL instead of the primary ambiguity guard retains more episodes in the larger windows, which shows that the primary guard is conservative and unvalidated.
\begin{table}[htbp]\centering\small
\caption{Episodes / serial groups under alternative ambiguity and thin-market filters (primary grouping).}\label{tab:guards2}
\resizebox{\linewidth}{!}{\begin{tabular}{lrrrrr}\toprule
Period & primary & no guard & full-curve guard & >=20 trades & >=50 trades \\ \midrule
2024-04 & 38/5 & 39/5 & 38/5 & 14/1 & 3/0 \\
2024-07 & 43/6 & 60/7 & 60/7 & 32/3 & 21/2 \\
2024-10 & 36/1 & 40/1 & 40/1 & 30/1 & 15/0 \\
2025-02 & 328/32 & 422/39 & 419/39 & 297/28 & 181/14 \\
2025-03 & 39/4 & 54/6 & 54/6 & 38/4 & 24/2 \\
2025-05 & 66/6 & 86/8 & 85/8 & 57/5 & 45/5 \\
2025-06 & 48/3 & 51/3 & 51/3 & 45/3 & 38/3 \\
2025-09 & 42/1 & 50/1 & 50/1 & 38/1 & 22/0 \\
2025-12 & 39/4 & 42/5 & 42/5 & 39/4 & 36/4 \\
\bottomrule\end{tabular}}
\par\smallskip\begin{minipage}{.98\linewidth}\small Full-curve guard drops episodes whose recorded reserve reached 84 SOL before the drop; the primary guard uses 70 SOL plus a residual-reserve signature. Trade filters require that many recorded trades of the token up to the drop.\end{minipage}
\end{table}

\subsection{Link between the theory and the data}\label{sec:economics_link}
The unsold-inventory value uses the sale-proceeds expression of Section 3 with virtual reserves $x=30+R$ and $y=k/x$ for a final real reserve $R$ and $k=30\times1.073\times10^{9}$, capped at $R$. These are the initial virtual reserves of the public protocol description and they reproduce the recorded prices of tokens at the start of their curves; later program versions may differ, and graduated tokens are valued conservatively at their residual curve reserve.
\subsection{SOL/USD covariate}\label{sec:solusd}
Table~\ref{tab:solusd} adds the window-average daily SOL/USD close (Binance SOLUSDT spot, public daily candles, USDT as a USD proxy). The window labelled low (June 2025) averages 140 USD per SOL and the one labelled high (September 2025) 240, so the labels agree with the market data. Incidence is 0.57 and 0.53 per 1,000 launches in these two windows, practically the same, and across all nine windows the Spearman correlation between mean SOL/USD and incidence is 0.03 ($p=0.93$, $n=9$). With nine purposively selected windows of unequal length this is a descriptive check, not a test: it gives no sign of a price level effect, and it cannot exclude one.
\begin{table}[htbp]\centering\small\caption{Window-average SOL/USD and episode incidence.}\label{tab:solusd}
\begin{tabular}{lrrrr}\toprule Period & Mean SOL/USD & First--last day & Per 1,000 & Serial groups \\ \midrule
2024-04 & 173 & 173--172 & 3.30 & 5 \\
2024-07 & 139 & 142--139 & 3.09 & 6 \\
2024-10 & 144 & 139--146 & 1.64 & 1 \\
2025-02 & 198 & 200--197 & 4.60 & 32 \\
2025-03 & 122 & 118--125 & 1.62 & 4 \\
2025-05 & 177 & 178--184 & 1.15 & 6 \\
2025-06 & 140 & 135--144 & 0.57 & 3 \\
2025-09 & 240 & 242--234 & 0.53 & 1 \\
2025-12 & 131 & 132--128 & 0.76 & 4 \\
\bottomrule\end{tabular}\par\smallskip\begin{minipage}{.98\linewidth}\small Daily closing price of the Binance SOLUSDT spot pair (USDT as a USD proxy), UTC days covering each window. The two windows labelled low and high in the supplied names are 2025-06 and 2025-09.\end{minipage}\end{table}

A finer check uses only the daily series (Table~\ref{tab:solusddaily}). Each window covers one to three UTC days, which gives 16 day segments of at least six hours in 9 windows. Launches per hour, episodes per 1,000 launches and serial-group episodes per 1,000 launches are centred within their window and correlated with the daily SOL/USD close and its daily log change; the $p$-values come from circular shifts of the daily series inside windows, and all 14 tests of the table are Holm-corrected together. No daily correlation is distinguishable from zero (smallest adjusted $p$ 0.13); with so few segments the shift test has little resolution, so this is again a descriptive check. For the 62 serial groups of the price screen we correlate group size (wallets), the number of launches, the share of launches that are episodes and the buy turnover with the mean SOL/USD close on the days of the group's episodes. Groups with more launches appear at higher SOL/USD levels (Spearman 0.33, unadjusted $p=0.009$), but this does not survive the correction ($p=0.13$) and weakens within windows (0.22, $p=0.08$), so it mainly reflects differences between windows, such as their length. Group size and capital show no relation to the price level.
\begin{table}[htbp]\centering\small\caption{Daily SOL/USD and the activity and characteristics of teams: Spearman correlations.}\label{tab:solusddaily}
\resizebox{\linewidth}{!}{\begin{tabular}{lllrrrr}\toprule Level & Outcome & Covariate & $n$ & $\rho$ & $p$ & $p$ (Holm) \\ \midrule
day & launches per hour & SOL/USD level & 16 & $-$0.09 & 0.509 & 1.00 \\
day & launches per hour & daily SOL/USD change & 16 & $-$0.29 & 0.512 & 1.00 \\
day & episodes per 1,000 launches & SOL/USD level & 16 & +0.03 & 1.000 & 1.00 \\
day & episodes per 1,000 launches & daily SOL/USD change & 16 & +0.43 & 0.514 & 1.00 \\
day & serial-group episodes per 1,000 launches & SOL/USD level & 16 & $-$0.02 & 1.000 & 1.00 \\
day & serial-group episodes per 1,000 launches & daily SOL/USD change & 16 & +0.47 & 0.513 & 1.00 \\
group & group wallets & SOL/USD level at the group's episodes & 62 & +0.00 & 0.974 & 1.00 \\
group, within window & group wallets & SOL/USD level at the group's episodes & 62 & +0.01 & 0.938 & 1.00 \\
group & launches per group & SOL/USD level at the group's episodes & 62 & +0.33 & 0.009 & 0.13 \\
group, within window & launches per group & SOL/USD level at the group's episodes & 62 & +0.22 & 0.079 & 0.95 \\
group & episode share of launches & SOL/USD level at the group's episodes & 62 & $-$0.32 & 0.012 & 0.15 \\
group, within window & episode share of launches & SOL/USD level at the group's episodes & 62 & $-$0.11 & 0.394 & 1.00 \\
group & buy turnover (SOL) & SOL/USD level at the group's episodes & 62 & +0.18 & 0.160 & 1.00 \\
group, within window & buy turnover (SOL) & SOL/USD level at the group's episodes & 62 & +0.15 & 0.240 & 1.00 \\
\bottomrule\end{tabular}}\par\smallskip\begin{minipage}{.98\linewidth}\small Day rows: UTC days clipped to the observation window, at least 6 hours each; outcomes and covariates are centred within each window, and $p$ comes from circular shifts of the daily series inside windows (2,000 shifts), which keeps the day-to-day dependence. Group rows: serial groups of the price screen; the first group row of each pair uses raw values, the second values centred within the window. Holm correction over all rows of the table.\end{minipage}\end{table}

\subsection{Effective trading costs from transactions}\label{sec:fees}
For 216 randomly drawn trades of at least 0.5 SOL (24 per window, half buys and half sells) the transaction was fetched from the Solana network (only the signatures were requested). The wallet's lamport balance change, net of the network fee, divided by the recorded SOL gives the effective cost of a buy and the effective proceeds of a sell. The median buy costs 1.0207 times the recorded amount (interquartile range 1.013--1.027) and the median sell returns 0.9799 (0.977--0.987); medians by window lie between 1.011 and 1.026 for buys and between 0.977 and 0.990 for sells. These include protocol and creator fees and any tips or account rent paid in the same transaction, so about two per cent per side is an empirical, not hypothetical, order of magnitude. Applying the sample medians to the serial groups' first-hour turnover changes April from 56.0 to -16.7 SOL and deepens the three negative windows (May -191.7 to -430.6, June -277.1 to -479.2, September -164.3 to -174.0); April turns negative and the other five windows stay positive. The sample is small and not stratified by group, so these are order-of-magnitude adjustments.
\begin{table}[htbp]\centering\small\caption{Median effective cost ratios from sampled transactions and serial groups' first-hour flow after applying the overall sample medians (SOL).}\label{tab:feeadj}
\begin{tabular}{lrrrr}\toprule Period & Buy cost ratio & Sell proceeds ratio & Recorded flow & Fee-adjusted \\ \midrule
2024-04 & 1.011 & 0.990 & 56.0 & -16.7 \\
2024-07 & 1.021 & 0.980 & 148.9 & 78.8 \\
2024-10 & 1.017 & 0.981 & 33.3 & 11.8 \\
2025-02 & 1.017 & 0.977 & 3,936.4 & 2,692.7 \\
2025-03 & 1.015 & 0.980 & 625.7 & 490.8 \\
2025-05 & 1.021 & 0.981 & -191.7 & -430.6 \\
2025-06 & 1.018 & 0.980 & -277.1 & -479.2 \\
2025-09 & 1.026 & 0.977 & -164.3 & -174.0 \\
2025-12 & 1.026 & 0.977 & 609.0 & 536.3 \\
\bottomrule\end{tabular}\par\smallskip\begin{minipage}{.98\linewidth}\small Twelve buys and twelve sells per window; ratios are relative to the recorded SOL amount. The adjustment uses the overall medians, not window-specific ones.\end{minipage}\end{table}
\subsection{Persistence across windows}
We do not link identities across windows, but a descriptive check is possible: of the wallets in primary serial groups, between 0 and 1.8\% of each window's wallets appear in any group of another window (largest: June 8 of 444, February 10 of 1,404), and none of the April, July, October, March or September wallets do. Because a wallet enters a window's groups only if it launches or trades in that window, and the windows are months apart, this shows little reuse of the same addresses. It cannot distinguish wallet rotation by the same operators from different operators, and it does not support a claim of persistent teams.

\section{Numerical validation}
We validate several claims separately. First, the simulation of Section~\ref{sec:sim} reproduces the original simulation trade by trade whenever the original does not fail, conserves SOL in every experiment run (largest residual about $10^{-12}$ SOL) and passes its unit tests. Second, method checks cover the difference between a finite sale and a marginal valuation, the rejection of dust-trade prices, the drawdown thresholds, the ambiguity guard and the equivalence of the optimized wallet-membership evaluation with a direct implementation. Third, aggregates recomputed independently from the transaction records reproduce the first-hour buys and sells of every primary serial group (62 group--window checks), the endpoints and flows of all 679 episode tokens, and the whole-window and curve-marked flows: whole-window flows agree to within $1.3\times10^{-4}$ SOL, marked flows within 0.5\% in aggregate, with larger differences for single groups (at most 20 SOL in February) where several trades share one block and the final reserve depends on their unobserved order. The single-precision representation of the source amounts bounds the accumulated rounding error of any flow at $2^{-24}$ of turnover, at most 0.004 SOL here. Fourth, the revised analysis reproduces exactly the 679 episodes, the 62 serial groups and the first-hour flows (to within $10^{-12}$ SOL) before any alternative construction is applied. These checks verify arithmetic and consistency with the records. They do not verify data completeness, protocol interpretation, beneficial ownership, intent or a true rug-pull label.

\section{Interpretation and limitations}
\subsection{What the observations support}
The records support a descriptive account of repeated launch-associated exits under the specified group and episode definitions. They support the statement that repeat concentration and recorded cash-flow outcomes differ materially between these selected windows. They also support an accounting warning: inspecting only a creator's sales or only qualifying collapse windows can omit linked purchases and other launches that reverse the consolidated sign.

The simulation supports comparative statements about a stylized mechanism, not prevalence: the team's result depends strongly on the share of other traders, the stopping reserve, fees and, above all, the order of the forced final sales, and the simulated exits reproduce only some descriptors of the empirical episodes. The empirical screen does not assume that real tokens follow the simulated sequence, and we do not estimate the simulation's parameters from the empirical aggregates.

\subsection{What remains unidentified}
\textbf{Profit and losses.} Observed first-hour SOL sales minus purchases omit unobserved inventories, token transfers, costs, creator income and other venues. A trade-only running token balance can indicate that the ledger is insufficient, but cannot establish actual holdings. Even a seemingly closed recorded trade ledger does not establish that all wealth changes are observed. We do not publish an operator ROI, victim-loss estimate or total extracted wealth as if these were identified.

\textbf{Ownership and intent.} A funding-connected component is not a legal entity. A service can connect unrelated creators; transfers can be incidental; sophisticated coordination can avoid the observed links. Our activity exclusion and fan-out cap are conservative design choices, not empirically validated ownership classifiers. The creator-scope check does not remove graph uncertainty. Manual attribution and alternative graph constructions remain necessary before naming an organization or estimating classification precision.

\textbf{Protocol transitions.} The reserve field is not independently reconciled to historical contract state. The ambiguity guard may remove genuine events and retain legitimate transitions. No matched migration ground truth or liquidity-provider withdrawal log is available. Calling the result a count of confirmed liquidity rug pulls would overstate the evidence.

\textbf{Retrospective selection.} Full-window group construction and episode selection use future observations relative to an early launch. The study does not evaluate a real-time warning service or an external trader's executable return. Its window-level comparisons require no training/test claim. A future predictive study would need frozen group information, forward-only labels and out-of-sample evaluation.

\textbf{Coverage and selection of time.} Collection completeness is not certified. Timestamps have limited within-second ordering, amounts include single-precision measurements, and selected periods have unequal length. The windows were chosen for analytical interest. Results cannot be extrapolated to market-wide annual prevalence, and the low/high labels do not identify a causal SOL-price relationship. A representative panel, historical price controls and stable collection coverage would be needed for that question.

\textbf{Controls and multiplicity.} The 486-setting grid describes detector dependence rather than selecting a best rule. The allocation comparator is conditional and partly dependent on the same group evidence used in labeling. Neither is an externally calibrated test of wrongdoing. Case selection emphasizes repeated examples and is unsuitable for estimating a population false-positive rate.

\textbf{Label validity.} No human-labelled precision estimate exists yet. A seeded sample of 200 primary episodes (half from serial groups) with charts is released; visual inspection of the first contact sheet shows many sampled episodes with few trades, consistent with the thin-market sensitivity above. Until labels are scored, the screen should be read as a candidate generator.

\subsection{Next empirical steps}
The most valuable additional evidence is transaction-level historical protocol state and migration/venue linkage; token-transfer histories and initial balances; explicit fee and creator-payment instructions; and independently reviewed positive and negative cases. These would allow a mechanism taxonomy, more complete wallet accounting and a real classification evaluation. Equal-length rebuilding of groups is now done for a 22-hour launch window (Section~\ref{sec:robust}); rebuilding with prospectively available information only, scoring the released audit sample, and a bridge rule defined against a reference transfer distribution remain open.

An extended period panel could then test association with SOL/USD, demand of other traders and broader attention using period-level covariates and a transparent sampling design. The present nine-window comparison is a foundation for that study, not its causal result. Within this dataset, more elaborate profit optimization would be less informative than closing the missing accounting and labeling gaps.

\section{Conclusion}
Repeated token exits can be measured without equating a price collapse with a proven liquidity rug pull. In nine independently processed windows, the specified screen identifies 679 token episodes and 62 serial group--period records, and episodes concentrate in groups beyond a within-hour label-shuffle benchmark in every window (weakly where only one serial group exists). The recorded own-token cash flow of the serial groups is negative in May, June and September 2025 under the price-based screen and positive elsewhere, and this holds beyond the first hour and with unsold inventory valued on the curve. Because token prices are not fully reliable, the screen was repeated using only the bonding-curve reserve and SOL amounts: the episodes largely coincide, but the May result turns positive, June stays negative within sampling error and September rests on one group.

The result is also fragile. The sign in May and June depends on the price criterion of the screen and on including funded repeat early buyers in the group definition, group-bootstrap intervals include zero, serial-group counts fall sharply in equal-length windows, and a thin-market filter removes most April episodes. Cross-period differences in the original comparison therefore partly reflect window length and team construction. The defensible conclusions are the existence of concentrated repeat-exit patterns and the need to report group-level flows with sensitivity; operator profit, ownership and intent remain unidentified, and the released audit sample is the next step towards a precision estimate.

\clearpage
\appendix
\section{Complete one-at-a-time simulation results}\label{app:simoat}
{\footnotesize\setlength{\tabcolsep}{3pt}\begin{longtable}{llrrrrrrr}\caption{All one-at-a-time sweep points: mean team profit and other traders' result (SOL), share of runs with positive team profit, mean duration, share stopped by the reserve cap, and share flagged by the empirical screen without and with the migration guard.}\label{tab:simoatfull}\\\toprule Parameter & Value & Team & Other traders & P$>$0 & Dur. (s) & Cap & Screen & Guarded \\ \midrule\endfirsthead
\toprule Parameter & Value & Team & Other traders & P$>$0 & Dur. (s) & Cap & Screen & Guarded \\ \midrule\endhead
team wallets & 6 & 22.8 & -41.4 & 100\% & 514 & 100\% & 100\% & 0\% \\
team wallets & 10 & 29.1 & -46.4 & 100\% & 454 & 100\% & 100\% & 0\% \\
team wallets & 20 & 33.4 & -46.8 & 100\% & 346 & 100\% & 100\% & 0\% \\
team wallets & 30 & 29.7 & -39.9 & 100\% & 262 & 100\% & 100\% & 0\% \\
team wallets & 50 & 14.1 & -19.2 & 100\% & 118 & 100\% & 100\% & 0\% \\
team wallets & 80 & 3.7 & -6.3 & 90\% & 45 & 100\% & 89\% & 0\% \\
team wallets & 110 & 2.2 & -4.4 & 77\% & 34 & 100\% & 70\% & 0\% \\
early buyers & 0 & 29.8 & -40.0 & 100\% & 265 & 100\% & 100\% & 0\% \\
early buyers & 1 & 29.6 & -39.8 & 100\% & 261 & 100\% & 100\% & 0\% \\
early buyers & 2 & 29.5 & -39.5 & 100\% & 260 & 100\% & 100\% & 0\% \\
early buyers & 3 & 29.4 & -39.6 & 100\% & 260 & 100\% & 100\% & 0\% \\
early buyers & 5 & 29.4 & -39.5 & 100\% & 259 & 100\% & 100\% & 0\% \\
early buyers & 8 & 29.5 & -39.7 & 100\% & 258 & 100\% & 100\% & 0\% \\
early buyers & 12 & 29.5 & -39.6 & 100\% & 256 & 100\% & 100\% & 0\% \\
early buyers & 20 & 29.8 & -40.1 & 100\% & 258 & 100\% & 100\% & 0\% \\
other-trader sell prob. & 0.0 & 36.0 & -41.0 & 100\% & 127 & 100\% & 100\% & 0\% \\
other-trader sell prob. & 0.2 & 34.4 & -40.6 & 100\% & 155 & 100\% & 100\% & 0\% \\
other-trader sell prob. & 0.4 & 32.8 & -40.3 & 100\% & 185 & 100\% & 100\% & 0\% \\
other-trader sell prob. & 0.6 & 31.3 & -40.1 & 100\% & 222 & 100\% & 100\% & 0\% \\
other-trader sell prob. & 0.8 & 29.7 & -39.9 & 100\% & 259 & 100\% & 100\% & 0\% \\
other-trader sell prob. & 0.9 & 29.0 & -39.8 & 100\% & 281 & 100\% & 100\% & 0\% \\
other-trader sell prob. & 1.0 & 27.5 & -39.0 & 100\% & 302 & 100\% & 100\% & 0\% \\
team sell prob. & 0.0 & 15.8 & -18.2 & 100\% & 109 & 100\% & 100\% & 0\% \\
team sell prob. & 0.1 & 24.7 & -30.9 & 100\% & 193 & 100\% & 100\% & 0\% \\
team sell prob. & 0.2 & 29.6 & -39.8 & 100\% & 263 & 100\% & 100\% & 0\% \\
team sell prob. & 0.4 & 32.5 & -49.0 & 100\% & 365 & 100\% & 100\% & 0\% \\
team sell prob. & 0.6 & 30.4 & -50.0 & 100\% & 427 & 100\% & 100\% & 0\% \\
team sell prob. & 0.8 & 27.5 & -48.4 & 100\% & 469 & 100\% & 100\% & 0\% \\
team sell prob. & 1.0 & 24.7 & -45.8 & 100\% & 499 & 100\% & 100\% & 0\% \\
fee per side & 0.0 & 37.6 & -37.6 & 100\% & 257 & 100\% & 100\% & 0\% \\
fee per side & 0.005 & 33.5 & -38.7 & 100\% & 257 & 100\% & 100\% & 0\% \\
fee per side & 0.01 & 29.7 & -39.8 & 100\% & 258 & 100\% & 100\% & 0\% \\
fee per side & 0.02 & 22.6 & -42.2 & 100\% & 263 & 100\% & 100\% & 0\% \\
fee per side & 0.03 & 16.0 & -44.4 & 100\% & 266 & 100\% & 98\% & 0\% \\
sell fee kept in reserve & yes & 29.8 & -39.9 & 100\% & 260 & 100\% & 100\% & 0\% \\
sell fee kept in reserve & no & 32.7 & -44.4 & 100\% & 296 & 100\% & 100\% & 0\% \\
final sale order & team\_first & 29.6 & -39.8 & 100\% & 263 & 100\% & 100\% & 0\% \\
final sale order & others\_first & -9.1 & -1.1 & 2\% & 260 & 100\% & 91\% & 0\% \\
final sale order & random & 14.1 & -24.3 & 100\% & 263 & 100\% & 100\% & 0\% \\
initial virtual SOL & 10.0 & 38.0 & -54.8 & 100\% & 357 & 100\% & 100\% & 0\% \\
initial virtual SOL & 20.0 & 33.6 & -45.8 & 100\% & 290 & 100\% & 100\% & 0\% \\
initial virtual SOL & 30.0 & 29.5 & -39.7 & 100\% & 261 & 100\% & 100\% & 0\% \\
initial virtual SOL & 45.0 & 24.9 & -33.5 & 100\% & 237 & 100\% & 100\% & 0\% \\
initial virtual SOL & 60.0 & 21.2 & -28.9 & 100\% & 223 & 100\% & 100\% & 0\% \\
cap on the share of other traders & 0.0 & -19.2 & 0.0 & 0\% & 780 & 0\% & 0\% & 0\% \\
cap on the share of other traders & 0.1 & 12.9 & -35.4 & 99\% & 672 & 69\% & 100\% & 1\% \\
cap on the share of other traders & 0.2 & 24.7 & -38.5 & 100\% & 368 & 100\% & 100\% & 0\% \\
cap on the share of other traders & 0.3 & 29.5 & -39.6 & 100\% & 260 & 100\% & 100\% & 0\% \\
cap on the share of other traders & 0.5 & 35.2 & -42.3 & 100\% & 176 & 100\% & 100\% & 0\% \\
cap on the share of other traders & 0.7 & 38.2 & -43.9 & 100\% & 138 & 100\% & 100\% & 0\% \\
share ramp (s) & 10.0 & 29.5 & -39.5 & 100\% & 255 & 100\% & 100\% & 0\% \\
share ramp (s) & 25.0 & 29.7 & -39.8 & 100\% & 259 & 100\% & 100\% & 0\% \\
share ramp (s) & 50.0 & 29.6 & -39.7 & 100\% & 260 & 100\% & 100\% & 0\% \\
share ramp (s) & 100.0 & 29.7 & -40.0 & 100\% & 265 & 100\% & 100\% & 0\% \\
share ramp (s) & 200.0 & 29.2 & -39.9 & 100\% & 281 & 100\% & 100\% & 0\% \\
new-other-trader prob. & 0.1 & 12.6 & -35.5 & 98\% & 706 & 57\% & 100\% & 2\% \\
new-other-trader prob. & 0.3 & 29.4 & -39.6 & 100\% & 261 & 100\% & 100\% & 0\% \\
new-other-trader prob. & 0.5 & 34.9 & -41.4 & 100\% & 148 & 100\% & 100\% & 0\% \\
new-other-trader prob. & 0.8 & 37.8 & -41.8 & 100\% & 83 & 100\% & 100\% & 0\% \\
other-trader wallet SOL & (0.1, 1.0) & 16.2 & -35.5 & 100\% & 687 & 73\% & 100\% & 0\% \\
other-trader wallet SOL & (0.2, 3.0) & 29.4 & -39.5 & 100\% & 258 & 100\% & 100\% & 0\% \\
other-trader wallet SOL & (1.0, 6.0) & 35.7 & -42.6 & 100\% & 126 & 100\% & 100\% & 0\% \\
other-trader wallet SOL & (2.0, 10.0) & 37.9 & -43.3 & 100\% & 74 & 100\% & 100\% & 0\% \\
creator launch buy & 0.0 & 30.4 & -41.4 & 100\% & 279 & 100\% & 100\% & 0\% \\
creator launch buy & 1.0 & 29.7 & -40.4 & 100\% & 269 & 100\% & 100\% & 0\% \\
creator launch buy & 2.0 & 29.5 & -39.7 & 100\% & 261 & 100\% & 100\% & 0\% \\
creator launch buy & 5.0 & 27.8 & -36.6 & 100\% & 230 & 100\% & 100\% & 0\% \\
creator launch buy & 10.0 & 24.0 & -31.1 & 100\% & 185 & 100\% & 100\% & 0\% \\
creator launch buy & 20.0 & 15.1 & -19.6 & 100\% & 112 & 100\% & 100\% & 0\% \\
stop reserve (SOL) & 20.0 & -0.3 & -0.2 & 16\% & 4 & 100\% & 0\% & 0\% \\
stop reserve (SOL) & 40.0 & 1.7 & -3.0 & 81\% & 24 & 100\% & 48\% & 48\% \\
stop reserve (SOL) & 60.0 & 13.2 & -17.6 & 100\% & 111 & 100\% & 100\% & 100\% \\
stop reserve (SOL) & 70.0 & 21.4 & -28.5 & 100\% & 183 & 100\% & 100\% & 0\% \\
stop reserve (SOL) & 80.0 & 29.6 & -39.8 & 100\% & 261 & 100\% & 100\% & 0\% \\
stop reserve (SOL) & 85.0 & 33.5 & -45.4 & 100\% & 301 & 100\% & 100\% & 0\% \\
max duration (s) & 120 & 13.0 & -17.6 & 100\% & 120 & 1\% & 100\% & 89\% \\
max duration (s) & 300 & 28.3 & -38.2 & 100\% & 254 & 74\% & 100\% & 1\% \\
max duration (s) & 780 & 30.0 & -40.3 & 100\% & 263 & 100\% & 100\% & 0\% \\
max duration (s) & 1500 & 29.5 & -39.6 & 100\% & 260 & 100\% & 100\% & 0\% \\
max tx per second & 3 & 29.2 & -39.2 & 100\% & 357 & 100\% & 100\% & 0\% \\
max tx per second & 5 & 29.5 & -39.6 & 100\% & 259 & 100\% & 100\% & 0\% \\
max tx per second & 8 & 30.2 & -40.7 & 100\% & 188 & 100\% & 100\% & 0\% \\
max tx per second & 12 & 30.3 & -41.0 & 100\% & 137 & 100\% & 100\% & 0\% \\
\bottomrule\end{longtable}}

\clearpage
\clearpage
\section{Additional simulation results}\label{app:simmore}
Detail figures and tables of the simulation sections: the other traders' result in the one-at-a-time study, the baseline distributions, the team-size grids, the calibration parameters, the exit rules, the campaigns, the sensitivity of the extended model and the policy sweeps.

\begin{figure}[htbp]\centering\includegraphics[width=\linewidth]{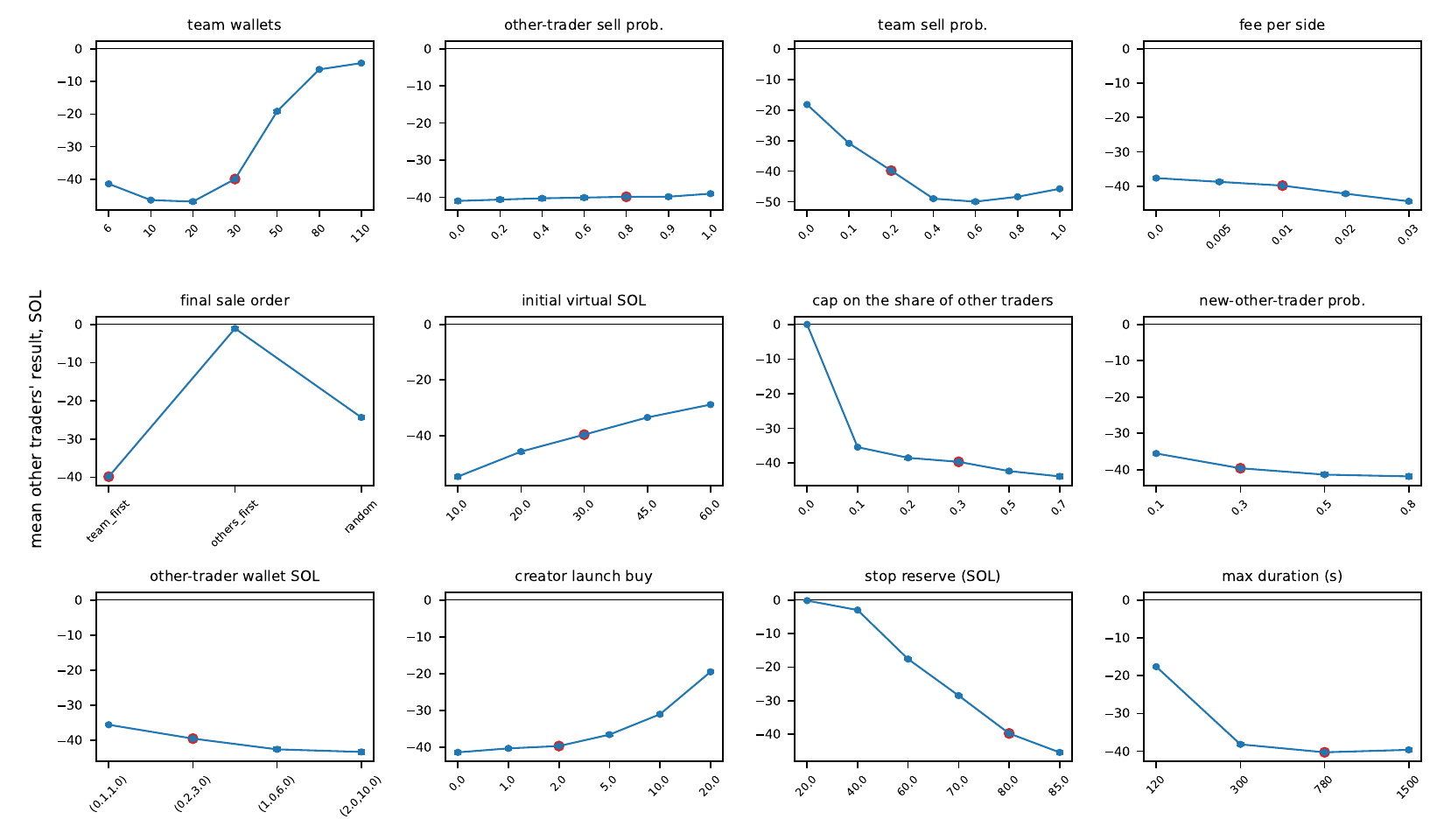}\caption{Mean other traders' result when one parameter is varied.}\label{fig:simoatoth}\end{figure}
\begin{figure}[htbp]\centering\includegraphics[width=\linewidth]{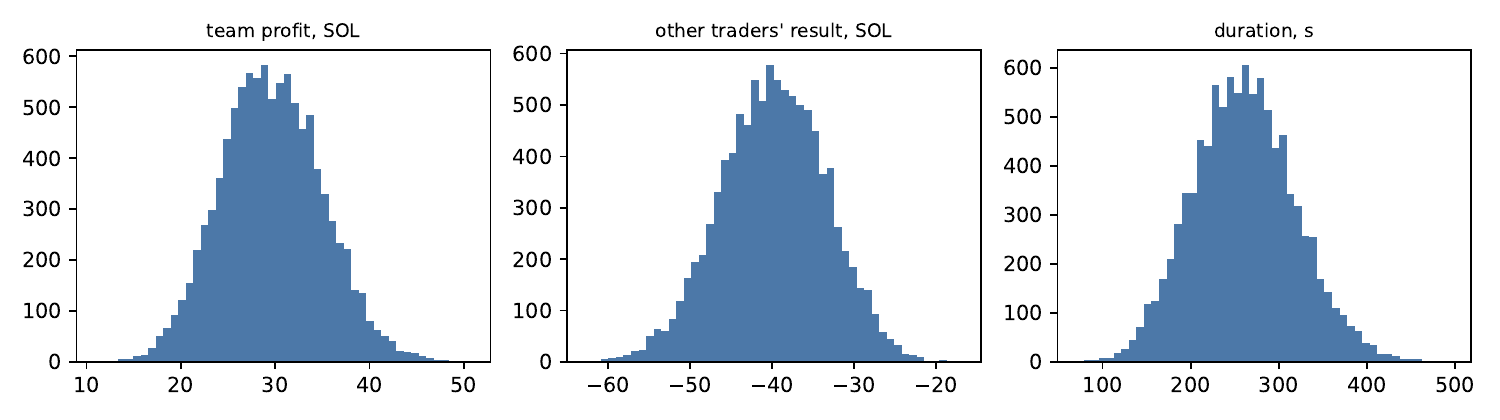}\caption{Baseline distributions of team profit, other traders' result and duration.}\label{fig:simhist}\end{figure}
\begin{figure}[htbp]\centering\includegraphics[width=\linewidth]{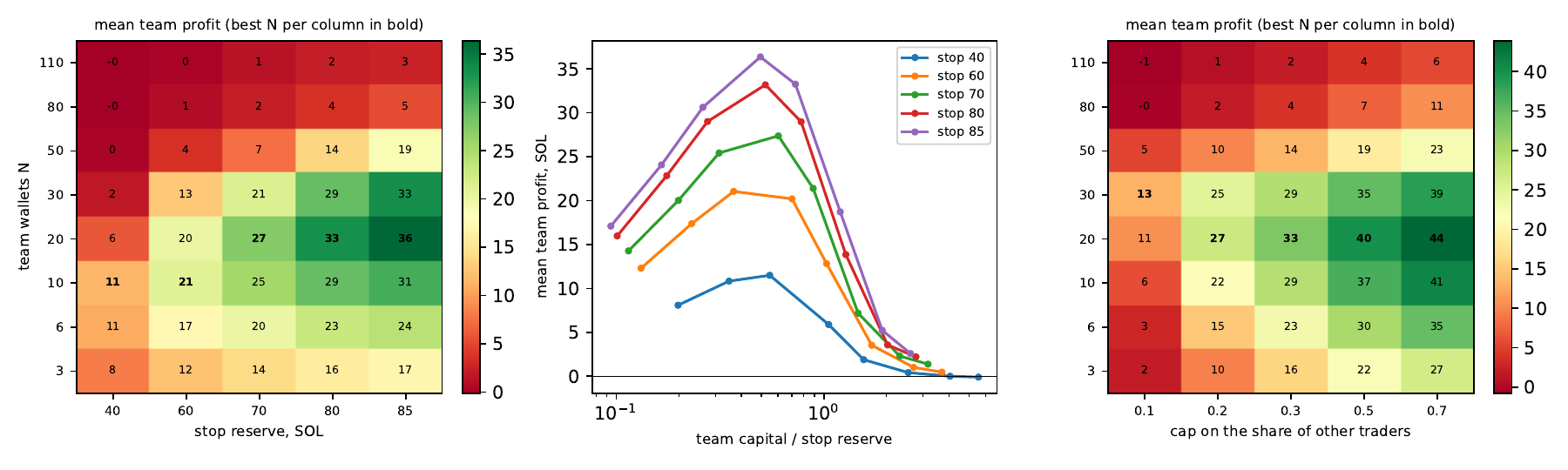}\caption{Left: mean team profit by team size and stopping reserve. Middle: the same profits against committed team capital divided by the stopping reserve. Right: team size against the cap on the share of other traders.}\label{fig:simteamgrid}\end{figure}
\begin{figure}[htbp]\centering\includegraphics[width=\linewidth]{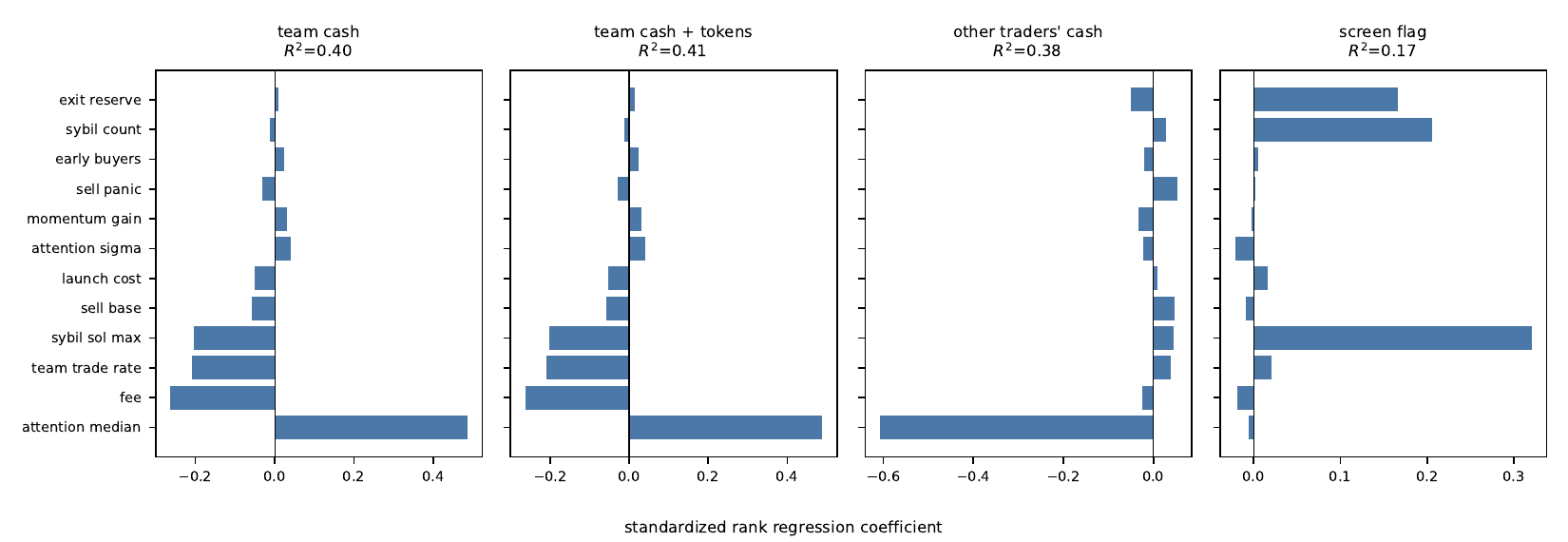}\caption{Standardized rank regression coefficients in the Latin-hypercube design of the extended model.}\label{fig:x2srrc}\end{figure}
\begin{figure}[htbp]\centering\includegraphics[width=\linewidth]{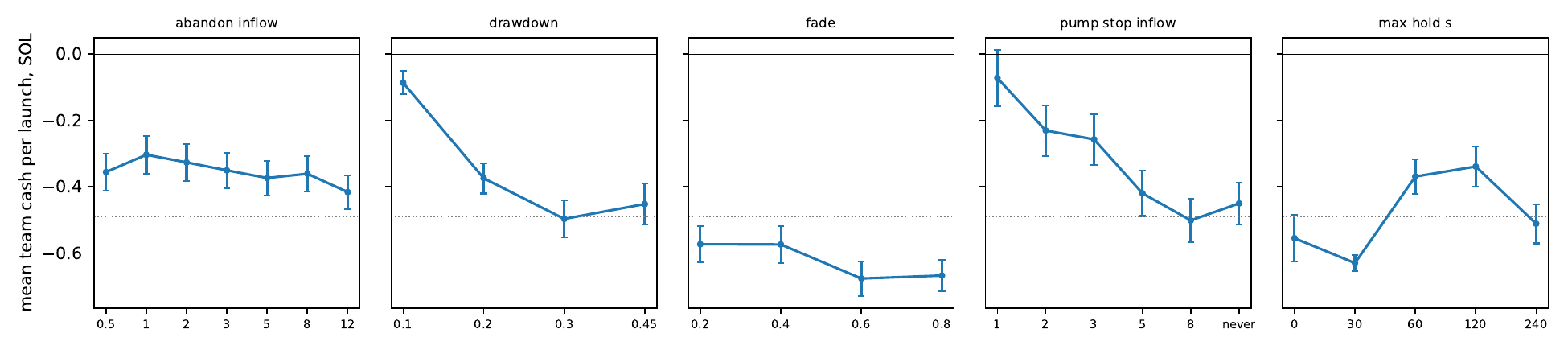}\caption{Mean team cash per launch when the parameter of one policy is varied (dotted line: static team with a 300 s limit).}\label{fig:x2pol}\end{figure}

\begin{table}[htbp]\centering\small\caption{One-at-a-time sweeps: values giving the lowest and highest mean team profit.}\label{tab:simoat}
\resizebox{\linewidth}{!}{\begin{tabular}{lrrrrrr}\toprule Parameter & Values & Lowest at & Mean (SOL) & Highest at & Mean (SOL) & Min.\ P(profit$>$0) \\ \midrule
team wallets & 7 & 110 & 2.2 & 20 & 33.4 & 77\% \\
early buyers & 8 & 5 & 29.4 & 0 & 29.8 & 100\% \\
other-trader sell prob. & 7 & 1.0 & 27.5 & 0.0 & 36.0 & 100\% \\
team sell prob. & 7 & 0.0 & 15.8 & 0.4 & 32.5 & 100\% \\
fee per side & 5 & 0.03 & 16.0 & 0.0 & 37.6 & 100\% \\
sell fee kept in reserve & 2 & yes & 29.8 & no & 32.7 & 100\% \\
final sale order & 3 & others\_first & -9.1 & team\_first & 29.6 & 2\% \\
initial virtual SOL & 5 & 60.0 & 21.2 & 10.0 & 38.0 & 100\% \\
cap on the share of other traders & 6 & 0.0 & -19.2 & 0.7 & 38.2 & 0\% \\
share ramp (s) & 5 & 200.0 & 29.2 & 100.0 & 29.7 & 100\% \\
new-other-trader prob. & 4 & 0.1 & 12.6 & 0.8 & 37.8 & 98\% \\
other-trader wallet SOL & 4 & (0.1, 1.0) & 16.2 & (2.0, 10.0) & 37.9 & 100\% \\
creator launch buy & 6 & 20.0 & 15.1 & 0.0 & 30.4 & 100\% \\
stop reserve (SOL) & 6 & 20.0 & -0.3 & 85.0 & 33.5 & 16\% \\
max duration (s) & 4 & 120 & 13.0 & 780 & 30.0 & 100\% \\
max tx per second & 4 & 3 & 29.2 & 12 & 30.3 & 100\% \\
\bottomrule\end{tabular}}\par\smallskip\begin{minipage}{.98\linewidth}\small Each point is the mean of 1\,000 runs with all other parameters at their baseline values.\end{minipage}\end{table}

\begin{table}[htbp]\centering\small\caption{Team size with the original fixed endowment (250 SOL): means over 1,000 runs per size.}\label{tab:simteam}
\resizebox{\linewidth}{!}{\begin{tabular}{rrrrrrrrr}\toprule $N$ & Capital & Funded & Team profit & P$>$0 & Return & Other traders & Duration (s) & Team trade share \\ \midrule
1 & 4 & 1 & 15.0 & 100\% & 3.88 & $-$26.3 & 608 & 0.60 \\
3 & 8 & 3 & 16.1 & 100\% & 2.06 & $-$33.3 & 567 & 0.61 \\
5 & 12 & 5 & 20.6 & 100\% & 1.75 & $-$39.3 & 537 & 0.62 \\
10 & 22 & 10 & 29.0 & 100\% & 1.33 & $-$46.4 & 458 & 0.62 \\
20 & 42 & 20 & 32.9 & 100\% & 0.79 & $-$46.2 & 343 & 0.63 \\
30 & 62 & 30 & 29.6 & 100\% & 0.48 & $-$39.8 & 260 & 0.64 \\
50 & 102 & 50 & 13.8 & 100\% & 0.14 & $-$18.9 & 118 & 0.70 \\
80 & 162 & 80 & 4.0 & 89\% & 0.02 & $-$6.6 & 46 & 0.79 \\
100 & 202 & 100 & 2.3 & 80\% & 0.01 & $-$4.6 & 36 & 0.82 \\
150 & 232 & 116 & 3.3 & 86\% & 0.01 & $-$5.5 & 43 & 0.77 \\
\bottomrule\end{tabular}}\par\smallskip\begin{minipage}{.98\linewidth}\small Capital: SOL committed by the creator and the funded sybil wallets; Funded: wallets that received SOL (the endowment runs out above about 115 wallets).\end{minipage}\end{table}

\begin{table}[htbp]\centering\small\caption{Calibrated parameters of the extended model (posterior medians of the best draws).}\label{tab:calparams}
\begin{tabular}{llr}\toprule Parameter & Stage & Value \\ \midrule
attention\_median & A & 0.026 \\
attention\_sigma & A & 1.888 \\
attention\_decay\_s & A & 443.288 \\
momentum\_gain & A & 4.076 \\
sell\_base & A & 0.292 \\
sell\_panic & A & 2.177 \\
other trader\_sol\_median & A & 0.403 \\
sniper\_prob & A & 0.556 \\
exit\_rule & B & reserve \\
exit\_reserve & B & 40.71 \\
exit\_multiple & B & 5.35 \\
sybil\_count & B & 51.00 \\
early\_buyers & B & 5.00 \\
sybil\_sol\_max & B & 1.95 \\
team\_trade\_rate & B & 0.85 \\
\bottomrule\end{tabular}\par\smallskip\begin{minipage}{.8\linewidth}\small Stage A: side of the other traders fitted to the distribution of the maximal reserve and of trades per launch; stage B: team side fitted to episode descriptors. Not fitted: fee, launch cost, graduation level, curve constants (taken from the data checks and the program description).\end{minipage}\end{table}

\begin{table}[htbp]\centering\small\caption{Team exit rules in the calibrated model (means over 4\,000 launches per rule).}\label{tab:x2exit}
\resizebox{\linewidth}{!}{\begin{tabular}{lrrrrrrr}\toprule Rule & Team cash & Team pooled & Cash $>$0 & Other traders cash & Other traders pooled & Exit reached & Screen flag \\ \midrule
no exit & $-$42.0 & $-$2.8 & 0\% & $-$3.9 & $-$1.8 & 0\% & 0.0\% \\
reserve 20 & $-$0.5 & $-$0.5 & 11\% & $-$1.4 & $-$0.1 & 100\% & 0.0\% \\
reserve 40 & $-$0.5 & $-$0.5 & 20\% & $-$2.1 & $-$0.8 & 100\% & 65.7\% \\
reserve 60 & $-$0.5 & $-$0.5 & 19\% & $-$3.2 & $-$1.7 & 100\% & 95.1\% \\
reserve 80 & $-$0.2 & $-$0.2 & 20\% & $-$3.5 & $-$2.1 & 100\% & 93.7\% \\
x2 & 0.5 & 0.5 & 43\% & $-$2.1 & $-$0.9 & 100\% & 0.0\% \\
x3 & $-$0.6 & $-$0.6 & 11\% & $-$1.4 & $-$0.1 & 100\% & 0.0\% \\
x5 & $-$0.5 & $-$0.5 & 21\% & $-$2.3 & $-$0.8 & 100\% & 62.5\% \\
x10 & $-$0.1 & $-$0.1 & 21\% & $-$3.7 & $-$2.1 & 100\% & 94.8\% \\
drawdown 0.2 & $-$0.3 & $-$0.2 & 19\% & $-$2.7 & $-$1.2 & 100\% & 68.4\% \\
drawdown 0.3 & $-$0.7 & $-$0.3 & 20\% & $-$3.3 & $-$1.7 & 99\% & 92.2\% \\
drawdown 0.5 & $-$0.9 & $-$0.3 & 20\% & $-$3.5 & $-$1.9 & 99\% & 93.9\% \\
time 60 s & $-$0.4 & $-$0.3 & 17\% & $-$1.9 & $-$0.6 & 100\% & 1.2\% \\
time 120 s & $-$0.4 & $-$0.1 & 21\% & $-$2.7 & $-$1.2 & 99\% & 57.4\% \\
time 300 s & $-$1.0 & $-$0.4 & 19\% & $-$3.3 & $-$1.8 & 99\% & 93.8\% \\
\bottomrule\end{tabular}}\par\smallskip\begin{minipage}{.98\linewidth}\small Results in SOL per launch, including the many launches that attract few other traders. Pooled: cash plus tokens valued as if all holders sold together at the end.\end{minipage}\end{table}

\begin{table}[htbp]\centering\small\caption{Campaigns of $K$ launches by one team in the calibrated model (realized cash after all launches; 600 campaigns per $K$).}\label{tab:x2camp}
\begin{tabular}{rrrrrrrr}\toprule $K$ & Mean & Median & 5\% & 95\% & Result $>$0 & Ruined & Mean pooled value \\ \midrule
1 & $-$0.3 & $-$1.0 & $-$2.4 & 5.2 & 24\% & 0\% & $-$0.3 \\
5 & $-$2.1 & $-$3.3 & $-$8.1 & 7.8 & 25\% & 0\% & $-$2.1 \\
20 & $-$10.0 & $-$11.0 & $-$23.5 & 7.7 & 14\% & 0\% & $-$10.0 \\
50 & $-$24.5 & $-$25.7 & $-$47.3 & 1.3 & 6\% & 0\% & $-$24.5 \\
100 & $-$46.8 & $-$48.4 & $-$82.4 & $-$9.3 & 2\% & 0\% & $-$46.8 \\
\bottomrule\end{tabular}\par\smallskip\begin{minipage}{.95\linewidth}\small SOL. Each launch starts from the team's current cash (at most the 250 SOL endowment); a campaign stops (ruin) when cash cannot fund another launch. Pooled value: sum of the launches' cash plus pooled token values.\end{minipage}\end{table}

\begin{table}[htbp]\centering\small\caption{Extended model: one-at-a-time sweeps around the calibrated baseline (mean team cash per launch).}\label{tab:x2oat}
\resizebox{\linewidth}{!}{\begin{tabular}{lrrrrrr}\toprule Parameter & Values & Lowest at & Mean (SOL) & Highest at & Mean (SOL) & Min.\ share $>$0 \\ \midrule
fee & 5 & 0.03 & $-$2.9 & 0 & 0.7 & 10\% \\
launch\_cost & 4 & 0.3 & $-$0.8 & 0 & $-$0.5 & 18\% \\
attention\_median & 5 & 0.00650253 & $-$1.2 & 0.104041 & 1.2 & 8\% \\
attention\_sigma & 5 & 0.5 & $-$1.2 & 2.5 & $-$0.0 & 6\% \\
momentum\_gain & 5 & 0 & $-$0.9 & 6 & $-$0.4 & 15\% \\
sell\_base & 4 & 0.4 & $-$0.6 & 0.02 & $-$0.1 & 19\% \\
sell\_panic & 4 & 4 & $-$0.6 & 0 & $-$0.4 & 20\% \\
sybil\_sol\_max & 5 & 3 & $-$1.0 & 0.6 & 0.7 & 13\% \\
sybil\_count & 5 & 35 & $-$0.6 & 5 & $-$0.1 & 17\% \\
early\_buyers & 4 & 10 & $-$0.5 & 2 & $-$0.4 & 21\% \\
team\_trade\_rate & 5 & 3 & $-$1.9 & 0.2 & 0.7 & 9\% \\
exit\_reserve & 5 & 20 & $-$0.5 & 10 & 0.8 & 12\% \\
graduation\_sol & 4 & 70 & $-$0.5 & 100 & $-$0.5 & 21\% \\
post\_exit\_s & 4 & 300 & $-$0.5 & 30 & $-$0.4 & 22\% \\
sniper\_prob & 3 & 1 & $-$0.9 & 0 & $-$0.1 & 17\% \\
exit\_fraction & 3 & 0.5 & $-$12.5 & 1 & $-$0.5 & 0\% \\
\bottomrule\end{tabular}}\end{table}

\begin{table}[htbp]\centering\small\caption{Latin-hypercube design: standardized rank regression coefficients.}\label{tab:simsrrc}
\begin{tabular}{llrrrr}\toprule Parameter & Range & Team profit & Other traders & Duration & Guarded flag \\ \midrule
stop reserve (SOL) & 20--85 & +0.44 & -0.70 & +0.72 & +0.13 \\
fee per side & 0--0.03 & -0.36 & -0.01 & +0.01 & -0.13 \\
team wallets & 5--100 & -0.32 & +0.36 & -0.34 & -0.21 \\
cap on the share of other traders & 0--0.7 & +0.22 & -0.08 & -0.07 & +0.06 \\
team sell prob. & 0--1 & +0.16 & -0.23 & +0.21 & +0.11 \\
creator launch buy & 0--20 & -0.13 & +0.18 & -0.21 & -0.09 \\
new-other-trader prob. & 0.1--0.8 & +0.12 & -0.04 & -0.08 & +0.05 \\
early buyers & 0--20 & -0.11 & +0.13 & -0.19 & -0.06 \\
initial virtual SOL & 10--60 & -0.08 & +0.08 & -0.03 & -0.01 \\
share ramp (s) & 10--200 & -0.08 & +0.05 & +0.03 & -0.05 \\
other-trader sell prob. & 0--1 & -0.06 & +0.02 & +0.03 & -0.02 \\
max tx per second & 3--12 & -0.03 & +0.02 & -0.09 & -0.04 \\
max duration (s) & 120--1500 & +0.00 & -0.01 & +0.01 & -0.01 \\
\midrule $R^2$ & & 0.56 & 0.75 & 0.79 & 0.11 \\
\bottomrule\end{tabular}\par\smallskip\begin{minipage}{.96\linewidth}\small 20\,000 design points, one run each; ranks of inputs and outputs.\end{minipage}\end{table}

\section{Primary-group role partition and exposure checks}
\begin{table}[htbp]\centering\small
\caption{Same primary serial groups, different roles: direct SQL decomposition of own-token first-hour flow (SOL).}\label{tab:roles}
\begin{tabular}{lrrr}\toprule
Period & Creator wallets & Additional linked & Consolidated \\ \midrule
2024-04 & 56.02 & 0.00 & 56.02 \\
2024-07 & 208.15 & -59.21 & 148.95 \\
2024-10 & 33.27 & 0.00 & 33.27 \\
2025-02 & 3\,495.11 & 441.24 & 3\,936.35 \\
2025-03 & 734.90 & -109.23 & 625.68 \\
2025-05 & 467.85 & -659.54 & -191.69 \\
2025-06 & 197.44 & -474.53 & -277.09 \\
2025-09 & -164.32 & 0.00 & -164.32 \\
2025-12 & 642.67 & -33.63 & 609.04 \\
\bottomrule\end{tabular}
\par\smallskip\begin{minipage}{.98\linewidth}\small The group set is fixed to the primary selection. This partition must not be confused with the creator-scope detector, which can select a different set of episodes and groups.\end{minipage}
\end{table}

\begin{table}[htbp]\centering\small
\caption{Partial fixed-exposure check and hypothetical additional friction.}\label{tab:exposure}
\begin{tabular}{lrrrrr}\toprule
Period & 12h tokens & 12h ep. & 12h groups & Net at 1\% & Net at 2\% \\ \midrule
2024-04 & 5\,962 & 21 & 4 & 20.41 & -15.21 \\
2024-07 & 4\,886 & 13 & 0 & 114.58 & 80.21 \\
2024-10 & 8\,202 & 21 & 1 & 22.73 & 12.20 \\
2025-02 & 26\,872 & 144 & 11 & 3\,326.50 & 2\,716.65 \\
2025-03 & 12\,798 & 22 & 2 & 559.53 & 493.38 \\
2025-05 & 11\,258 & 13 & 0 & -308.69 & -425.69 \\
2025-06 & 11\,365 & 5 & 1 & -376.03 & -474.97 \\
2025-09 & 9\,778 & 2 & 0 & -169.01 & -173.71 \\
2025-12 & 9\,012 & 7 & 1 & 573.29 & 537.54 \\
\bottomrule\end{tabular}
\par\smallskip\begin{minipage}{.98\linewidth}\small 12h uses tokens whose complete first hour ends within the first twelve recorded hours, holding full-period groups fixed. Friction columns use full-period selected groups, in SOL, with an additional deduction on each trade side; they are not historical fee estimates.\end{minipage}
\end{table}

\clearpage
\section{Selected group inventory}
\begin{table}[htbp]\centering\small
\caption{Groups selected by episode count for detailed audit; not selected by cash-flow sign.}\label{tab:cases}
\begin{tabular}{lrrrrr}\toprule
Case & Local ID & Creators & Launches & Episodes & Net SOL \\ \midrule
2024-04\_S1 & G03889 & 2 & 11 & 10 & 31.31 \\
2024-04\_S2 & G03760 & 1 & 7 & 7 & 13.08 \\
2024-07\_S1 & G00248 & 12 & 12 & 5 & 62.65 \\
2024-07\_S2 & G00538 & 7 & 7 & 4 & 60.58 \\
2024-10\_S1 & G06815 & 1 & 10 & 9 & 33.27 \\
2025-02\_S1 & G00852 & 84 & 88 & 20 & -334.40 \\
2025-02\_S2 & G00201 & 105 & 107 & 14 & 544.29 \\
2025-03\_S1 & G00286 & 17 & 17 & 6 & -203.36 \\
2025-03\_S2 & G00696 & 33 & 33 & 5 & -68.19 \\
2025-05\_S1 & G00128 & 66 & 69 & 6 & -99.76 \\
2025-05\_S2 & G00833 & 86 & 86 & 5 & -312.36 \\
2025-06\_S1 & G00404 & 98 & 98 & 9 & -124.41 \\
2025-06\_S2 & G00170 & 124 & 124 & 6 & -120.19 \\
2025-09\_S1 & G06220 & 2 & 5 & 3 & -164.32 \\
2025-12\_S1 & G03568 & 1 & 36 & 6 & 77.08 \\
2025-12\_S2 & G00388 & 24 & 25 & 4 & 301.26 \\
\bottomrule\end{tabular}
\par\smallskip\begin{minipage}{.98\linewidth}\small Net SOL is first-hour own-token recorded cash flow. Complete group identifiers and all primary group results are supplied separately. The selection includes negative-flow cases.\end{minipage}
\end{table}

\FloatBarrier
\section{Token-level diagnostic examples}
The first two examples are qualifying tokens from an early positive-flow group and a later negative-flow group. The latter two are detector contrasts. These deterministic examples illustrate observables and screening boundaries, not a validated sample of fraudulent or legitimate tokens.
\begin{figure}[htbp]\centering\includegraphics[width=.80\linewidth]{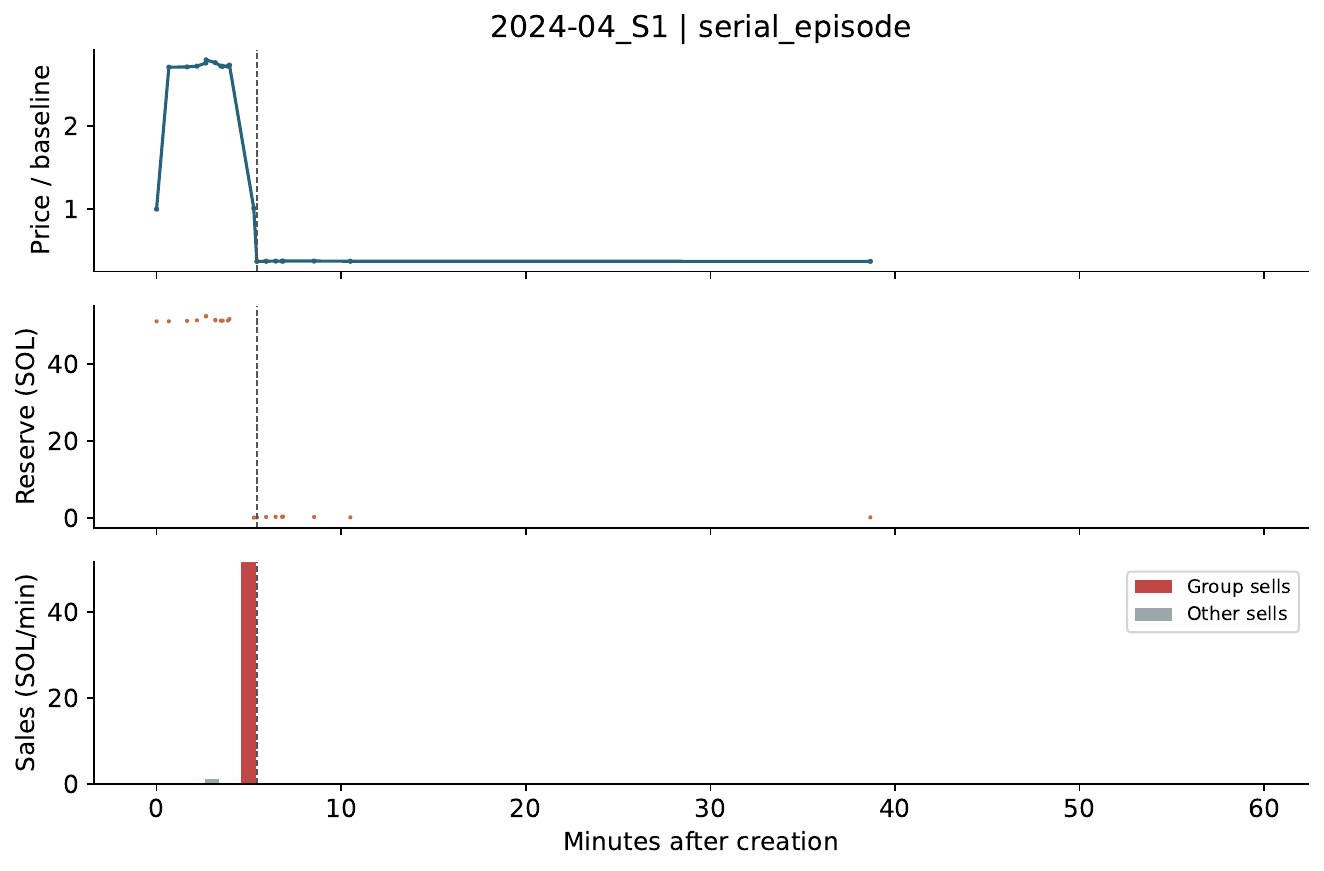}
\caption{2024-04\_S1: serial episode. Price lines connect observed valid candles without establishing prices between observations; reserve values are recorded measurements. Dashed lines mark primary episode drops when a primary label is present. For unlabelled and ambiguity contrasts, the absence of a line is not a zero-price observation. Mint prefix: FNzChPgu.}\label{fig:diagnostic-data_2024_04_10_11-serial_episode}\end{figure}

\FloatBarrier

\begin{figure}[htbp]\centering\includegraphics[width=.80\linewidth]{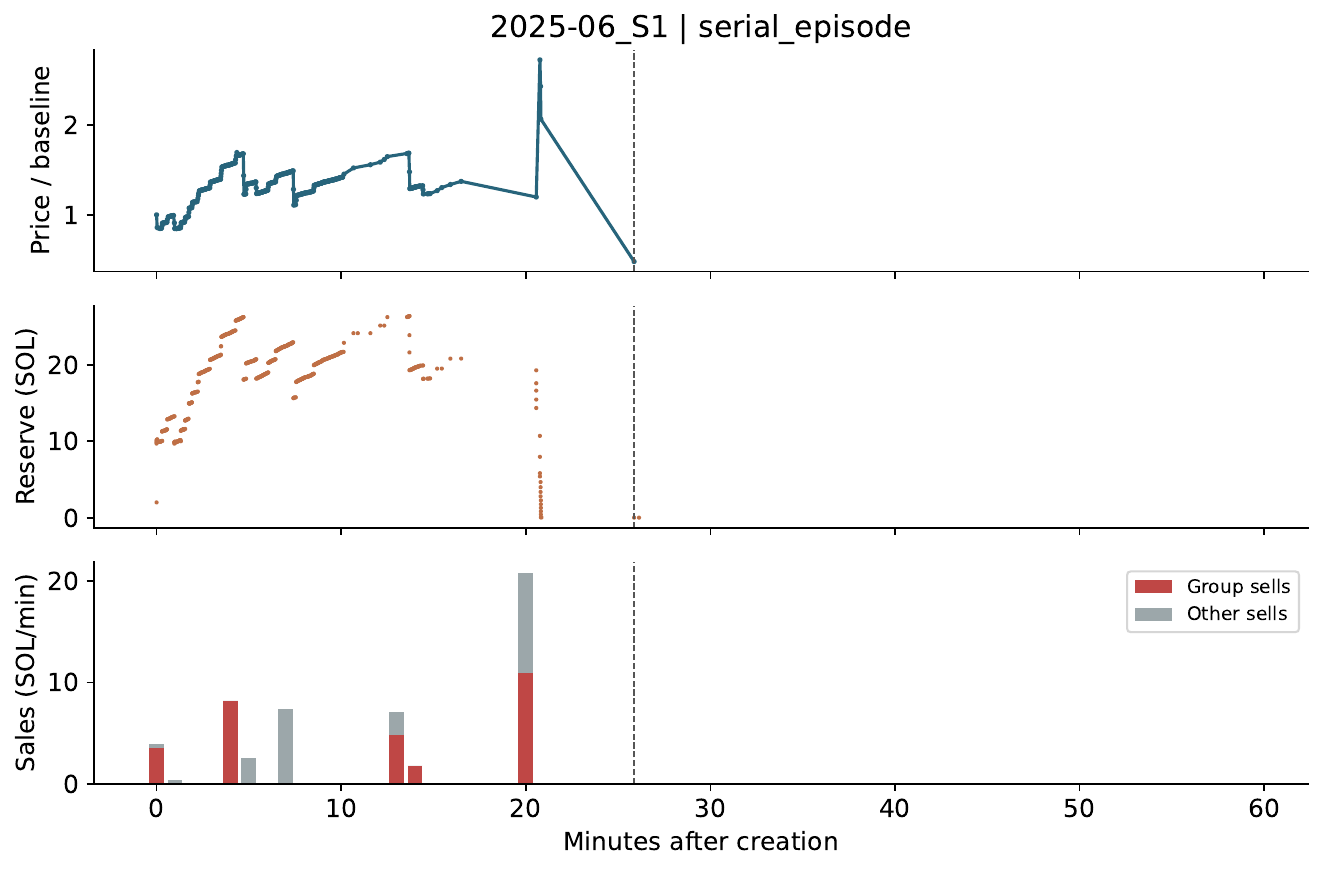}
\caption{2025-06\_S1: serial episode. Price lines connect observed valid candles without establishing prices between observations; reserve values are recorded measurements. Dashed lines mark primary episode drops when a primary label is present. For unlabelled and ambiguity contrasts, the absence of a line is not a zero-price observation. Mint prefix: 9takZxuh.}\label{fig:diagnostic-data_2025_06_21_low-serial_episode}\end{figure}

\FloatBarrier

\begin{figure}[htbp]\centering\includegraphics[width=.80\linewidth]{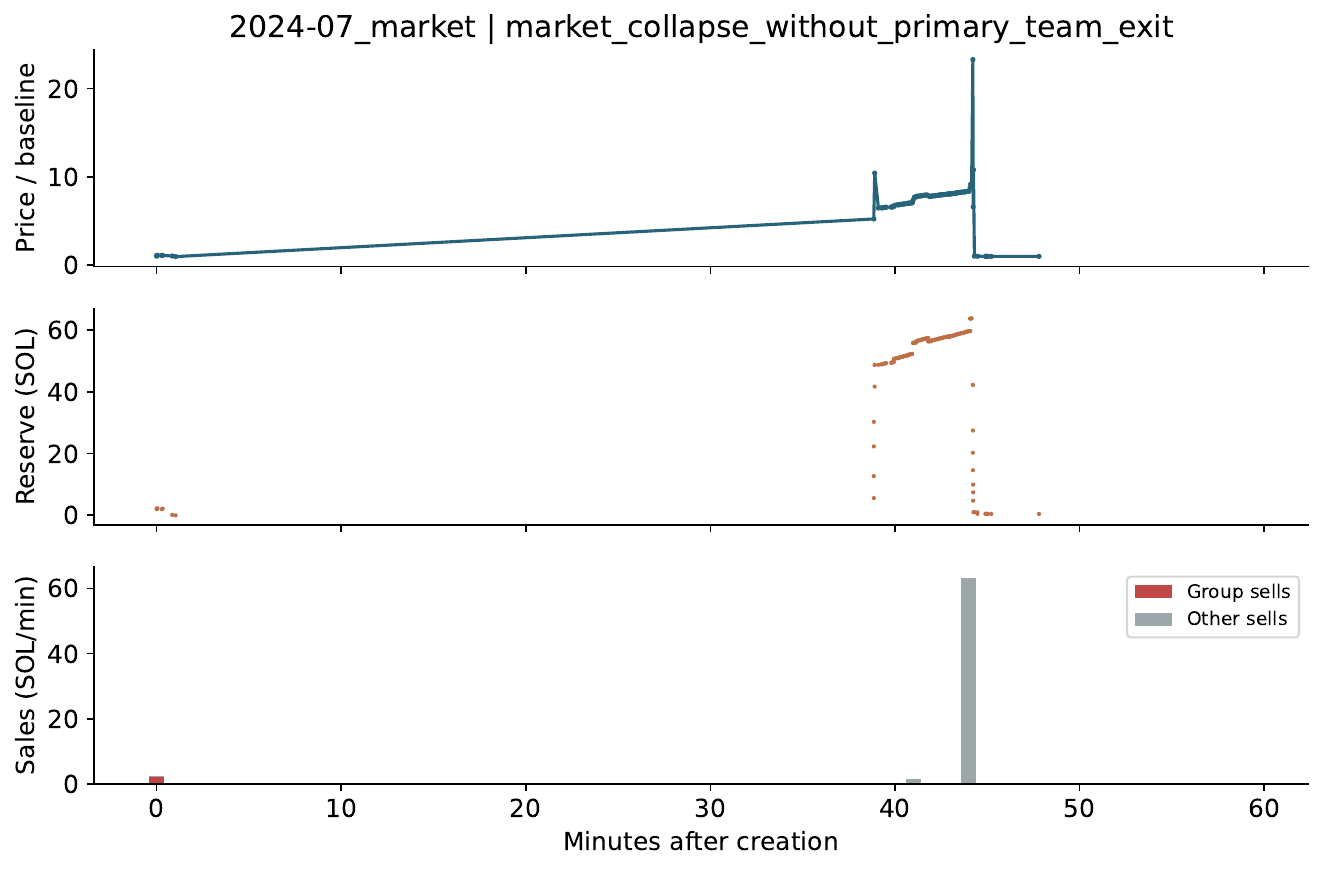}
\caption{2024-07\_market: market collapse without primary team exit. Price lines connect observed valid candles without establishing prices between observations; reserve values are recorded measurements. Dashed lines mark primary episode drops when a primary label is present. For unlabelled and ambiguity contrasts, the absence of a line is not a zero-price observation. Mint prefix: 9BEronw7.}\label{fig:diagnostic-data_2024_07_10_12-market_collapse_without_primary_team_exit}\end{figure}

\FloatBarrier

\begin{figure}[htbp]\centering\includegraphics[width=.80\linewidth]{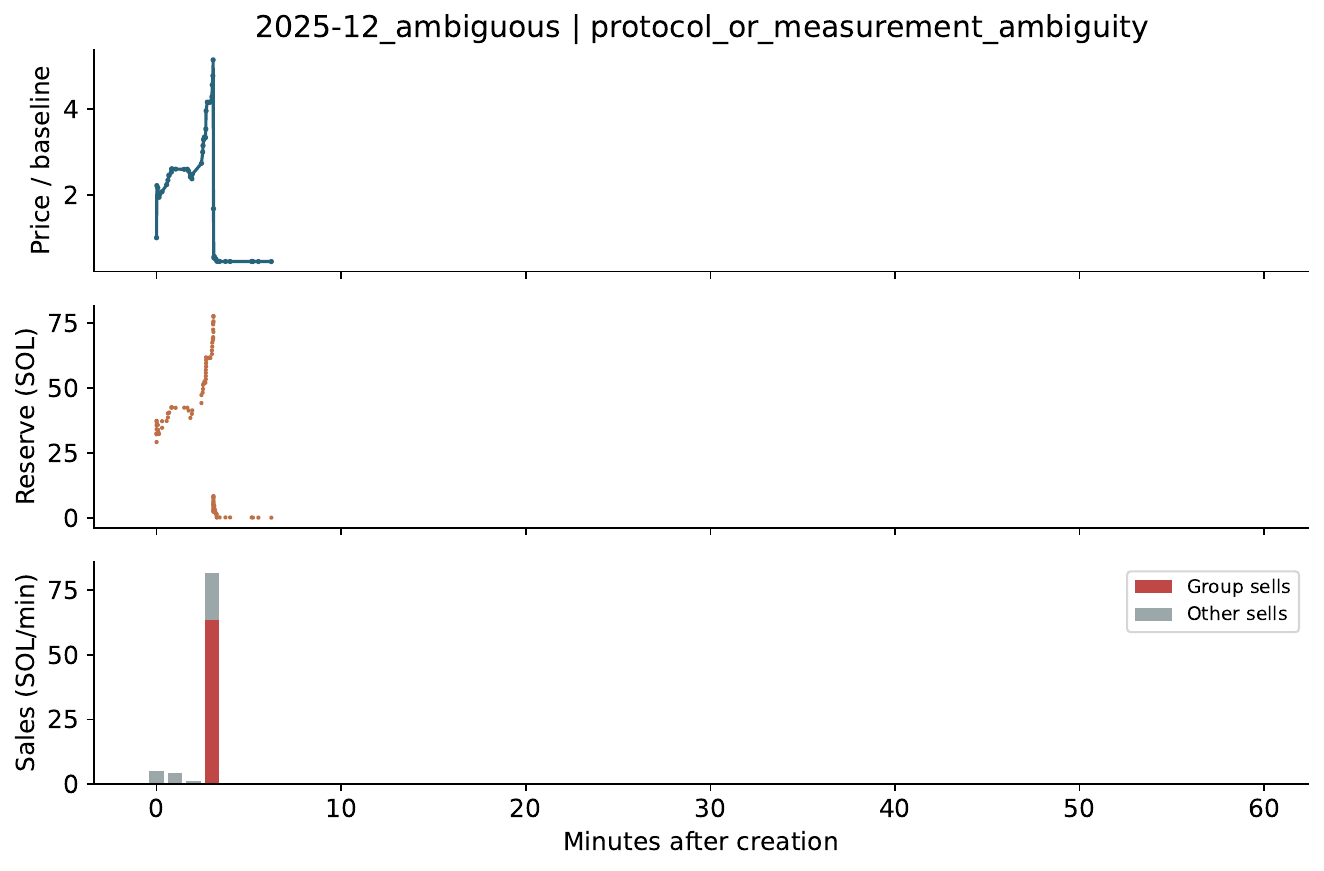}
\caption{2025-12\_ambiguous: protocol or measurement ambiguity. Price lines connect observed valid candles without establishing prices between observations; reserve values are recorded measurements. Dashed lines mark primary episode drops when a primary label is present. For unlabelled and ambiguity contrasts, the absence of a line is not a zero-price observation. Mint prefix: FosbN6wP.}\label{fig:diagnostic-data_2025_12_12-protocol_or_measurement_ambiguity}\end{figure}

\FloatBarrier

\section{Audit sample protocol}
The audit sample contains 200 primary episodes drawn with a fixed seed (100 from serial groups, 100 from other groups), with the group sale share, the creator's sales inside the window and a chart of the trades of each token (green buys, red sells; grey and black lines mark peak and drop). A reviewer assigns 1 (team exit pattern with a plausible rug-pull-like mechanism), 0 (other or benign) or ? (cannot tell), and precision is reported with Wilson 95\% intervals for all, serial and other episodes. Labels should be assigned blind to the group and window.

\end{document}